\documentclass[%
 amsmath,
 amssymb,
 showkeys,
 pra,
 aps,
 floatfix,
 twocolumn,
 10pt,longbibliography,nofootinbib
]{revtex4-2}

\usepackage{xcolor}
\definecolor{WeakGreen}{HTML}{52BD7C}
\definecolor{Red}{HTML}{D7333B}
\definecolor{Green}{HTML}{00973E}
\definecolor{Orange}{HTML}{DE8900}
\definecolor{Purple}{HTML}{9B3992}
\definecolor{XanaBlue}{HTML}{4D53C8}
\definecolor{LightBlue}{HTML}{44ACE8}

\usepackage[colorlinks=true,linkcolor=Orange,citecolor=Red,urlcolor=Red]{hyperref}
\hypersetup{
    pdftitle={Parameter-optimal unitary synthesis with flag decompositions},
    pdfauthor={Korbinian Kottmann, David Wierichs, Guillermo Alonso-Linaje, Nathan Killoran},
}
\usepackage{footnotebackref}
\usepackage[capitalize]{cleveref}
\usepackage{booktabs}
\usepackage{algorithm}
\usepackage{algpseudocode}
\usepackage{array}
\usepackage{graphicx} 
\usepackage{here}
\usepackage{bbold}
\usepackage{tikz}
\usepackage{quantikz}
\usepackage{amsthm}
\usepackage{thmtools,mathtools}
\usepackage{thm-restate}
\usepackage[justification=centerlast]{caption}
\usetikzlibrary{circuits.logic.US}
\usetikzlibrary{shapes.geometric}
\usepackage{comment}
\usepackage{nicematrix}
\usepackage{fontspec}

\tikzset{
    diagonal/.style={
        draw,
        diamond,
        minimum width=1.67em,
        inner sep=0.5pt,
    }
}
\DeclareExpandableDocumentCommand{\diagonal}{O{}{m}}{|[diagonal]| {#2} \qw}

\newcommand{\multidiagonal}[2]{\gate[#1,style={color=white}]{#2}\gategroup[#1, steps=1,style={rounded corners=0.3cm, inner xsep=-3.5pt, inner ysep=-5.2pt}]{}}

\newcommand{\cnot}{\operatorname{CNOT}}
\newcommand{\cy}{\operatorname{CY}}
\newcommand{\cz}{\operatorname{CZ}}

\newcommand{\smallflag}{\includegraphics[width=0.2cm]{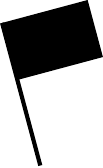}}
\newcommand{\mltplx}[1]{\ctrl[style={shape=rectangle, fill=white}]{#1}}

\crefname{thm}{Thm.}{Thms.}

\crefname{lemma}{Lemma}{Lemmas}

\crefname{definition}{Def.}{Defs.}

\crefname{remark}{Remark}{Remarks}

\crefname{prop}{Prop.}{Props.}

\crefname{algorithm}{Alg.}{Algs.}
\crefname{appendix}{App.}{Apps.}
\crefname{section}{Sec.}{Secs.}
\crefname{table}{Tab.}{Tabs.}

\newcommand*{\id}{\mathchoice
  {\openone}
  {\openone}
  {\scalebox{.7}{\openone}} 
  {\scalebox{.5}{\openone}} 
}

\renewcommand{\det}[1]{\mathrm{det}(#1)}
\newcommand{\Ad}{\operatorname{Ad}}

\tikzset{
    glow/.style={%
    preaction={#1, draw, line join=round, line width=0.5pt, opacity=0.04,
    preaction={#1, draw, line join=round, line width=1.0pt, opacity=0.04,
    preaction={#1, draw, line join=round, line width=1.5pt, opacity=0.04,
    preaction={#1, draw, line join=round, line width=2.0pt, opacity=0.04,
    preaction={#1, draw, line join=round, line width=2.5pt, opacity=0.04,
    preaction={#1, draw, line join=round, line width=3.0pt, opacity=0.04,
    preaction={#1, draw, line join=round, line width=3.5pt, opacity=0.04,
    preaction={#1, draw, line join=round, line width=4.0pt, opacity=0.04,
    preaction={#1, draw, line join=round, line width=4.5pt, opacity=0.04,
    preaction={#1, draw, line join=round, line width=5.0pt, opacity=0.04,
}}}}}}}}}}}}

\begin{document}
\title{Parameter-optimal unitary synthesis with flag decompositions}
\author{Korbinian Kottmann}
\author{David Wierichs}
\author{Guillermo Alonso-Linaje}
\author{Nathan Killoran}
\affiliation{Xanadu, Toronto, ON, M5G 2C8, Canada} 

\begin{abstract}
    We introduce the \textit{flag decomposition} as a central tool for unitary synthesis.
    It lets us carve out a diagonal unitary with $2^n$ degrees of freedom in such a way that the remaining \textit{flag circuit} is parametrized by the optimal number of $4^n-2^n$ rotations.
    This enables us to produce parameter-optimal quantum circuits for generic unitaries and matrix product state preparation. Our approach improves upon the state of the art, both when compiling down to the \{Clifford + Rot\} gate set with what we call \textit{selective de-multiplexing}, and when using phase gradient resource states together with quantum arithmetic to implement multiplexed rotations.
    All of our synthesis methods are efficiently implementable in terms of recursive Cartan decompositions realized by standard linear algebra routines, making them applicable to all practically relevant system sizes.
\end{abstract}

\maketitle

\section{Introduction}
Unitary synthesis is an important subroutine in quantum compilation. It allows us to decompose arbitrary matrices of modest size into a sequence of gates that can be executed on a quantum computer.
Leading techniques for (numerically) exact unitary synthesis are based on recursive Cartan decompositions~\cite{shende2006synthesis,Khaneja-Glaser,vartiainen2003efficient,tucci1999rudimentary,krol2021efficient,De-Vos-De-Baerdemacker,mansky2023near,Krol-Al-Ars,wierichs2025recursive}, including the widely-used quantum Shannon decomposition (QSD)~\cite{shende2006synthesis} and most advanced variant thereof~\cite{Krol-Al-Ars}.
These techniques have been developed with a focus on minimizing the required number of static entanglers like $\cnot$ or $\cz$ gates.
However, non-Clifford gates such as $T$ and Toffoli gates dominate the cost in fault-tolerant quantum computing (FTQC)~\cite{Fowler2012,Litinski2019}. 

We revisit unitary synthesis from the perspective of minimizing parametrized rotation gates, which are the main contributors of non-Clifford gates. Each rotation gate can be either discretized into a series of \{Clifford + T\} operations \cite{Dawson2005solovaykitaevalgorithm,ross2016gridsynth,Bocharov2015RUS,Kliuchnikov2023},  or realized via a combination of phase gradient resource states, quantum read-only memory (QROM) \cite{LKS2024}, and bit-wise quantum addition using Toffoli gates \cite{Gidney2018halving}\footnote{We refer to this as the phase gradient decomposition (see \cref{sec:main_flag_decomposition} for details) and note that this approach is sometimes also simply referred to as QROM in the literature.}. We develop unitary synthesis schemes that improve upon the best known resource counts for either decomposition strategy. An overview of the methods is provided in \cref{fig:one}. The crucial ingredient is the parameter optimality, which is unlocked by the \textit{flag decomposition}. This decomposition lets us split a unitary $V$ into a diagonal operator $\Delta$ (including a global phase) and its ``remainder" $\smallflag$ in a parameter-optimal fashion:

\begin{equation*}
\resizebox{.75\linewidth}{!}{%
\includegraphics{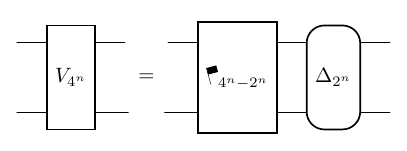}
}.
\end{equation*}

The remainder $\smallflag$ of the unitary lives on a so-called complete flag manifold of dimension $4^n-2^n$, complementing the diagonal gate that has $2^n$ parameters, as indicated by the subscripts in the circuit diagram. Thus, the optimum of $4^n$ parameters\footnote{We consider the full unitary group $U(2^n)$, including a global phase, throughout this manuscript.} for general unitary matrices is maintained. The elements of the (complete) flag manifold are equivalence classes, where equivalence is defined as equality up to left multiplication with a diagonal phase matrix. The particular representative of each class will be fixed through the decomposition of the flag circuit itself, which we construct recursively from the two known base cases for one and two qubits in \cref{sec:flag_decomposition}. It turns out that the resulting circuit is equivalent to an old result in~\cite[Fig.~3]{Mottonen2004quantumgeneralmultiqubit} 
that already achieved parameter-optimality via recursive cosine-sine decompositions (CSDs) but has been underappreciated in the literature since.
The flag decomposition is a valid unitary synthesis scheme on its own and well-suited for the phase gradient decomposition, as discussed in \cref{sec:flag_decomposition}. Further, we use it as a subroutine in \cref{sec:selective_de_multiplexing} to derive \textit{selective de-multiplexing} (SDM), a parameter-optimal decomposition that builds on the QSD and improves upon the $\cnot$ gate count compared to~\cite{Mottonen2004quantumgeneralmultiqubit,bergholm2005quantumuniformlycontrolled} (see \cref{tab:known_synthesis_techniques}).

These parameter-optimal unitary synthesis schemes have direct implications for matrix product state (MPS) preparation, which is a crucial subroutine in many quantum algorithms~\cite{fomichev2024initial,LKS2024,berry2024rapid}.
There is a wide variety of preparation algorithms for MPS of different flavors~\cite{Schoen2005,Lu2022,Smith2024,Wierichs2024ConstantDepth,Wei2023,Malz2024}.
In these preparation algorithms, the crucial subroutine that determines the overall cost of the circuit is unitary synthesis. In \cref{sec:mps_state_prep}, we tailor our synthesis techniques to become parameter-optimal with respect to the manifold spanned by the MPS matrices. In particular, we ensure that the reductions due to the isometric properties of MPS matrices, as well as the gauge freedom of MPS, is respected by the constructed circuits. We do this specifically for the MPS preparation scheme from~\cite{Schoen2005}, but note that the employed tricks apply to any MPS preparation scheme. Additionally, the reductions in the flag decompositions together with the optimizations for the MPS preparation circuits lead to improvements over the currently best-known quantum resources in terms of Toffoli gates in~\cite{berry2024rapid} when targeting phase gradient decompositions.

Implementations of all our synthesis techniques using SciPy~\cite{virtanen2020scipy} and PennyLane~\cite{pennylane} can be found in~\cite{our_repo}.


\section{Results overview}

\begin{figure*}
    \centering
    \includegraphics[width=0.9\linewidth]{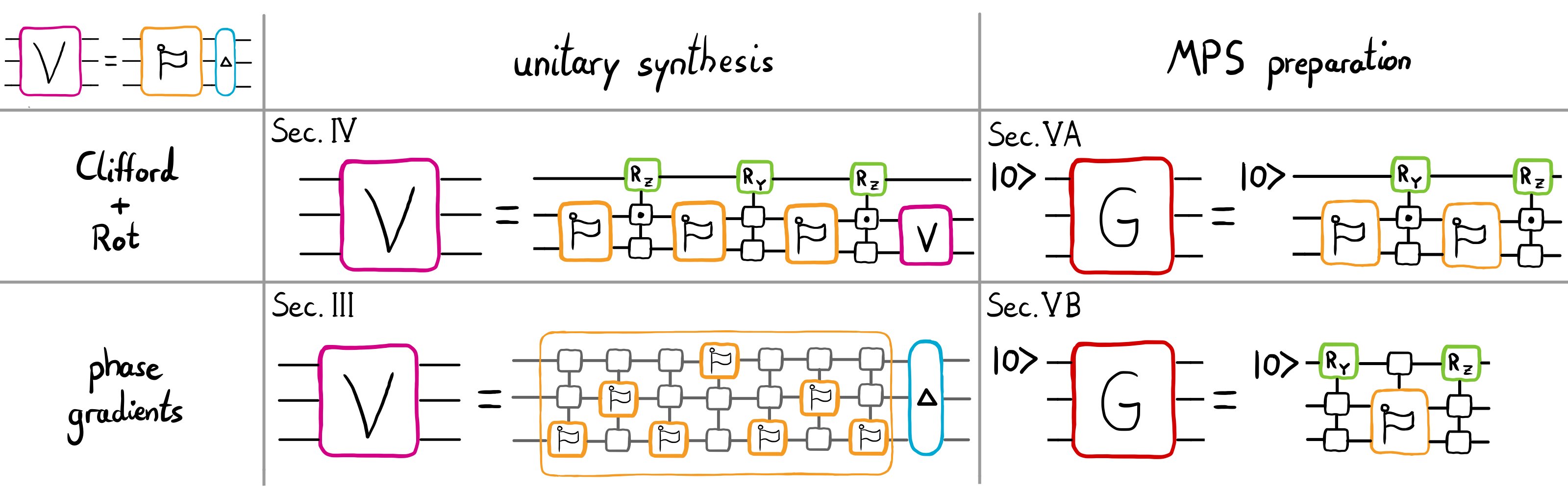}
    \caption{In this work we discuss the so-called flag decomposition (top left) as a central tool for parameter-optimal unitary synthesis. It decomposes a unitary (magenta) with $4^n$ parameters into a flag circuit (orange) with $4^n-2^n$ parameters and a diagonal (blue) with $2^n$ parameters. Using this tool, we derive parameter-optimal synthesis results for unitaries in general (left column) and MPS matrices (right column). The shown circuit skeletons can then be decomposed further into a target gate set. We consider \{Clifford + Rot\} decompositions (top row), as well as phase gradient decompositions (bottom row). The multiplexer node with a dot inside refers to a symmetrized Möttönen decomposition; see \cref{eq:symmetrized_Möttönen}.\hspace*{\fill}}
    \label{fig:one}
\end{figure*}

We derive the flag decomposition and use it to design parameter-optimal unitary synthesis methods. As indicated in \cref{fig:one}, we differentiate two distinct decomposition methods:
\begin{itemize}
    \item \{Clifford + Rot\} decomposition: Decomposing the circuit skeleton into Clifford gates and rotations.
    \item Phase gradient decomposition: Decomposing the circuit skeleton into a QROM and adders using many auxiliary qubits and a reusable phase gradient resource state. 
\end{itemize}
We tailor our schemes to each decomposition method for general-purpose unitary synthesis. Further, we adapt the optimized synthesis circuits to MPS preparation to optimally take advantage of the inherent reduced dimensionality of the tensors in an MPS. These careful design choices lead to improvements upon the best known gate counts in all cases considered in \cref{fig:one}. We continue below by providing an overview of the results for the two decomposition schemes and MPS preparation.

\subsection{\{Clifford + Rot\} decomposition}
For noisy intermediate-scale quantum computing (NISQ), where CNOT gates are the main contributors to noise and therefore the costliest resource, the quantum Shannon decomposition (QSD)~\cite{shende2006synthesis} and variants thereof like the Block-ZXZ decomposition~\cite{Krol-Al-Ars} provide the best unitary synthesis technique with the lowest CNOT gate count (see \cref{tab:known_synthesis_techniques}).

For FTQC, we prioritize optimality in terms of the number of parametrized rotation gates. This was already achieved by an underappreciated decomposition in 2004~\cite{bergholm2005quantumuniformlycontrolled}, using a recursive cosine-sine decomposition (CSD). In terms of CNOT cost, the best QSD variant has a leading-order factor of $\tfrac{22}{48} 4^n$ CNOTs as indicated in \cref{tab:known_synthesis_techniques}, whereas \cite{bergholm2005quantumuniformlycontrolled} has a leading-order factor of $\tfrac{1}{2} 4^n$ CNOTs (see \cref{sec:recursive_csd}). Hence, \cite{bergholm2005quantumuniformlycontrolled} is parameter-optimal while at the same time being only $\tfrac{1}{24}4^n$ away from the best-known CNOT count.

We improve upon this result by introducing what we call selective de-multiplexing (SDM) with the resources indicated in \cref{tab:known_synthesis_techniques}. It provides the lowest CNOT count while at the same time being parameter-optimal.

The resulting SDM circuit has the following recursive structure,
\begin{equation}\label{sec2:POQSD}
\resizebox{\linewidth}{!}{%
\includegraphics{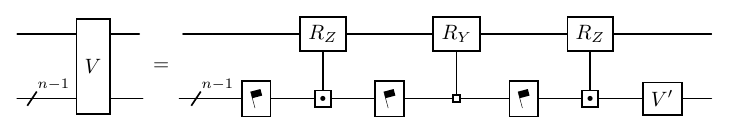},
}
\end{equation}
where each $\smallflag$ indicates the reduction of a $(n-1)$-qubit unitary to the so-called complete flag manifold with $4^{n-1} - 2^{n-1}$ parameters, making the circuit overall parameter-optimal. The multiplexer node with a dot denotes a symmetrized multiplexer decomposition that removes one entangling gate; see \cref{eq:symmetrized_Möttönen}.
The full derivation with all optimization tricks to reduce the CNOT count is provided in \cref{sec:selective_de_multiplexing}. A prerequisite of selective de-multiplexing is the flag decomposition, which we derive in \cref{sec:flag_decomposition}. Its dense multiplexer structure is well-suited for phase gradient decompositions, as we discuss next.

\begin{table}
    \centering
    \bgroup
    \def\arraystretch{1.5}
    \begin{tabular}{lccc}
        Reference & $R_Y/R_Z$ count & $\cnot$ count \\
        \hline
        QR~\cite{vartiainen2003efficient} '03 & $\lesssim 25\cdot 4^n$ & $\approx 8.7\cdot 4^n$ \\
        CSD~\cite{tucci1999rudimentary,krol2021efficient} '99 & $\tfrac{3}{2}\cdot 4^n - \tfrac{1}{2} \ 2^n$ & $\tfrac{n}{2}\cdot 4^n-\tfrac{1}{2}\cdot 2^{n}$ \\
        QSD~\cite{De-Vos-De-Baerdemacker} '15 & $\tfrac{3}{2}\cdot 4^n-\tfrac{3}{2}\cdot 2^n$ & $\tfrac{3}{4}\cdot 4^n - \tfrac{3}{2}\cdot 2^n$ \\
        QSD~\cite{wierichs2025demo_unitary} '25 & $\tfrac{21}{16}\cdot 4^n-\tfrac{3}{2}\cdot 2^n$ & $\tfrac{9}{16}\cdot 4^n - \tfrac{3}{2}\cdot 2^n$ \\
        QSD~\cite{Krol-Al-Ars} '24 & $\tfrac{5}{4}\cdot 4^n - \tfrac{3}{2}\cdot 2^n+1$ & \color{Green}{$\tfrac{22}{48}\cdot 4^n - \tfrac{3}{2}\cdot 2^n + \tfrac{5}{3}$} \\
        CSD~\cite{Mottonen2004quantumgeneralmultiqubit} '04 & \color{Green}{$4^n-1$} & $4^n - 2\cdot 2^n$ \\
        CSD~\cite{bergholm2005quantumuniformlycontrolled} '04 & \color{Green}{$4^n-1$} & $\tfrac{1}{2}\cdot 4^n - \tfrac{1}{2}\cdot 2^n -1$\\
        \textbf{\smallflag-decomp} & \color{Green}{$\mathbf{4^n-1}$} & $\mathbf{\tfrac{1}{2}\cdot 4^n - \tfrac{3}{4}\cdot 2^n -1}$ \\
        \textbf{SDM} & \color{Green}{$\mathbf{4^n-1}$} & $\mathbf{\tfrac{1}{2}\cdot4^n - \tfrac{3}{8}(n+2)\cdot2^n+n-1}$ \\
        BWC \cite{wierichs2025unitarysynthesisoptimalbrick} '25 & $4^n-1$ & $\left\lceil\tfrac{1}{4}\cdot4^n-\tfrac{3n-1}{4}\right\rceil$ \\ \hline
        Optimum & $4^n-1$ & $\left\lceil\tfrac{1}{4}\cdot4^n-\tfrac{3n-1}{4}\right\rceil$
    \end{tabular}
    \egroup
    \caption{Known auxiliary-free unitary synthesis techniques into the \{Clifford + Rot\} gate set $\{R_Y,R_Z, \cnot \}$, as well as the contributions of this paper: the flag decomposition ($\smallflag$-decomp), best suited for phase gradient decompositions, and selective de-multiplexing (SDM), which achieves parameter-optimality and $\cnot$ counts close to those in \cite{Krol-Al-Ars}. Based on~\cite[Tab.~1]{Krol-Al-Ars}. 
    The rows are sorted by decreasing $R_Y/R_Z$ count and then by decreasing $\cnot$ count (for $n>3$), with lowest achieved counts marked in green. We refer to the Block-ZXZ decomposition \cite{Krol-Al-Ars} as quantum Shannon decomposition (QSD) \cite{shende2006synthesis} as they only differ by a choice of subalgebra~\cite{wierichs2025recursive}, and all considerations can be transferred between them, leading to the same gate counts. For brevity, we leave out gate counts for Khaneja-Glaser decompositions derived in \cite{mansky2023near,wierichs2025demo_unitary}. Although we consider the full unitary group $U(2^n)$ with $4^n$ degrees of freedom (including a global phase), the optimal number of $R_Y$/$R_Z$ is $4^n-1$. The lower bound for the CNOT gates in the last row~\cite{shende2003minimal} is attained by the brick wall circuits (BWC) in \cite{wierichs2025unitarysynthesisoptimalbrick}. However, this comes at a significant classical computational cost, practically restricting the method to few qubits and hence rendering it unsuitable for MPS preparation.\hspace*{\fill}}
    \label{tab:known_synthesis_techniques}
\end{table}

\subsection{Phase gradient decomposition}\label{sec:sec2phase_gradients}

We consider the phase gradient decomposition that realizes the multiplexer of two consecutive $Z$ rotation gates, potentially with an intermediate basis change given by $A$ and $A^\dagger$:

\begin{align*}
\resizebox{\linewidth}{!}{%
\includegraphics{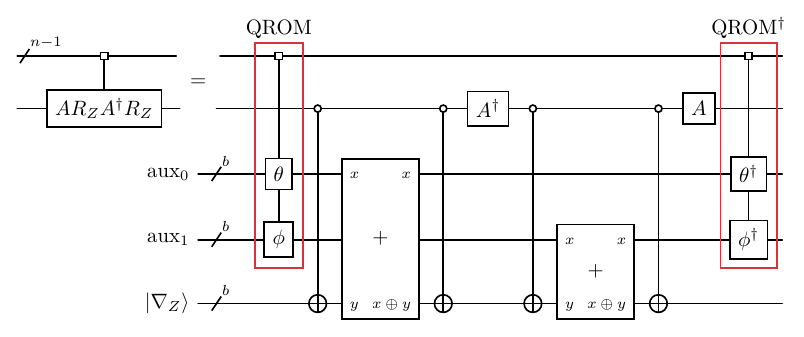}
}.
\end{align*}
The first operation is a QROM~\cite{LKS2024} that loads the digitized angles (with $b$ decimals) of the two rotations, $\theta$ and $\phi$, on two auxiliary qubit registers. As detailed in \cref{sec:main_flag_decomposition}, these rotations are then performed via phase kickback when arithmetically adding the loaded angles with the phase gradient register $|\nabla_Z\rangle$.

Such phase gradient decompositions are particularly well-suited for a dense arrangement of multiplexed rotation gates. 
This fact is used in~\cite{berry2024rapid}, where the authors use a unitary synthesis skeleton based on QR decomposition, taken from optical interferometry~\cite{Clements16,Pastor21}. 
The resulting circuit is a series of phase gradient blocks that realize two consecutive multiplexed phase shift gates, interleaved with incrementer and decrementer operations ($\pm1$)\footnote{The very last decrementer can be removed; not shown here explicitly.}, and a final diagonal gate ($\Delta)$:
\begin{align}\label{eq:berry_phase_gradient}
\resizebox{\linewidth}{!}{%
\includegraphics{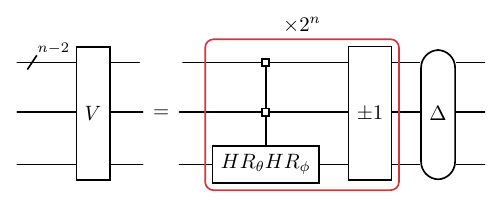}
}.
\end{align}
In this work, we apply phase gradient decompositions to the flag decomposition circuit instead. This entirely eliminates the need for $\pm1$ blocks via a Gray-code ordering. Additionally, it removes one of the phase gradient blocks such that the circuit optimally makes use of multiplexed rotation gates:
\begin{align}\label{eq:flag_decomp_phase_gradient}
\resizebox{\linewidth}{!}{%
\includegraphics{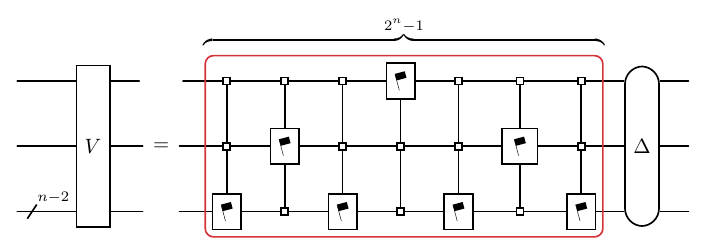}.
}
\end{align}
The flag circuit $\smallflag$ on a single qubit is just a $R_Y R_Z = H_YR_Z H_Y^\dagger R_Z$ block\footnote{Note that we denote the order of the unitary matrices here, which is reversed compared to gates in circuit diagrams.}, where $H_Y = S H$, which makes it easy to compare the circuits in \cref{eq:berry_phase_gradient} and \cref{eq:flag_decomp_phase_gradient}. In particular, we can see that we save one multiplexer and all incrementer and decrementer operations.

\subsection{Matrix product state preparation}

The flag-based unitary synthesis strategy, with either decomposition method, directly translates to improved resources for matrix product state preparation, a crucial subroutine for many quantum algorithms. An MPS on $L$ qubits is defined in terms of its tensors $\{A^{\sigma_j}\}_{j=1}^L$ that are arranged on a one-dimensional chain \cite{Schollwock2011MPS} (see \cref{sec:mps_state_prep} for more):

\begin{equation}\label{eq:finite_dim_MPS}
\resizebox{\linewidth}{!}{%
\includegraphics{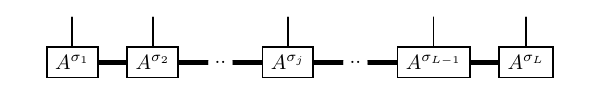}
}.
\end{equation}
The thick horizontal connections are limited in size to what is called the maximum \textit{bond dimension} $\chi$\footnote{Note that the tensors at the boundaries have slightly different dimensions. We treat these effects separately in \cref{sec:boundary-effects}.}. This bond dimension is a hyper-parameter that determines the degree of approximation of the state. We choose it to be $\chi=2^n$ such that those degrees of freedom can fit on a register of $n$ auxiliary qubits.

Each of the tensors defines an isometry between $n$ and $n+1$ qubits. These, in turn, can be completed into unitaries $G_j$ on $n$ auxiliary qubits and an additional one that is initialized in $|0\rangle$:

\begin{equation}\label{eq:mps_as_circuit_diagram}
\resizebox{\linewidth}{!}{%
\includegraphics{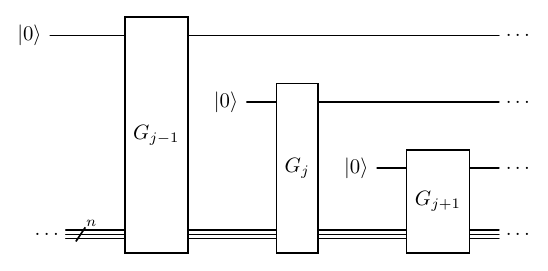}
}.
\end{equation}
The horizontal connections in \cref{eq:finite_dim_MPS} are the so-called \textit{virtual bonds}. They have a direct correspondence to the $n$ auxiliary qubits in  \cref{eq:mps_as_circuit_diagram}, which are shared between all the unitaries $G_j$. The thin vertical lines in \cref{eq:finite_dim_MPS} are the so-called physical bonds of dimension $d=2$ for qubits. The fixed input state $|0\rangle$ in \cref{eq:mps_as_circuit_diagram} can be interpreted as a second physical bond that is trivially of dimension $1$.

We can use our new unitary synthesis techniques for the $G_j$ unitaries with $4^{n+1}=4\chi^2$ real parameters each. Due to the specific structure of MPS circuits, we can perform additional optimizations. These manifest in two crucial reductions:
\begin{itemize}
    \item We use the fixed input state $|0\rangle$ to simplify the circuit structure by removing multiplexer nodes and $R_Z$ rotations, effectively removing $\chi^2$ parameters. This reduction stems from the fact that the tensors $A^{\sigma_j}$ define isometries with $3\chi^2$ parameters.
    \item We can merge a unitary with $\chi^2$ parameters on the shared auxiliary $n$-qubit register from one isometry into its neighbor. This corresponds to the so-called gauge degree of freedom of MPS; see \cref{sec:mps_state_prep} and \cite[Sec. 4.1.3, iv)]{Schollwock2011MPS}.
\end{itemize}
Overall, this leaves $2\chi^2$ as the optimal number of parameters for each of the $G_j$ in the MPS, which we achieve with our parameter-optimal circuits indicated in \cref{fig:one} and discussed in detail in \cref{sec:mps_state_prep}. The structural improvements of our unitary synthesis circuits and the parameter-optimality with respect to the MPS manifold lead to improved resources compared to the current state of the art in~\cite{berry2024rapid}. Further reductions due to boundary effects that have not been previously discussed are provided in \cref{sec:boundary-effects}.

\section{Flag decomposition}
\label{sec:flag_decomposition}
In this section we will derive the so-called \textit{flag decomposition} of a unitary $V$ into a diagonal unitary $\Delta$ and a \textit{flag circuit} $\smallflag$. It serves as an essential building block for all the (recursive) parameter-optimal decompositions illustrated in \cref{fig:one}. Further, it improves upon the state of the art in terms of Toffoli counts when utilizing a phase gradient decomposition.
We note that the circuit resulting from the flag decomposition derived in this section is equivalent to the circuit obtained from a recursive CSD by Bergholm et al.~\cite{bergholm2005quantumuniformlycontrolled}, which we discuss in detail in \cref{sec:recursive_csd}.

We will show in the following that any unitary $V \in U(2^n)$ with $4^n$ degrees of freedom can be factorized into a ``flag circuit" with $4^n-2^n$ degrees of freedom and a diagonal matrix with $2^n$ degrees of freedom:

\begin{equation}\label{eq:flag_decomposition}
\resizebox{.75\linewidth}{!}{%
\includegraphics{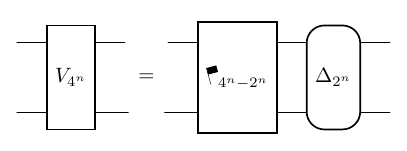}.
}
\end{equation}
The component before the diagonal lives in a so-called complete flag manifold\footnote{We shall refer to it as just the flag manifold for brevity.} 

\begin{equation}
    \smallflag_{4^n-2^n} \in U(2^n)/\mathbb{T}^{2^n},
\end{equation}
where $\mathbb{T}$ refers to the maximal torus, i.e., the maximal Abelian subgroup of diagonal matrices. Each element of the manifold thus is an equivalence class, rather than a specific unitary matrix. A representative of the class can be chosen by fixing one phase per row. For example, we could force the flag unitary to have a real diagonal. Instead, we will fix the representative through the \textit{decomposition} of the flag circuit itself, building on the notion of unitary synthesis ``up to a diagonal". Note, that this makes the representative depend on the particular decomposition that is chosen, leading to a dependency on the target gate set.
Alternatively, we can view it as the manifold generated as $\exp(\mathfrak{h})$, where $\mathfrak{h}$ is spanned by the \textit{hollow} matrices with no diagonal entries\footnote{These hollow matrices form the root spaces of $\mathfrak{u}(2^n)$, so we have the root-space decomposition $\mathfrak{u}(2^n) = \mathfrak{t} \oplus \mathfrak{h} = \mathfrak{t} \oplus \bigoplus_\alpha g_\alpha$, where $\mathfrak{t}$ is the Lie algebra of the maximal torus $\mathbb{T}$.}.

Note that throughout this work, we will consider the unitary group $U(2^n)$ as the total group, including a global phase. This simplifies the presented recursive constructions, as it allows us to attach multiplexing nodes without having to manually track ``global" phases that become local phases in the process. Removing the genuinely global phase at the end of the synthesis is a simple task, reducing the parameter count (but not the number of rotation gates) by one.

The simplest flag decomposition for a single qubit comes in the form of the Euler decomposition,

\begin{align}\label{eq:flag_decomp_n-1}
\resizebox{0.9\linewidth}{!}{%
\includegraphics{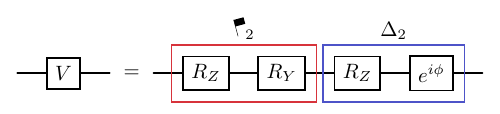}
},
\end{align}
where we interpret the first two gates as the flag circuit with $4^1-2^1=2$ parameters, and the latter two form a diagonal unitary with the remaining $2$ parameters.

For two qubits, we obtain a flag decomposition by modifying Thm. 14 in~\cite{shende2006synthesis}:
\begin{align}\label{eq:flag_decomp_n-2}
\resizebox{\linewidth}{!}{%
\includegraphics{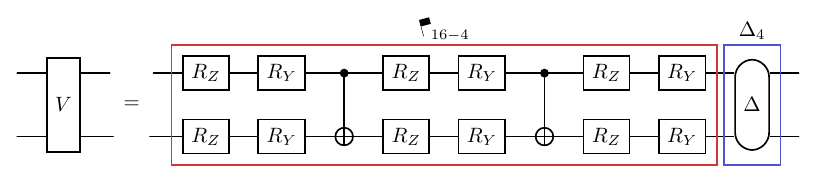}.%
}
\end{align}
Compared to~\cite[Thm.~14]{shende2006synthesis}, we extracted the diagonal to the right end of the circuit, and applied Euler decompositions to all $SU(2)$ gates to reduce the parameter count and distribute the remaining gates into a more regular structure. We note also that the flag circuit here contains six smaller single-qubit flag circuits.

Having seen the one- and two-qubit examples, we now generalize the flag decomposition to arbitrary qubit counts. This decomposition is going to be recursively repeated until one of the two base cases, with one or two qubits, is reached. In the following derivation, we are going to extensively use multiplexers, so recall the following definition for two square matrices $M_0, M_1 \in U(2^{n-1})$, multiplexed on a single qubit:


\begin{align*}
\vcenter{\hbox{%
\resizebox{0.65\linewidth}{!}{%
\includegraphics{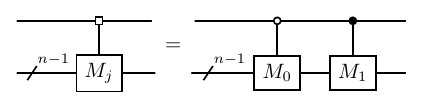}%
}}}
=
\begin{pmatrix}
    M_0 & 0 \\ 0 & M_1
\end{pmatrix}.
\end{align*}

To derive the general flag decomposition, suppose we are given a unitary matrix $V \in U(2^{n})$.
First, we perform a cosine-sine decomposition (CSD), which can conveniently be expressed as
\begin{align}\label{eq:AIII_flag}
\resizebox{\linewidth}{!}{%
\includegraphics{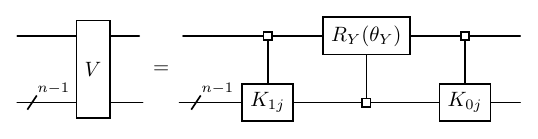}%
}.
\end{align}
Here, operators attached by a single-qubit multiplexer get an index $j\in\{0,1\}$, for the respective control value.
Numerically, the CSD functionality to compute the angles $\theta_Y$ for $R_Y$ and the four matrices $K_{ij}$ is readily available in standard linear algebra packages, such as LAPACK~\cite{lapack_users_guide} or \texttt{SciPy}~\cite{virtanen2020scipy}. For our purposes, \href{https://docs.scipy.org/doc/scipy/reference/generated/scipy.linalg.cossin.html}{\texttt{scipy.linalg.cossin}} with the option \texttt{separate=True} does exactly what we need.

Next, we perform a flag decomposition on each of the two matrices $K_{10}$ and $K_{11}$, which yields

\begin{align}\label{eq:smallflagdecomposition}
\scalebox{0.95}{
\includegraphics{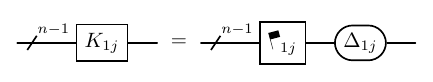}
}.
\end{align}
This is where recursion comes into play, as we are performing an $(n-1)$-qubit flag decomposition inside the $n$-qubit flag decomposition.
We can then use the multiplexer extension property (MEP)~\cite[Sec.~3.2]{shende2006synthesis} that allows us to simply attach multiplexers to circuit identities, which yields

\begin{align}\label{eq:MEP_flag}
\resizebox{0.9\linewidth}{!}{%
\includegraphics{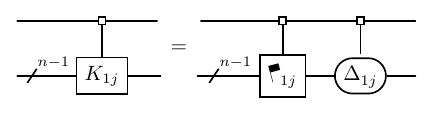}
}.
\end{align}

We then split a diagonal operator $\Delta'$ on the lower $n$ qubits off the multiplexed diagonal~\cite{shende2006synthesis,PennyLane-diagonal-unitary-decomp}:

\begin{align}\label{eq:balance_diagonal_circ}
\scalebox{0.95}{
\includegraphics{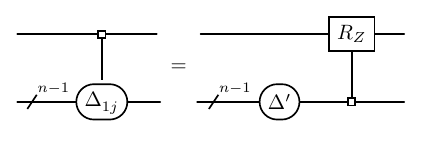}
}.
\end{align}
The new diagonal and the $R_Z(\theta)=\exp(-i\tfrac\theta2Z)$ rotation angles are given by
\begin{align}\label{eq:balance_diagonal_0}
    \Delta' &= \sqrt{\Delta_{10}\Delta_{11}} 
    = \exp\left(\tfrac{i}2(\arg(\Delta_{10}) + \arg(\Delta_{11}))\right),\\\label{eq:balance_diagonal_1}
    \theta_Z &= \arg(\Delta_{11}\Delta_{10}^\ast)=\arg(\Delta_{11}) - \arg(\Delta_{10}).
\end{align}
We then insert all this back into \cref{eq:AIII_flag}, commute $\Delta'$ through the controls of the multiplexed $R_Y$ rotation, and merge it into the right multiplexer. This yields

\begin{align*}
\resizebox{\linewidth}{!}{%
\includegraphics{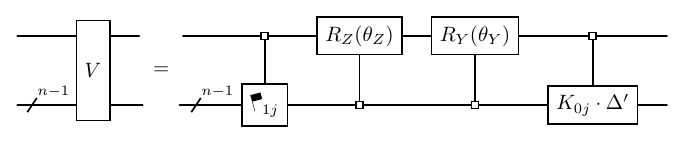}%
}.
\end{align*}
Note that the multiplexed $R_Z$ and $R_Y$ gates together already constitute a multiplexed single-qubit flag circuit (see \cref{eq:flag_decomp_parametrized} below).

In a last decomposition step, we apply \cref{eq:MEP_flag} to $K_{0j}\cdot\Delta'$\footnote{We will not make a notational difference between diagonal matrices and their diagonal, and standard matrix multiplication is implied throughout.} to obtain $\Delta_{0j}$ and $\smallflag_{0j}$. This time, we leave the multiplexed diagonal intact. For that, note that a multiplexed diagonal is equivalent to a diagonal, i.e.

\begin{align}\label{eq:MEP_flag2}
\scalebox{0.95}{
\includegraphics{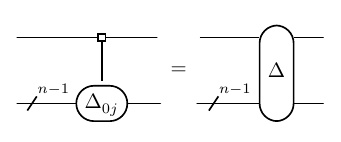}
},
\end{align}
where the larger diagonal matrix is simply a concatenation of the two smaller diagonals,
\begin{equation}\label{eq:diagonal_split_formula}
\Delta = \Delta_{00}\oplus\Delta_{01}.
\end{equation}
This then leaves us with the recursive flag decomposition:

\begin{align}\label{eq:flag_decomp_parametrized}
\resizebox{\linewidth}{!}{%
\includegraphics{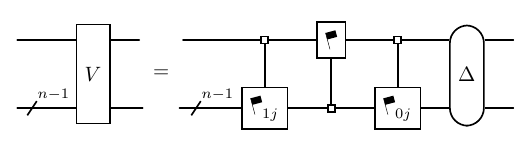}%
}.
\end{align}
Note that everything before the diagonal operator constitutes the flag sub-circuit of the decomposition.

Repeating the previous steps on more qubits explicitly, we see the following regular Gray code structure:

\begin{align}
\resizebox{\linewidth}{!}{%
\begin{quantikz}
&& \gate[3]{V} & \midstick[3,brackets=none]{$\stackrel{(\ref{eq:AIII_flag})}{=}$} && \mltplx{1} & \gate{R_Y} & \mltplx{1} & \midstick[3,brackets=none]{$\stackrel{(\ref{eq:MEP_flag})}{=}$} && \mltplx{1} & \mltplx{1} & \gate{R_Y} & \mltplx{1} &
\\
&& & && \gate[2]{K} & \mltplx{-1} & \gate[2]{K} && & \gate[2]{\smallflag} & \multidiagonal{2}{\Delta} & \mltplx{-1} & \gate[2]{K} & 
\\
& \qwbundle{n-2} & & & \qwbundle{n-2} & & \mltplx{-1} & & & \qwbundle{n-2} & & & \mltplx{-1} & &
\end{quantikz}
}%
\nonumber
\\
\resizebox{.75\linewidth}{!}{%
$\stackrel{(\ref{eq:balance_diagonal_circ})}{=}$ 
$\vcenter{\hbox{\includegraphics{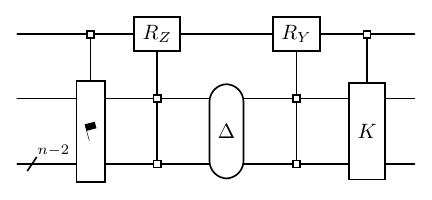}}}$
}
\nonumber
\\
\resizebox{.75\linewidth}{!}{%
$\stackrel{(\ref{eq:MEP_flag}, \ref{eq:MEP_flag2})}{=}$ 
$\vcenter{\hbox{\includegraphics{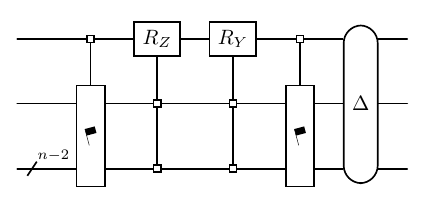}}}$
}
\nonumber
\\
\resizebox{\linewidth}{!}{%
$\stackrel{(\ref{eq:balance_diagonal_circ}, \ref{eq:MEP_flag2}, \ref{eq:flag_decomp_parametrized})}{=}$ 
$\vcenter{\hbox{\includegraphics{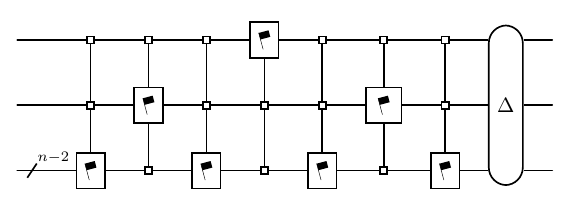}}}$
}.\label{eq:flag_decomp_n}
\end{align}

The flag sub-circuits have the same Gray code structure. The recursion is continued until the base case of either $n_b=1$ (\cref{eq:flag_decomp_n-1}) or $n_b=2$ (\cref{eq:flag_decomp_n-2}) qubits is reached.

Hence, for $n=3$ qubits, we simply obtain the circuit in \cref{eq:flag_decomp_n} with a single qubit in the bottom register for the base case $n_b=1$:
\begin{align*}
\resizebox{\linewidth}{!}{%
\includegraphics{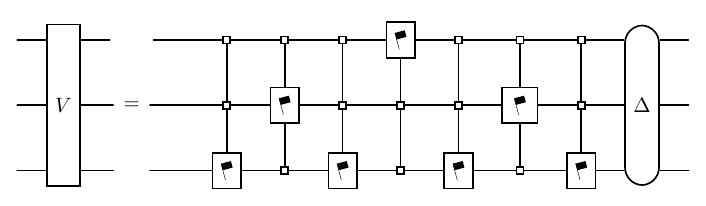}.
}
\end{align*}
For $n_b=2$, we can utilize \cref{eq:flag_decomp_n-2}, which leads to
\begin{align}\label{eq:flag_decomp_n-3_b-2}
\resizebox{\linewidth}{!}{%
\includegraphics{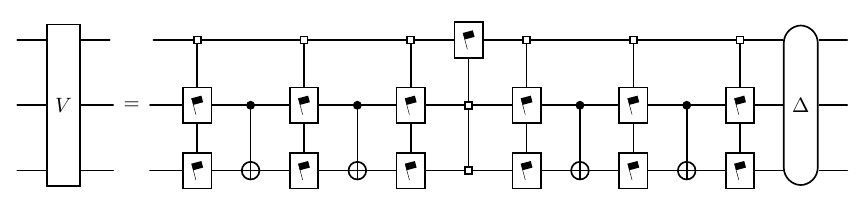}.
}
\end{align}

The two base cases lead to different control structures. Depending on how we choose to further decompose the multiplexers, either case can have advantages. The regular structure for the base case $n_b=1$ is used in the phase gradient decomposition (see \cref{sec:main_flag_decomposition}) and reproduces the recursive CSD (see \cref{sec:recursive_csd}). The base case $n_b=2$ has fewer CNOT gates in the \{Clifford + Rot\} decomposition, improving upon the recursive CSD, and we detail its decomposition to this gate set in \cref{sec:recursive_csd:details,sec:gate_count_calculations:mux-one-two-qubit-ops}. This decomposition will be superseded by SDM in \cref{sec:selective_de_multiplexing} due to the improved CNOT counts. In SDM, we will employ the $n_b=2$ recursive flag decomposition as a subroutine; see also \cref{sec:gate_count_calculations} for details on this analysis.

The recursive algorithm with both base cases is summarized in \cref{algo:recursive_flag_decomp} in \cref{app:impl_details:main_blocks}, including the optional decomposition of multiplexed single-qubit flags into \{Clifford+Rot\} gates, and implemented as \texttt{flagsynth.mux\_multi\_qubit\_decomp} in~\cite{our_repo}.

\subsection{Phase gradient decomposition}
\label{sec:main_flag_decomposition}

\begin{figure*}
    \centering
    \includegraphics[width=0.8\linewidth]{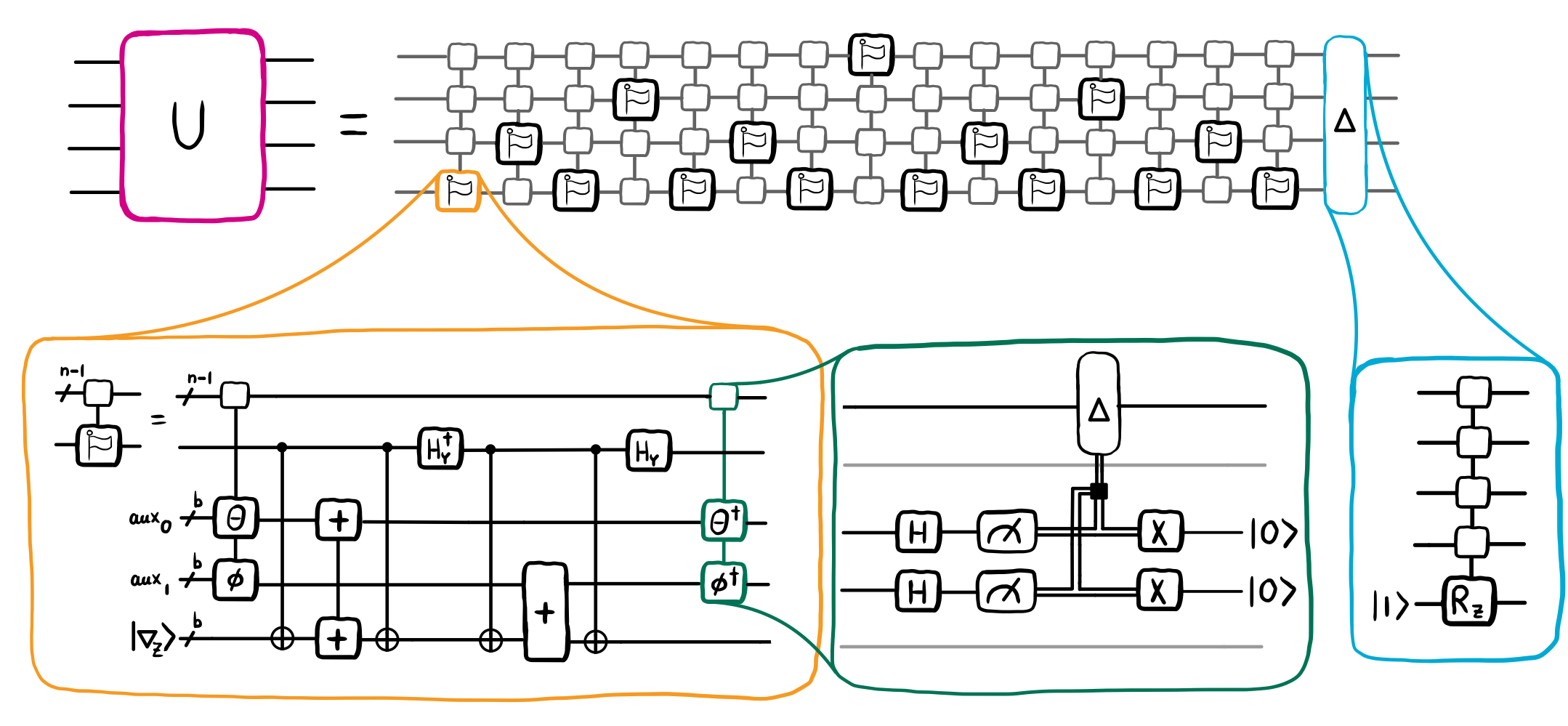}
    \caption{Phase gradient decomposition of our unitary synthesis circuit. The unitary is first broken down into $(n-1)$-multiplexed single-qubit flags and a trailing diagonal via a recursive flag decomposition (top circuit). Each multiplexed flag can be implemented via loading of the flag angles $\theta$ and $\phi$, adders onto a resource phase gradient state $|\nabla_Z\rangle$ that are control-flipped by the flag target into subtractors, and angle unloading (orange, bottom left). The latter is implemented via measurement in the $X$ basis and a corrective diagonal, which depends on the measurement outcomes and can be merged into subsequent flags (dark green, bottom mid). The terminal diagonal is implemented as a multiplexed $R_Z$ gate on a fixed-state qubit (blue, bottom right); see \cref{appendix:diagonal} for details.\hspace*{\fill}}
    \label{fig:two}
\end{figure*}

For the phase gradient decomposition, we use \cref{eq:flag_decomp_n} with the base case $n_b=1$, resulting in $2^n - 1$ multiplexed single-qubit flag operators followed by a diagonal unitary. 
We provide an overview of the complete decomposition pipeline in \cref{fig:two}.
The phase gradient decomposition of the multiplexed single-qubit flag yields the following circuit with additional auxiliary qubit registers onto which the angles are loaded, as well as another register carrying a phase gradient state $|\nabla_Z\rangle$:
\begin{align*}
\resizebox{\linewidth}{!}{%
\includegraphics{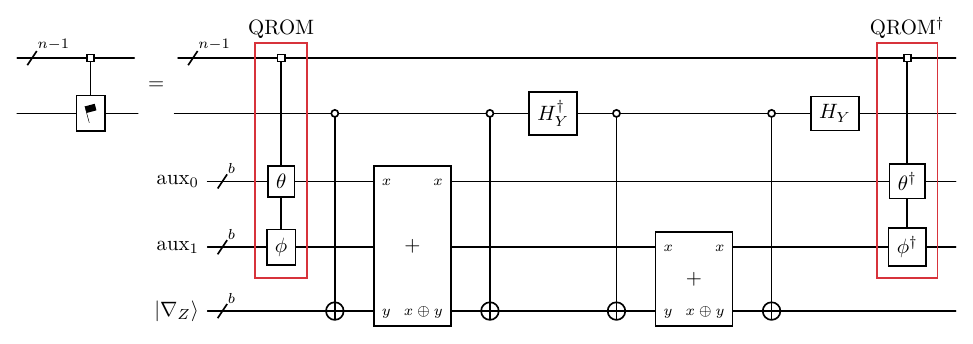}
}.
\end{align*}
Here, $b$ is the number of digits we use to encode the angles in their binary representation. More details and a full derivation can be found in~\cref{appendix:flag_mul} and~\cite{PennyLane-PhaseGradient}. The second QROM, which uncomputes the loaded angles, can be performed with measurements and a classically controlled diagonal operator,

\begin{align*}
\resizebox{\linewidth}{!}{%
\includegraphics{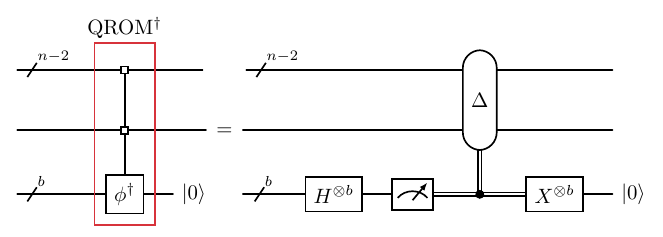}%
}.
\end{align*}
This diagonal operator can be merged into the subsequent multiplexed flag of the circuit skeleton in \cref{eq:flag_decomp_n}, such that it does not invoke any additional cost:

\begin{align}
\resizebox{\linewidth}{!}{%
\includegraphics{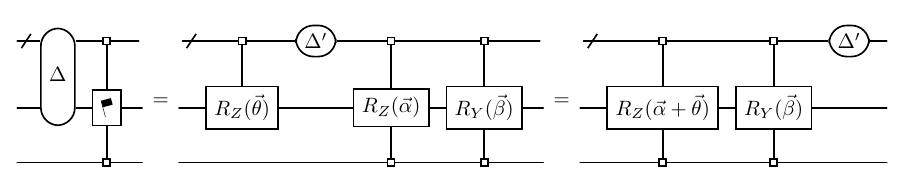}.%
}
\label{eq:diagonal_push}
\end{align}
Here, $\vec{\alpha}$ and $\vec{\beta}$ represent the rotation angles of the multiplexer we aim to merge the diagonal into. The vector $\vec{\theta}$ is given by \cref{eq:balance_diagonal_1} and \cref{eq:diagonal_split_formula}\footnote{Note that the angles in $\vec{\theta}$ need to be extended to match the dimension of $\vec{\alpha}$, which is additionally multiplexed by the lower qubit register. This can be achieved by $\vec{\theta} \otimes (1,1)$.}.

When uncomputing the QROM of the next multiplexer, a new classically controlled diagonal is generated. The previously computed diagonal \(\Delta'\) is then merged into the new diagonal, allowing the process to be repeated until the last multiplexer is reached.
The final classically controlled diagonal operation can be merged with the very final diagonal gate on the right in \cref{fig:two}.

After this optimization, the Toffoli cost for each of the $2^n - 1$ multiplexers is given by 
\begin{equation}
C_\text{Tof}^\text{mux-\smallflag} = \frac{2^n}{2\Lambda} + 2\Lambda b - 5,
\end{equation}
where $b$ is the number of bits used to represent the rotations and $\Lambda$\footnote{Given $k$ available auxiliary qubits in the circuit, we can estimate $\Lambda$ as the closest power of two to $  \min\left(\left\lceil \sqrt{2^m / b} \right\rceil, k / b\right),$ where $m$ is the number of QROM control qubits.} is a tuneable parameter that represents the number of rows defined in the QROM structure. A detailed derivation of this cost can be found in~\cref{appendix:flag_mul}.

The final diagonal $\Delta$ can also be decomposed using phase gradient states and leads to the cost
\begin{equation}
    C_\text{Tof}^\Delta = \frac{2^n}{2\Lambda} + 2\Lambda b + \frac{2^n}{\Lambda'} + \Lambda' - 5.
\end{equation}
The two different values $\Lambda$ and $\Lambda'$ are due to the usage of two separate QROMs: one that stores bitstrings of size $b$ with $2\Lambda$ rows, and a second one that stores bitstrings of size $1$ with $\Lambda'$ rows. A detailed description is provided in~\cref{appendix:diagonal}. Therefore, the total cost of the unitary synthesis technique is:

\begin{align}
    C_\text{Tof}^{\smallflag-\text{decomp}} = &(2^n-1) \ C_\text{Tof}^\text{mux-\smallflag} \text{ (flag multiplexers)} \nonumber\\
    &+ C_\text{Tof}^\Delta \text{ (final diagonal)}.
\end{align}
This is an improvement over the decomposition in \cite{berry2024rapid} with cost\footnote{See \cref{eq:berry_phase_gradient} for the corresponding circuit structure.}:

\begin{align*}
C^\text{\cite{berry2024rapid}}_\text{Tof} = &2^n \ C_\text{Tof}^\text{mux-\smallflag}
\text{ (phase multiplexers)}
\\
&+ C_\text{Tof}^\Delta
\text{ (final diagonal)}
\\
&+ C_\text{Tof}^{\pm1}
\text{ (incrementers/decrementers)}.
\end{align*}
Here, the additional cost of the incrementers and decrementers is given by $C_\text{Tof}^{\pm1} = (2^n - 1)(n - 2)$, since the cost of an adder with a classical input is $n-2$ \cite{Gidney2018halving}. The cost for multiplexed phases is the same as for the multiplexed single-qubit flag circuit.
These savings are due to the structure of the flag decomposition circuit in \cref{fig:two}. The parameters are distributed in an optimal fashion among the $2^n-1$ multiplexers. On the other hand, the Gray code structure removes the necessity for incrementers and decrementers, and still allows us to merge the classically controlled diagonals from the QROM uncomputations with measurements.
Compared to~\cite{berry2024rapid}, we overall reduce the Toffoli gate count by
\begin{equation}
C_\text{Tof}^\text{mux-\smallflag} + C_\text{Tof}^{\pm1} = \left(\frac{2^n}{2 \Lambda} + 2 \Lambda b - 5\right) + (n-2)(2^n-1).
\end{equation}

\section{Selective De-Multiplexing}\label{sec:selective_de_multiplexing}

\begin{figure*}
    \centering
    \includegraphics[width=0.8\linewidth]{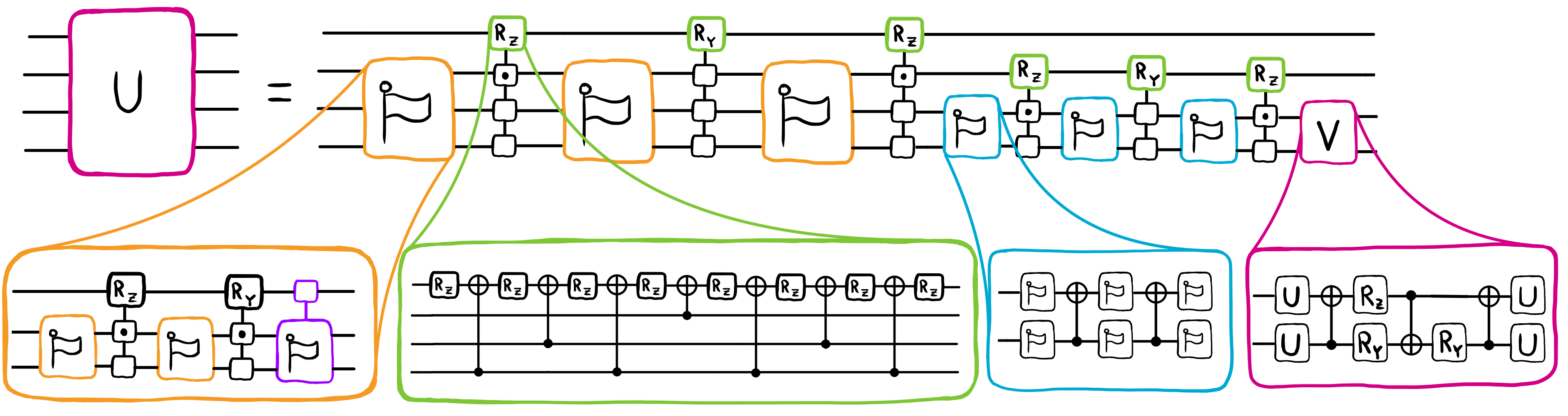}
    \caption{Selective de-multiplexing (SDM) proposed in this work. A unitary operator is decomposed with a single QSD step in a parameter-optimal fashion (\cref{sec:po_qsd:single_qsd_step}, top circuit), producing flag circuits, multiplexed rotations, and a smaller unitary. These constituents are decomposed by selectively de-multiplexing the flag circuit (orange, lower left, \cref{sec:selective_de-mux}), a (symmetrized) M\"ott\"onen decomposition (green, lower left mid, \cref{eq:symmetrized_Möttönen}) and a next QSD step using SDM. Ultimately, we use known base case decompositions for two-qubit flags (blue, lower right mid, \cref{eq:flag_decomp_n-2}) and two-qubit unitaries (magenta, lower right, \cref{eq:two-qubit-unitary}). Multiplexed flags (purple) are decomposed with the recursive flag decomposition/CSD from \cref{sec:flag_decomposition} (not shown here).\hspace*{\fill}}
    \label{fig:three}
\end{figure*}

As we have seen in the previous section, the flag decomposition is well-suited for phase gradient decompositions. In the context of \{Clifford + Rot\} decompositions, it is already parameter-optimal and provides a reasonable $\cnot$ count of $\tfrac12 4^n - \tfrac34 2^n-1$, as reported in \cref{tab:known_synthesis_techniques}. We can further improve this to a $\cnot$ count of $\tfrac12 4^n - \tfrac38(n+2) 2^n + n -1$, matching the best known $\cnot$ count reported in Krol et al. \cite{Krol-Al-Ars} for $n\in \{3, 4\}$. This is achieved by what we call \textit{selective de-multiplexing (SDM)}, illustrated in \cref{fig:three}. The procedure builds on the quantum Shannon decomposition, but in order to remain parameter-optimal after the first recursion step, we introduce selective de-multiplexing for flag circuits in \cref{sec:selective_de-mux}.
Thus, we are selective both across the depth of the recursion and across the circuit length about which multiplexers to de-multiplex.

We start out just like the original QSD~\cite{shende2006synthesis}, performing a CSD followed by de-multiplexing in the first recursive step (a type-AIII + type-A recursive Cartan decomposition~\cite{wierichs2025recursive}). However, we then modify the circuit using the flag decomposition from \cref{sec:flag_decomposition} to achieve and maintain parameter optimality; see \cref{sec:po_qsd:single_qsd_step}.
This results in a richer recursion structure and forces us to decompose multiple different multi-qubit operators from the second step onwards, which we analyze in \cref{sec:gate_count_calculations}. Finally, we selectively de-multiplex flag circuits during the recursion in order to lower the $\cnot$ count; see \cref{sec:selective_de-mux}.
We put all pieces of SDM together in \cref{sec:gate_counts_po_qsd}.
As the phase gradient decomposition does not benefit from SDM, we target the \{Clifford + Rot\} decomposition for the rest of this section.

\subsection{Making a single QSD step parameter-optimal}\label{sec:po_qsd:single_qsd_step}
As in \cref{eq:AIII_flag}, we first decompose a given unitary from $V\in U(2^n)$ via a CSD:

\begin{align}\label{eq:first_CSD}
\resizebox{\linewidth}{!}{%
\includegraphics{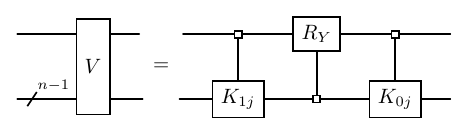}%
}.
\end{align}
Instead of decomposing the matrices $K_{ij}$ inside the multiplexers, we now de-multiplex these operators,

\begin{align}\label{eq:de-multiplexing-qsd}
\resizebox{\linewidth}{!}{%
\includegraphics{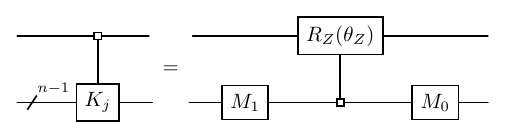}.%
}\nonumber
\end{align}
Note that we now omit the first subscript of $K$ and $M$ for better readability.
The matrices for this type-$A$ decomposition are computed as follows. We perform an eigenvalue decomposition of $K_0 K_1^\dagger$ as $M_0 D^2 M_0^\dagger$ and set $M_1 = D M_0^\dagger K_1$~\cite{shende2006synthesis}. The $Z$-rotation angles are given by the phases of the diagonal, $\theta_Z = -2 \arg(D)$.
Putting the CSD and the de-multiplexing together, we arrive at

\begin{align}
\resizebox{\linewidth}{!}{%
\includegraphics{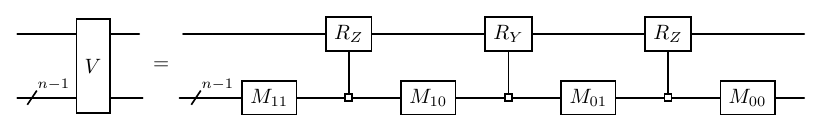},%
}
\end{align}
where $M_{jk}\in U(2^{n-1})$.
Before decomposing these $M_{jk}$, we perform a small optimization that was introduced in~\cite{shende2006synthesis} and extended in~\cite{Krol-Al-Ars}. For this, we re-assemble $M_{10}$, $M_{01}$ and the $R_Y$ multiplexer into a multiplexed unitary, where the multiplexing qubit now acts in the Pauli-$Y$ basis, and insert identities in the form of $\cy^2$:

\begin{align}\label{eq:po_qsd_cnot_trick}
\resizebox{\linewidth}{!}{%
\includegraphics{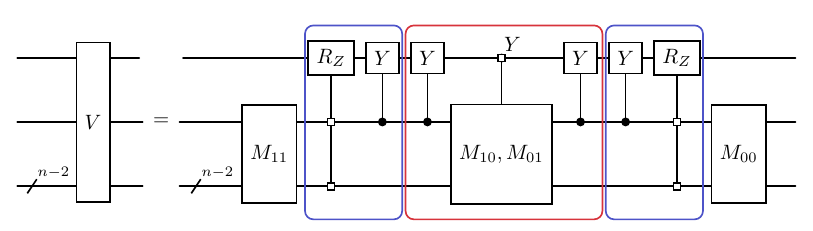}
}.
\end{align}
The gates in the center red box can then be merged and de-multiplexed again, while the decomposition of the $R_Z$ multiplexers can be chosen to produce a $\cy$ gate to the right or left, such that it cancels with the created gate in each blue box.
The result is a symmetrized variant of the M\"ott\"onen decomposition, which we mark with a dot in the corresponding multiplexing node:
\begin{align}\label{eq:symmetrized_Möttönen}
\resizebox{0.95\linewidth}{!}{%
\includegraphics{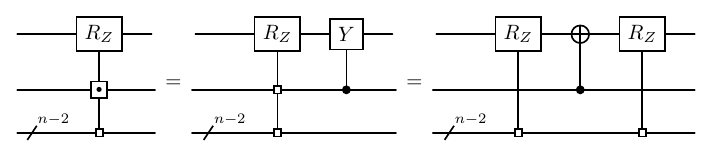}.%
}
\end{align}
The crucial part of this decomposition is that we save one entangling gate compared to the standard multiplexer decomposition~\cite{mottonen2004transformation,PennyLane-SelectU2Decomp}; see \cref{sec:gate_count_calculations:mux-one-two-qubit-ops} for further details. Note that we could instead choose to symmetrize any other two of the three multiplexed rotations.
 
We continue with a flag decomposition $M_{11} = \Delta_{11}\smallflag_{11}$, commute its diagonal $\Delta_{11}$ through the left-most multiplexed rotation (and $\cy$ gate), and compute $M_{10}'=M_{10}\Delta_{11} \in U(2^{n-1})$.
Consecutively repeating these steps for the other matrices, we arrive at

\begin{align}\label{eq:parameter_optimal_qsd}
\resizebox{\linewidth}{!}{%
\includegraphics{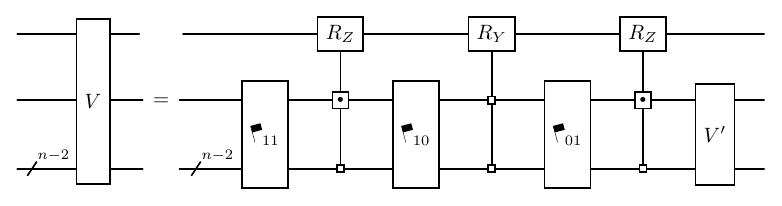}.%
}
\end{align}
We are left with the unitary $V' = M_{00}\Delta_{01}$, which can be decomposed recursively by applying all steps described in this section again, beginning with the CSD.

What remains are flag circuits, which we could decompose with the strategy from \cref{sec:flag_decomposition}. However, there is an alternative inspired by the QSD, which we term \textit{selective de-multiplexing} and detail in the next section. For brevity, we move the discussion about why selective de-multiplexing is preferred over flag circuits in the context of \{Clifford + Rot\} decompositions to \cref{sec:gate_count_calculations:flags}.

\subsection{Selective de-multiplexing of a flag circuit}\label{sec:selective_de-mux}
When decomposing a flag circuit recursively with the flag decomposition from \cref{sec:flag_decomposition}, we may take a slight detour that is inspired by the de-multiplexing in the QSD above.
Namely, after executing the first CSD like in \cref{eq:first_CSD}, we may apply the de-multiplexing to the first of the two multiplexed flags only, leading to the following circuit\footnote{Note that we may apply the CSD to the flag circuit as we would to any unitary matrix. This causes overparametrization which is fixed subsequently by merging diagonals.}:

\begin{align}\label{eq:selective_de-mux}
\resizebox{\linewidth}{!}{%
\includegraphics{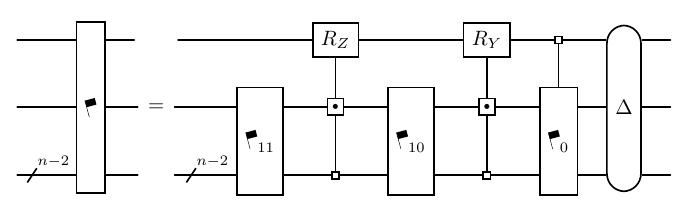}.%
}
\end{align}
Here, we already commuted a diagonal through the circuit, and used the trick in \cref{eq:po_qsd_cnot_trick} to symmetrize both multiplexed rotations, by producing a $\cy$ gate between the multiplexed $R_Z$ and $\smallflag_{10}$ as well as a $\cz$ between the multiplexed $R_Y$ and $\smallflag_0$.
This selective de-multiplexing step maintains parameter optimality\footnote{Again considering the diagonal produced by the CSD step as ``mitigable overparametrization".}, and we will employ it to obtain our cheapest decomposition; see the following section and \cref{sec:gate_count_calculations} for details.

\subsection{Assembling the selective de-multiplexing synthesis}\label{sec:gate_counts_po_qsd}
We now have collected all pieces to assemble the complete synthesis procedure using SDM. 
For unitaries, we use a parameter-optimal QSD step with a symmetrized Möttönen decomposition for multiplexed rotations (see \cref{sec:po_qsd:single_qsd_step}), until we reach the well-known two-qubit base case:
\begin{align}\label{eq:two-qubit-unitary}
\resizebox{\linewidth}{!}{%
\includegraphics{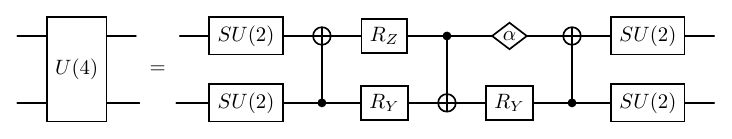}.%
}
\end{align}
Here, the diamond-shaped gate denotes a global phase $e^{i\alpha}$.
Non-multiplexed flag circuits are broken down via a CSD and selective de-multiplexing (see \cref{eq:selective_de-mux}) whereas multiplexed flag circuits are decomposed with pure recursive CSDs.
For both multiplexed and non-multiplexed flags, we make use of the decomposition on $n_b=2$ qubits, shown in \cref{eq:flag_decomp_n-2}, as the base case.
Ultimately, multiplexed single-qubit flags are expressed in the elementary gate set with a technique from~\cite{bergholm2005quantumuniformlycontrolled}, shown here for $n=3$ qubits:
\begin{align*}
    \resizebox{\linewidth}{!}{%
\includegraphics{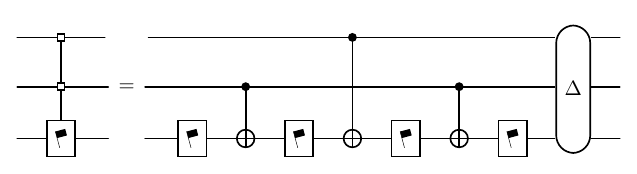}.
    }
\end{align*}
We provide a detailed analysis for these optimized decompositions in \cref{sec:gate_count_calculations}, illustrate the SDM procedure in \cref{fig:three} for $n=4$ qubits, and summarize it in \cref{algo:po_qsd,algo:recursive_flag_decomp_cliff_rz} in \cref{app:impl_details:main_blocks}.

Regarding the gate counts, we first consider the parameter count to verify parameter-optimality.
In \cref{eq:parameter_optimal_qsd}, we see three flag sub-circuits with $4^{n-1} - 2^{n-1}$ parameters each, as we have removed $3 \cdot 2^{n-1}$ parameters via diagonal merging. These removed parameters exactly offset the $3 \cdot 2^{n-1}$ parameters contributed by the three multiplexed rotations. Together with the $4^{n-1}$ parameters from the final $V'$, we overall have $4 \cdot 4^{n-1}=4^n$ parameters, which is optimal for an $n$-qubit unitary.

Considering the $\cnot$ count associated with a single QSD step, we obtain the relation
\begin{align}
    C_{\cnot}^{\text{QSD}}(n) = &C_{\cnot}^{}(n-1) \nonumber\\
    &+ 3 (C^{\text{flag}}_{\cnot}(n-1) + 2^{n-1})-2,
\end{align}
which is based on \cref{eq:parameter_optimal_qsd} and the technique from \cref{eq:po_qsd_cnot_trick,eq:symmetrized_Möttönen} to reduce the $\cnot$ cost for two of the multiplexed rotations by one.

In \cref{sec:gate_count_calculations:flags}, we compute the minimal $\cnot$ cost of a flag circuit to be
\begin{align}
    C_{\cnot}^{\text{flag}}(n)&=\frac{1}{2}4^n-\frac{n+6}{4}2^n+1,
\end{align}
using selective de-multiplexing for non-multiplexed flags and the recursive flag decomposition from \cref{sec:flag_decomposition} for multiplexed flags.
This is then inserted into the recursion relation above, together with the base case $C_{\cnot}^{}(2)=3$ from the standard two-qubit unitary decomposition in \cref{eq:two-qubit-unitary} to obtain the final $\cnot$ count of SDM,
\begin{align}
    C_{\cnot}^{\text{SDM}}(n) =\frac{1}{2}4^n - \frac{3}{8}(n+2)2^n+n-1.
\end{align}

\section{Matrix product state preparation}
\label{sec:mps_state_prep}

\begin{figure*}
    \centering
    \includegraphics[width=.8\linewidth]{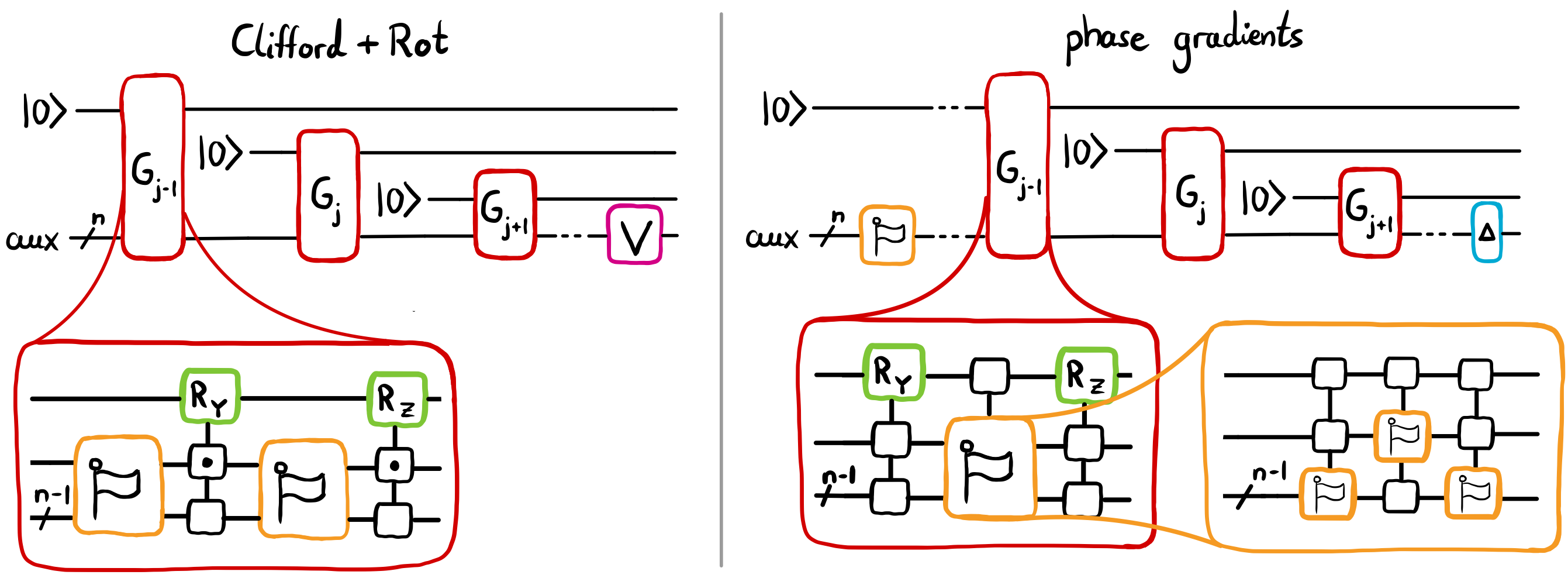}
    \caption{Decomposition of matrix product state (MPS) preparation using our unitary synthesis results and further optimizations. For each tensor of the MPS (red), we obtain a multiplexed $R_Y$ gate and a subcircuit representing a reduced multiplexed unitary operator. The exact structure of this subcircuit depends on the target gate set, i.e., on whether we compile for the \{Clifford + Rot\} decomposition (left) or the phase gradient decomposition (right). The resulting (multiplexed) flags and rotations, as well as the trailing diagonal gate $\Delta$ are then decomposed with the respective techniques from \cref{fig:three,fig:two}. Note that for the phase gradient decomposition, the unitary degrees of freedom on the auxiliary register are split into a flag circuit and a diagonal on either side.\hspace*{\fill}}
    \label{fig:mps-complete}
\end{figure*}

Preparing matrix product states (MPS) on a quantum computer is a crucial subroutine for many quantum algorithms. A prominent example is quantum phase estimation, where the overlap of the initial state with the target state is crucial \cite{fomichev2024initial,LKS2024,berry2024rapid}.
The improved unitary synthesis techniques introduced in \cref{fig:two,fig:three} translate to improved resources for MPS preparation. An overview of the resulting circuits for both the \{Clifford + Rot\} and the phase gradient decomposition is provided in \cref{fig:mps-complete}.

We consider an MPS with open boundary conditions on $L$ physical sites of local dimension $d=2$ (for qubits). This state is formally defined in terms of the tensors $\{A^{\sigma_j}\}_{j=1}^L$ as 
\begin{equation}\label{eq:mps_definition_sec2}
    |\psi\rangle = \sum_{\sigma_1 \cdots \sigma_L \in \{0,1\}} A^{\sigma_1} \cdots A^{\sigma_L} |\sigma_1 \cdots \sigma_L\rangle,
\end{equation}
where each of the $A^{\sigma_j}$ is interpreted as a matrix for the concrete value of $\sigma_j \in \{0,1\}$. The matrix dimensions of $A^{\sigma_j}$ are slightly different at the left and right boundaries of the MPS. We are going to discuss these boundary effects separately in \cref{sec:boundary-effects}, and here focus on the central \textit{bulk} tensors of the MPS:
\begin{equation*}
\resizebox{1.0\linewidth}{!}{%
\includegraphics{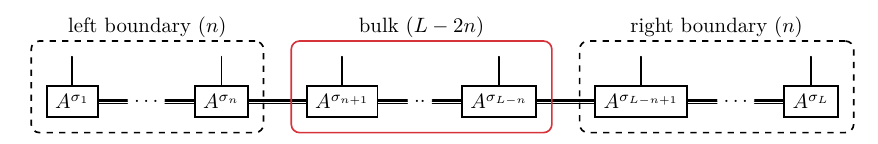}.
}
\end{equation*}
In the bulk, all matrix dimensions (horizontal connections) are truncated to the so-called bond dimension $\chi$. This is a hyper-parameter that determines the degree of approximation of the MPS. We conveniently choose it to be $2^n$, such that we can map it to an $n$-qubit register. This is highly desired for MPS preparation because there are significant overparametrizations otherwise. This also determines the boundary region to span exactly $n$ tensors on either side of the bulk, which itself comprises $L-2n$ tensors.

We assume the tensors to define isometries, implying $\sum_{\sigma_j\in\{0,1\}} \left(A^{\sigma_j}\right)^\dagger A^{\sigma_j} = \id$ and that they map from $n$ to $n-1$ qubits. This can always be achieved by transforming the tensors into the so-called left-canonical form; see \cite{Kottmann2024MPS}. To prepare an MPS on a quantum computer, we translate the isometry tensors

\begin{equation*}
\resizebox{\linewidth}{!}{%
\includegraphics{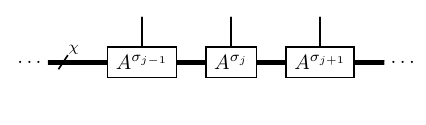}
}
\end{equation*}
into unitary matrices $G_j$ via unitary completion.
The resulting circuit then has the following structure:

\begin{equation}\label{eq:mps_prep_barebone_skeleton}
\resizebox{\linewidth}{!}{%
\includegraphics{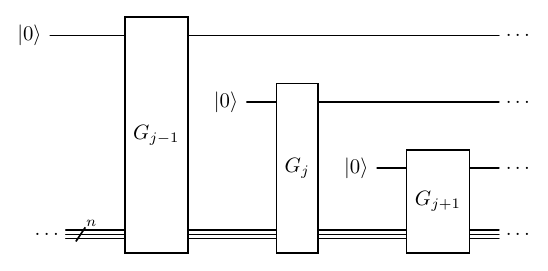}
}.
\end{equation}
In general, an isometry from $n$ to $n+1$ qubits has $3\nolinebreak\cdot\nolinebreak4^n$ degrees of freedom, as it lives on a so-called Stiefel manifold\footnote{The Stiefel manifold dimension is $2 \cdot 2^n \cdot 2^m - (2^m)^2$, for an isometry from $m$ to $n$ qubits with $m\leq n$.}. In particular for our case with $\chi = 2^n$, we need to reduce the degrees of freedom of the unitary $G_j$ from $4 \chi^2$ to $3\chi^2$. As we will see, this is achieved by removing multiplexer nodes and $R_Z$ gates due to the fixed $|0\rangle$ input.

Further, we will be able to remove another $\chi^2$ degrees of freedom by merging unitaries on the shared auxiliary $n$-qubit register in \cref{eq:mps_prep_barebone_skeleton} between the isometries. This corresponds to the so-called gauge degree of freedom of MPS. It is due to the ambiguity of the $A^{\sigma_j}$ isometries, as we can always insert arbitrary unitary matrices $W_j$ and $W^\dagger_j$ on the horizontal connections (the so-called virtual bonds):
\begin{equation*}
\resizebox{\linewidth}{!}{%
\includegraphics{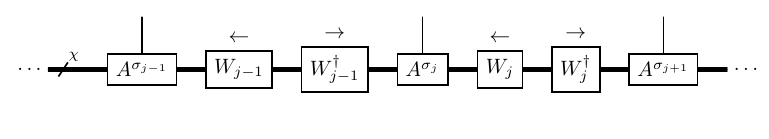}.
}
\end{equation*}
By multiplying the unitary matrices in the direction indicated by the arrows, we can arrive at altered isometries $A^{\sigma_j}$ that define the same state. This ambiguity reduces the free parameters of the isometries by another $\chi^2$. For more details on the gauge degree of freedom we refer to Sec. 4.1.3, iv) in \cite{Schollwock2011MPS}.

In summary, we attain the optimal number of parameters,
\begin{equation}\label{eq:mps_parameters}
    2\chi^2
\end{equation}
for each unitary $G_j$ with the following steps:
\begin{enumerate}
    \item Parameter-optimal unitary synthesis ($4\chi^2$): Synthesis of $G_j$ using the circuits in \cref{fig:one}.
    \item Isometry condition ($3\chi^2$): Use the fixed input state $|0\rangle$ on the top qubit to remove multiplexer nodes and $R_Z$ rotations.
    \item Gauge freedom ($2\chi^2$): Merge a unitary on the shared auxiliary $n$-qubit register between isometries.
\end{enumerate}
We will go through these steps for each of the considered decomposition methods in the following two subsections.

\subsection{\{Clifford + Rot\} decomposition}
\label{subsec:mps_clifford_rot}

For \{Clifford + Rot\} decomposition, we substitute SDM from \cref{fig:three} for the unitaries $G_j$. We apply the symmetrization trick in \cref{eq:po_qsd_cnot_trick,eq:symmetrized_Möttönen} to the second and third multiplexers instead:

\begin{align}
&\resizebox{\linewidth}{!}{%
\includegraphics{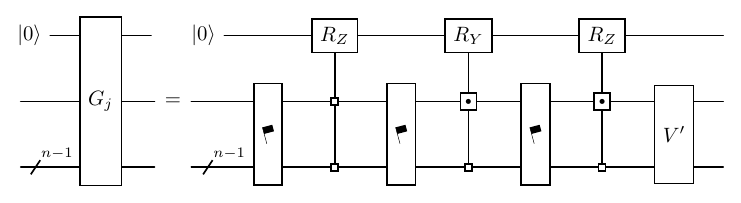}}
\nonumber \\
\label{eq:synth_G_j_clifford_rot}
&\hspace{1cm}\resizebox{.75\linewidth}{!}{%
\includegraphics{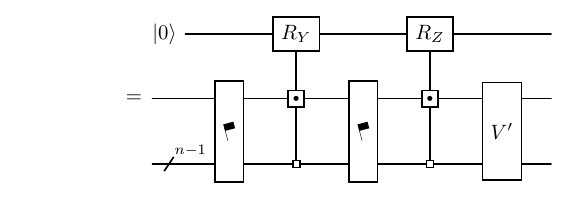}}.
\end{align}
The reductions left of the $R_Y$ multiplexer compared to \cref{eq:parameter_optimal_qsd} are due to the fixed initial state on the first qubit. The multiplexed $R_Z$ is reduced to a diagonal on the lower $n$ qubits,
\begin{align}\label{eq:remove_RZ}
&\resizebox{.5\linewidth}{!}{%
\includegraphics{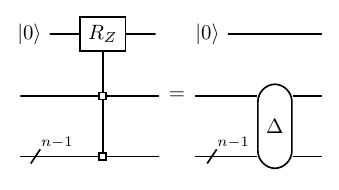}}
\end{align}
and can be merged together with the two flags into a new $n$-qubit unitary. This new unitary can then be flag-decomposed and its diagonal merged to the right again.
Because a flag circuit has $4^n - 2^n = \chi^2 - \chi$ parameters, this circuit matches the expected $2 (\chi^2 -\chi) + 2 \ \chi + \chi^2 = 3\chi^2$ parameters for a single isometry.

Further, we can merge the unitary $V'$ into the consequent isometry. This corresponds to the gauge freedom of the MPS, which subtracts another $\chi^2$ parameters, overall matching the optimal parameter count in \cref{eq:mps_parameters}. Thus, we arrive at the following circuit structure\footnote{The dotted multiplexer on $n$ qubits here is to be interpreted like above, with one dotted multiplexer node and $(n-1)$ usual multiplexer nodes.}:

\begin{equation}\label{eq:mps_clifford_rot_skeleton}
\resizebox{\linewidth}{!}{%
\includegraphics{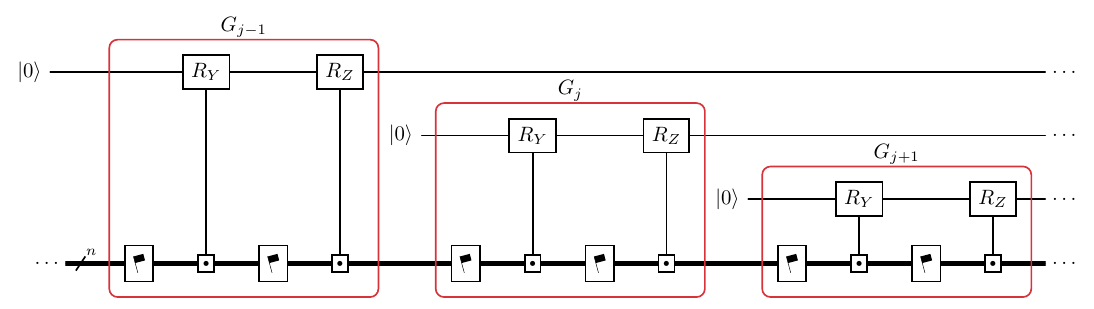}.
}
\end{equation}
Note that this synthesis procedure requires a causal ordering from left to right. We first synthesize $G_{j-1}$ like in \cref{eq:synth_G_j_clifford_rot}, then merge its $V'$ into $G_{j}$, and then continue to synthesize it.
Not shown here are additional savings at the boundary; see \cref{sec:boundary-effects}. 

Since we target the \{Clifford + Rot\} decomposition in this section, let us determine the $\cnot$ cost. The cost for each constituent of the $G_j$ circuit is provided in \cref{tab:gate_counts_subroutines}. In particular, each symmetrized multiplexed single-qubit rotation costs $2^n-1$ $\cnot$s, and each $n$-qubit flag circuit costs another $\tfrac12 4^n - \tfrac{n+12}{8} 2^n + 1$. In total we obtain a $\cnot$ cost of
\begin{equation}
C^{G_j}_{\cnot} = 4^n - \frac{n+4}{4} 2^n.
\end{equation}

\subsection{Phase gradient decomposition}
\label{subsec:mps_phase_gradient}

In this section, we will use the unitary synthesis circuit from \cref{fig:two} for a phase gradient decomposition in the MPS preparation circuit \cref{eq:mps_prep_barebone_skeleton}. We again start by removing $\chi^2$ parameters due to the fixed initial state, corresponding to the isometry condition of $A^{\sigma_j}$ (and, hence, $G_j$):

\begin{align}
\resizebox{\linewidth}{!}{%
\includegraphics{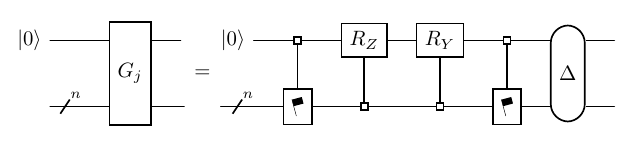}%
} \nonumber \\
\label{eq:mps_phase_grad_reduction}
\resizebox{.85\linewidth}{!}{%
\includegraphics{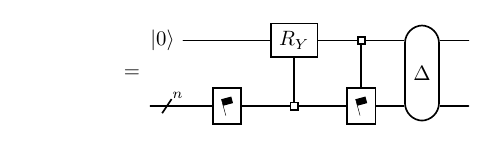}%
}.
\end{align}
The remaining $n$-qubit flag left of the $R_Y$ multiplexer is the one that triggers on $0$ on the first qubit. Like in \cref{eq:remove_RZ}, the $R_Z$ multiplexer with $|0\rangle$ as its input on the target qubit yields a diagonal on the multiplexing qubits, which can be merged to the right.
We confirm that we now have $(4^n-2^n) + 2^n + 2\cdot(4^n -2^n) + 2^{n+1} = 3 \cdot 4^n = 3\chi^2$ parameters.

We now remove another $\chi^2$ parameters by merging unitaries on the auxiliary $n$-qubit register (note that this corresponds to removing the gauge degree of freedom of the MPS). To achieve this, we first split the diagonal gate (see \cref{eq:balance_diagonal_circ,eq:MEP_flag2}),
\begin{align}\label{eq:diagonal_split_mps}
\resizebox{.6\linewidth}{!}{%
\includegraphics{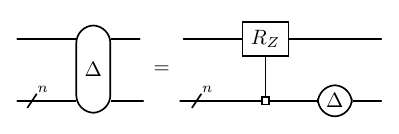}%
}.
\end{align}
This then leaves us with
\begin{align}\label{eq:isometry_before_merging_left_right}
\resizebox{\linewidth}{!}{%
\includegraphics{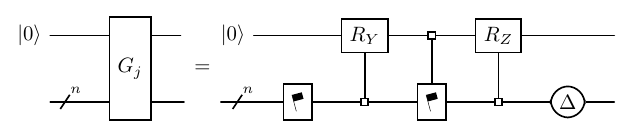}%
}
\end{align}
for the individual isometries. The diagonal on the auxiliary $n$-qubit register can be merged to the right, and the $n$-qubit flag to the left in the MPS preparation circuit \cref{eq:mps_prep_barebone_skeleton}. We note that because we merge to both left and right, the causal ordering in which the decompositions are performed for each $G_j$ is non-trivial and detailed for the full synthesis process in \cref{sec:mps_causal_ordering}. Overall, we obtain the following circuit structure:

\begin{equation}\label{eq:mps_phase_gradient_skeleton}
\resizebox{\linewidth}{!}{%
\includegraphics{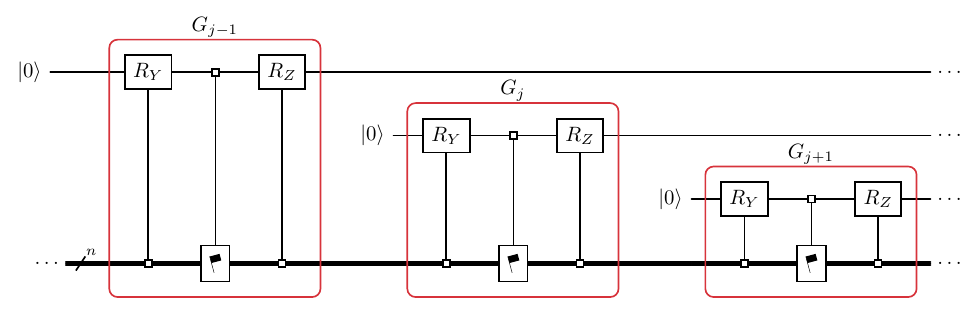}
}.
\end{equation}
Compared to \cref{eq:mps_clifford_rot_skeleton}, we have one multiplexed flag instead of two individual $n$-qubit flags, which is better suited for a phase gradient decomposition.
Overall, we achieve the desired optimal $2\chi^2$ parameters from \cref{eq:mps_parameters}.

Let us now determine the Toffoli cost for each isometry $G_j$. The multiplexed flag contributes $2^n-1$ multiplexed single-qubit flags,

\begin{align*}
\resizebox{\linewidth}{!}{%
\includegraphics{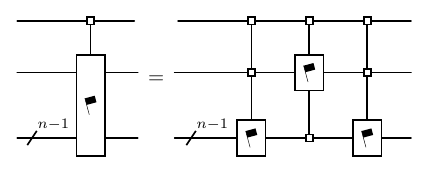}
}.
\end{align*}
Each of the multiplexers is now implemented using the phase gradient decomposition from \cref{sec:main_flag_decomposition}. The classically controlled diagonal gates that arise in the uncomputation of the QROM for each multiplexer can again be merged and pushed through all layers as done in \cref{eq:diagonal_push}, then merged with the final multiplexed $R_Z$ rotation.

Further, we can combine the final multiplexed $R_Z$ with the subsequent multiplexed $R_Y$ from the next isometry, creating the following circuit building block:

\begin{equation}
\resizebox{.6\linewidth}{!}{%
\includegraphics{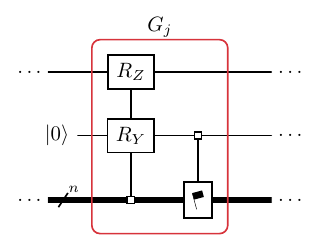}
}.
\end{equation}
This combined $R_Z$ and $R_Y$ multiplexer on two separate qubits has the same cost as a multiplexed single-qubit flag. So we overall have $(2^n-1) + 1 = 2^n$ multiplexers. Each of them contribute the Toffoli cost $\frac{2^{n+1}}{2\Lambda} + 2\Lambda b - 5$ of a single qubit flag circuit, as detailed in \cref{appendix:flag_mul}.

The total cost for an isometry $G_j$ is then given by
\begin{align*}
    C^{G_j}_\text{Tof} &= 
    2^n\left(\frac{2^{n+1}}{2\Lambda} + 2\Lambda b - 5\right) 
    \text{ (flag multiplexers)}.
\end{align*}
When we compare this with the Toffoli cost provided in~\cite{berry2024rapid},

\begin{align*}
    C^{\text{\cite{berry2024rapid}}}_\text{Tof} &= 2^n \left(\frac{2^{n+1}}{2 \Lambda} + 2 \Lambda b - 5\right) \text{ (phase multiplexers)}
    \\&+ \frac{2^{n+1}}{2 \Lambda} + 2 \Lambda b + \frac{2^{n+1}}{ \Lambda'} + \Lambda' - 5 \text{ (final diagonal)}
    \\&+ ((n+1)-3)(2^{n+1}/2-1) \text{ (incrementers)}
    \\&+ \frac{2^{n+1}}{2\Lambda} + \Lambda b - 2 \text{ (controlled qubit rotation)},
\end{align*}
we see that we save the final diagonal, the multiplexed $n$-qubit rotation, as well as not requiring incrementers and decrementers due to the improved synthesis method. This is achieved by utilizing flag decompositions to ensure parameter-optimality with respect to the MPS manifold.

\section{Conclusion}
In this work, we introduce the \textit{flag decomposition} as a central tool for synthesis of unitaries and isometries.
It enables us to produce parameter-optimal quantum circuits for generic unitaries and matrix product state (MPS) preparation that improve upon the state of the art, both when compiling down to the \{Clifford+Rot\} gate set and when using the phase gradient decomposition.


We find that the recursive flag decomposition yields the same circuit as a synthesis technique from 2004 \cite{bergholm2005quantumuniformlycontrolled}, that, to our knowledge, has been mostly underappreciated, if not forgotten.
Striving for a reduction in $\cnot$ counts, the literature dismissed it in favour of the quantum Shannon decomposition (QSD, 2004 \cite{shende2003minimal}) and, in a recent upgrade, the Block-ZXZ decomposition (2024, \cite{Krol-Al-Ars}). While they lowered the $\cnot$ count, these techniques introduced a significant overparametrization, which presents an unfavourable tradeoff for fault-tolerant quantum computation.

For \{Clifford + Rot\} decompositions, we further improve upon the recursive flag decomposition. For this, we draw inspiration from a key innovation of the QSD, called de-multiplexing.
While it leads to overparametrization when applied greedily, we are able to avoid this problem after a careful cost analysis, by applying it \textit{selectively}, across both the recursion levels and the circuit depth. This leads us to \textit{selective de-multiplexing} (SDM), our second parameter-optimal synthesis method, with improved $\cnot$ counts over the recursive flag decomposition for \{Clifford + Rot\} decompositions.

We then use our new synthesis techniques to compile MPS preparation circuits, a crucial subroutine for a number of promising quantum algorithms, in particular in the context of quantum chemistry and material design applications.
Our techniques prove well-suited for the adjustment to the isometry synthesis needed for this task, and the flag decomposition allows us to further optimize the resulting circuits.
In addition, we exploit the mathematical framework of recursive Cartan decompositions to adjust our synthesis to the conditions at the boundary of the MPS, maintaining parameter-optimality where existing simplifications cannot.
Our detailed cost analysis shows that we again improve over the best known methods both for the \{Clifford + Rot\} decomposition and the phase gradient decomposition.

Once recognized, flag decompositions---and the resulting ``flag circuits"---can be found across the unitary synthesis literature.
This underlines their usefulness for both unifying and advancing the field, and we believe that developing a systematic (Lie) group-based framework and computational toolbox will continue to prove fruitful for quantum compilation of parametric gate families.

Future research questions on unitary synthesis span a wide range, from mathematical questions about whether lower bounds on the $\cnot$ count and circuit depth can be achieved, to practical analyses of approximate compilation and hybrid decompositions between the \{Clifford + Rot\} gate set and phase gradient rotations motivated by space constraints.
Regarding the minimal achievable $\cnot$ count, recent progress on small-scale synthesis suggests that saturating both the $\cnot$ and the rotation lower bounds simultaneously is possible, but an efficient linear algebra synthesis technique remains elusive~\cite{wierichs2025unitarysynthesisoptimalbrick}.
A systematic extension of the group-based perspective to more specific circuit structures would be another interesting direction to consider, bridging unitary synthesis to other fields in quantum compilation.

\bibliographystyle{quantum}
\bibliography{main}

\appendix
\crefalias{section}{appendix}
\crefalias{subsection}{appendix}
\numberwithin{equation}{section}

\renewcommand{\theHequation}{\thesection.\arabic{equation}}

\section{Implementation details}\label{app:impl_details}

Here we discuss the implementation of the various algorithmic components for the recursive flag decomposition and selective de-multiplexing. We follow a top-down approach, starting with the coarse-grained building blocks and working our way down to base cases of the used recursions. The implementation can be found in the \texttt{flagsynth} github repository \cite{our_repo}.

\subsection{Main algorithms}\label{app:impl_details:main_blocks}

First, we show in \cref{algo:recursive_flag_decomp} the core recursive decomposition of (multiplexed) multi-qubit operators.
We use it as the recursive flag decomposition for phase gradient implementations, described in \cref{sec:flag_decomposition}, and as a subroutine in selective de-multiplexing, described in \cref{sec:selective_de_multiplexing}.

We distinguish two base cases when the recursion reaches either a single qubit or two-qubit unitaries, indicated by $n_b\in\{1, 2\}$; see \cref{algo:base_case_flag_decomps} below. \textsc{CSD} is the cosine-sine decomposition, the output of which is processed through a recursive call to \textsc{core\_decomp} in \cref{algo:recursive_flag_decomp}. \textsc{balance\_diagonal} implements \cref{eq:balance_diagonal_circ} and \textsc{mux} simply refers to attaching multiplexer nodes to a pair of gate sequences, effectively merging them into a single gate sequence. $\smallflag(\theta_Z, \theta_Y, c, t)$ denotes a multiplexed single-qubit flag with angles $\theta_{Y,Z}$, multiplexing qubits $c$ and target qubit $t$. 
$\kappa\in\{\text{True}, \text{False}\}$ encodes whether we break down multiplexed single-qubit gates via \textsc{dec\_mux\_1QF}, which is described in \cref{sec:recursive_csd:details} and is used for SDM. We do not define the sub-diagonals $\Delta_2^{(i,j)}$ but refer to the implementation for details~\cite{our_repo}. We use $\cup$ to denote list concatenation, and Pythonic slicing notation.

\begin{algorithm}[H]
\begin{algorithmic}[1]
\Procedure{core\_decomp}{$\{V_i\in U(2^{n-k})|0\leq i < 2^k\}, q_\text{mux}\in \mathbb{Z}^k,q_\text{targ}\in\mathbb{Z}^{n-k}, n_b\in\{1, 2\},\kappa\in\{0,1\}$}
    \If{$n-k=n_b$}
        \If{$n_b=1$}
            \State $F_i, \Delta_i \gets \textsc{one\_qubit\_flag}(V_i, q_\text{targ})$
        \Else
            \State $F_i, \Delta_i \gets \textsc{two\_qubit\_flag}(V_i, q_\text{targ})$
        \EndIf
        \State $F \gets \textsc{mux}(\{F_i\}, q_\text{mux})$
        \State $\Delta \gets \bigoplus_{i=0}^{2^k-1} \Delta_i$
        \If{$\kappa$}
            \State $F, \tilde{\Delta} \gets \textsc{dec\_mux\_1QF}(F)$
            \State $\Delta \gets \Delta \cdot \tilde{\Delta}$
        \EndIf
        \State \textbf{return} $F, \Delta$
    \EndIf
    \State $K^{(i)}_{00}, K^{(i)}_{01}, \theta_Y^{(i)}, K^{(i)}_{10}, K^{(i)}_{11}\gets \textsc{CSD}(V_i)$
    \State $q_\text{mux}' \gets q_\text{mux} \cup q_\text{targ,[:1]}$
    \State $q_\text{targ}' \gets q_\text{targ,[1:]}$
    \State $F_1, \Delta_1 \gets \textsc{core\_decomp}(\{K_{1j}^{(i)}\}, q_\text{mux}', q_\text{targ}', n_b, \kappa)$
    \State $\theta_Z,\Delta_2 \gets \textsc{balance\_diagonal}(\Delta_1, k)$
    \State $F_A \gets \smallflag(\theta_Z, \theta_Y, q_\text{mux} \cup q_\text{targ}' \cup q_\text{targ,[:1]})$
    \State $\Delta_2\gets \Delta_2 \underset{k}{\otimes}(1, 1)$
    \If{$\kappa$}
        \State $F_A, \tilde{\Delta} \gets \textsc{dec\_mux\_1QF}(F_A)$
        \State $\Delta_2 \gets \Delta_2 \cdot \tilde{\Delta}$
    \EndIf
    \State $K_{0j}^{(i)} \gets K_{0j}^{(i)} \cdot \Delta_2^{(i,j)}$
    \State $F_0, \Delta_0 \gets \textsc{core\_decomp}(\{K_{0j}^{(i)}\}, q_\text{mux}', q_\text{targ}', n_b, \kappa)$
    \State \textbf{return} $F_1 \cup F_A \cup F_0 , \Delta_0$
\EndProcedure
\end{algorithmic}
\caption{The recursive flag decomposition of (multiplexed) multi-qubit operators. See the appendix text for details about notation and used subroutines.\hspace*{\fill}\label{algo:recursive_flag_decomp}}
\end{algorithm}

Next, we show in \cref{algo:po_qsd} the SDM algorithm described in \cref{sec:selective_de_multiplexing}. It uses a standard two-qubit decomposition \textsc{two\_qubit\_decomp}  with $15$ rotations and $3$ $\cnot$s as base case,~\cref{eq:two-qubit-unitary}.
\textsc{de\_mux} de-multiplexes a $1$-multiplexed unitary—see \cref{eq:de-multiplexing-qsd}—\textsc{rec\_flag\_dec} is given in \cref{algo:recursive_flag_decomp_cliff_rz} below, and \textsc{möttönen} implements the Möttönen decomposition (see \cref{eq:Möttönen-generic}), symmetrized according to the last argument to the right or left. To enable the symmetrization, we need to re- and de-multiplex the central two unitaries via \textsc{re\_de\_mux}. This is a combination of simple matrix multiplication and \textsc{de\_mux}; see the implementation in \cite{our_repo} for more details.

\begin{algorithm}[H]
\begin{algorithmic}[1]
\Procedure{sdm}{$V\in U(2^{n}),q, n\geq 2$}
    \If{$n=2$}
        \State \textbf{return} \textsc{two\_qubit\_decomp}(V)
    \EndIf
        \State $K_{00}, K_{01}, \theta_Y, K_{10}, K_{11}\gets \textsc{CSD}(V)$
        \State $M_{10}, \theta^{(1)}_Z, M_{11} \gets \textsc{de\_mux}(K_{10}, K_{11})$
        \State $M_{00}, \theta^{(0)}_Z, M_{01} \gets \textsc{de\_mux}(K_{00}, K_{01})$
        \State $M_{01}',\theta_Y',M_{10}' \gets \textsc{re\_de\_mux}(M_{01}, M_{10}, \theta_Y, q)$

        \State $q'\gets q_{[1:]}$
        \State $F_{11}, \Delta_{11}\gets \textsc{rec\_flag\_dec}(M_{11}, q', n_b=2)$
        \State $F_{10}, \Delta_{10}\gets \textsc{rec\_flag\_dec}(M_{10}'\cdot \Delta_{11}, q', n_b=2)$
        \State $F_{01}, \Delta_{01}\gets \textsc{rec\_flag\_dec}(M_{01}'\cdot \Delta_{10}, q', n_b=2)$

        \State $U_{00} \gets \textsc{sdm}(M_{00}\cdot \Delta_{01}, q')$

        \State $C_{Z,1}=\textsc{möttönen}(\theta_Z^{(1)},q', q_0, Z,\text{sym}=R)$
        \State $C_{Y}=\textsc{möttönen}(\theta_Y',q', q_0, Z,\text{sym}=R)$
        \State $C_{Z,0}=\textsc{möttönen}(\theta_Z^{(0)},q', q_0, Z,\text{sym}=L)$
        \State \textbf{return} $F_{11}\cup C_Z^{(1)}\cup C_{10}\cup C_Y\cup C_{01}\cup C_Z^{(0)}\cup C_{00}$
\EndProcedure
\end{algorithmic}
\caption{SDM algorithm using an adapted recursive flag decomposition, see \cref{algo:recursive_flag_decomp_cliff_rz}, and symmetrized Möttönen decompositions. See the appendix text for more details.\hspace*{\fill}}\label{algo:po_qsd}
\end{algorithm}

Next, \cref{algo:recursive_flag_decomp_cliff_rz} shows the adapted recursive flag decomposition that we use in the previous SDM algorithm. It combines calls to itself and to the first variant of the (multiplexed) flag decomposition in \cref{algo:recursive_flag_decomp}. Other subroutines are the same as used above in SDM. The modified matrices $\tilde{K}_{0j}$ result from the re- and de-multiplexing step and the $CZ$ gate to be absorbed from the symmetrized M\"ott\"onen decomposition of the $R_Y$ multiplexer.

\begin{algorithm}[H]
\begin{algorithmic}[1]
\Procedure{rec\_flag\_dec}{$V\in U(2^{n}),q, n_b$}
    \If{$n=n_b=1$}
        \State \textbf{return} $\textsc{one\_qubit\_flag}(V, q)$
    \ElsIf{$n=n_b=2$}
        \State \textbf{return} $\textsc{two\_qubit\_flag}(V, q)$
    \EndIf
    
    \State $K_{00}, K_{01}, \theta_Y, K_{10}, K_{11}\gets \textsc{CSD}(V)$
    \State $M_{10}, \theta_Z^{(1)}, M_{11}\gets \textsc{de\_mux}(K_{10}, K_{11})$

    \State $q'\gets q_{[1:]}$
    \State $F_{11},\Delta_{11} \gets \textsc{rec\_flag\_dec}(M_{11}, q', n_b)$
    \State $F_{11}, \tilde{\Delta} \gets \textsc{dec\_mux\_1QF}(F_{11})$
    \State $\Delta_{11}\gets \Delta_{11}\cdot \tilde{\Delta}$

    \State $C_{Z,1}=\textsc{möttönen}(\theta_Z^{(1)}, q', q_0, Z,\text{sym}=R)$
    \State $M_{00}, \theta_Z^{(0)}, M_{01}\gets \textsc{de\_mux}(K_{00}, K_{01})$
    \State $M_{01}',\theta_Y',M_{10}' \gets \textsc{re\_de\_mux}(M_{01}, M_{10}, \theta_Y, q)$
    \State $F_{10},\Delta_{10} \gets \textsc{rec\_flag\_dec}(M_{10}'\cdot \Delta_{11}, q', n_b)$
    \State $F_{10}, \tilde{\Delta} \gets \textsc{dec\_mux\_1QF}(F_{10})$
    \State $\Delta_{10}\gets \Delta_{10}\cdot \tilde{\Delta}$
    \State $C_{Y}=\textsc{möttönen}(\theta_Y', q', q_0, Y,\text{sym}=R)$
    \State $\tilde{K}_{00}=M_{00} \cdot \exp(-\frac{i}{2}\theta_Z^{(0)}) \cdot M_{01}' \cdot \Delta_{10}$
    \State $\tilde{K}_{01}=M_{00} \cdot \exp(\frac{i}{2}\theta_Z^{(0)}) \cdot M_{01}' \cdot Z_0 \cdot \Delta_{10}$
    
    \State $F_0,\Delta \gets \textsc{core\_decomp}(\{\tilde{K}_{00},\tilde{K}_{01}\}, q_0,q_{[1:]}, n_b, 1)$
    \State \textbf{return} $F_{11}\cup C_{Z,1} \cup F_{10} \cup C_Y \cup F_0, \Delta$
\EndProcedure
\end{algorithmic}
\caption{Recursive flag decomposition adjusted to be used in \textsc{sdm} (\cref{algo:po_qsd}).\hspace*{\fill}}\label{algo:recursive_flag_decomp_cliff_rz}
\end{algorithm}

\subsection{Base case algorithms}\label{app:impl_details:base_cases}
With the main algorithmic building blocks covered, we now turn to the base case algorithms used above, beginning with the base case $n_b=2$ flag decomposition.
The solution is based on~\cite[Thm.~VI.3]{shende2003minimal}, given by:
\begin{align}\label{eq:asymmetric_two_qubit_decomp}
    \resizebox{0.85\linewidth}{!}{%
\includegraphics{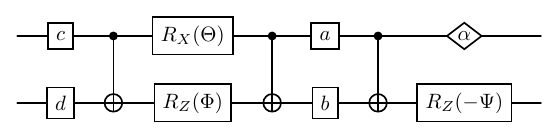},%
    }
\end{align}
where $a,b,c,d\in SU(2)$, the diamond-shaped gate denotes a global phase $e^{i\alpha}$, 
and we report the adjoint of the original decomposition in order to move the diagonal $\Delta$ to the right rather than the left later on.
This decomposition is used by~\cite[Thm.~14]{shende2006synthesis} to arrive at the two-qubit template
\begin{align}\label{eq:two_qubit_flag_pre}
    \resizebox{0.85\linewidth}{!}{%
\includegraphics{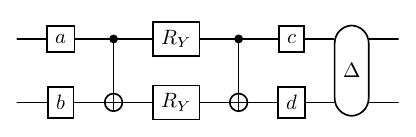}.%
    }
\end{align}
These references prove the existence of the ``asymmetric" decomposition \cref{eq:asymmetric_two_qubit_decomp} and the decomposition up to a diagonal in \cref{eq:two_qubit_flag_pre} constructively, only requiring standard linear algebra tools like matrix multiplication, eigenvalue decomposition\footnote{Strictly speaking, we require eigenvalue decompositions of symmetric unitary matrices into real eigenbases. Those can be achieved either through modification of a standard eigenvalue decomposition~\cite{fuhr2018note} or through simultaneous diagonalization of the real and imaginary parts.} and computing determinants. We implemented the asymmetric decomposition method in \texttt{flagsynth.asymmetric\_two\_qubit\_decomp} and summarize its algorithm for convenience in \cref{algo:asymmetric_two_qubit_decomp}.

\begin{algorithm}[H]
\begin{algorithmic}[1]
\Procedure{asymmetric\_decomp}{$V\in U(4)$}
    \State $V \gets \cnot\cdot V$
    \State $\alpha \gets \tfrac14 \arg(\det{V})$
    \State $V \gets V e^{-i\alpha}$
    \State $\Psi, \Theta, \Phi \gets \text{Prop.V2}(V)$
    \State $V' \gets \Ad_{\cnot} \left( R_Z^{(1)}(\Psi) \right)\cdot V$
    \State $W\gets \Ad_{\cnot}\left(R_X^{(0)}(\Theta)R_Z^{(1)}(\Phi)\right)$
    \State $a, b, c, d\gets \text{Prop.IV.3}(V', W)$
    \State \textbf{return} $a, b, c, d, \alpha, \Psi, \Theta, \Phi$
\EndProcedure
\end{algorithmic}
\caption{Computing the variables for \cref{eq:asymmetric_two_qubit_decomp}. We do not explicitly explain the algorithms detailed in~\cite[Props.~V.2,~IV.3]{shende2003minimal}.\hspace*{\fill}}\label{algo:asymmetric_two_qubit_decomp}
\end{algorithm}

The decomposition in~\cite{shende2006synthesis} yields a constructive description of the two-qubit flag decomposition, but with only two degrees of freedom---that of an Ising-interaction rotation $R_{ZZ}$, and that of the global phase---in the diagonal.
In order to arrive at the flag decomposition with the stated numbers of degrees of freedom, we perform a combination of Euler decompositions and basis transformations of the single-qubit operators, and commute $R_Z$ gates through (temporarily created) $\cz$ gates. This ultimately allows us to absorb two $R_Z$ rotations into $\Delta$, yielding $2^n=2^2=4$ degrees of freedom in $\Delta$ and $4^n-2^n=16-4=12$ degrees of freedom in the remaining flag circuit.
We implemented the technique from~\cite[Thm.~14]{shende2006synthesis} together with this postprocessing step in \texttt{flagsynth.two\_qubit\_flag} (see \cref{algo:base_case_flag_decomps}), which will be used by \cref{algo:recursive_flag_decomp}. The returned variables are to be used in a circuit as follows:

\begin{align}\label{eq:flag_decomp_su4_with_variables}
\resizebox{0.9\linewidth}{!}{%
\includegraphics{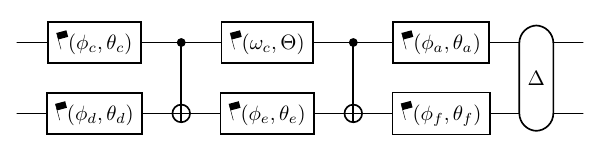}.%
}
\end{align}

Here we used repeated basis changes via rotations about $Y$ or $Z$ with angles $\pm\tfrac\pi2$ to modify the circuit obtained from Euler decompositions into the shown regular arrangement of single-qubit flags.

\begin{algorithm}[H]
\begin{algorithmic}[1]
\Procedure{one\_qubit\_flag}{$V\in U(2), q\in \mathbb{Z}$}
    \State $\theta_Z, \theta_Y, \omega, \phi \gets \textsc{Euler}(V)$
    \State $F \gets [\smallflag(\theta_Z, \theta_Y, q)]$
    \State $\Delta \gets (e^{-i(\phi+\omega/2)}, e^{-i(\phi-\omega/2)})$
    \State \textbf{return} $F, \Delta$
\EndProcedure
\Procedure{two\_qubit\_flag}{$V\in U(4), q\in \mathbb{Z}^2$}
    \State $a, b, c, d, \alpha, \Psi, \Theta, \Phi \gets \textsc{asymmetric\_decomp}(V)$
    \State $\phi_a,\theta_a,\omega_a \gets \textsc{Euler}(a)$
    \State $\phi_c,\theta_c,\omega_c \gets \textsc{Euler}(c)$
    \State $\phi_d,\theta_d,\omega_d \gets \textsc{Euler}(R_Y(-\tfrac{\pi}{2})\cdot d)$
    \State $\phi_e,\theta_e,\omega_e \gets \textsc{Euler}(R_X(-\Phi)\cdot R_Z(\omega_d))$
    \State $\phi_f,\theta_f,\omega_f \gets \textsc{Euler}(b \cdot R_X(\omega_e)\cdot R_Y(\tfrac{\pi}{2}))$
    \State $F,\Delta \gets \textsc{convert\_to\_flags}(C)$
    \State $\omega_c\gets\omega_c+\tfrac{\pi}{2}$
    \State $\phi_a\gets\phi_a-\tfrac{\pi}{2}$
    \State $\Delta\gets \operatorname{diag}\left(R^{(0)}_Z(\omega_a)\cdot R^{(1)}_Z(\omega_f)\cdot R_{ZZ}(\Psi)e^{i\alpha}\right)$
    \State \textbf{return} $\phi_c, \theta_c,\phi_d, \theta_d, \omega_c, \Theta, \phi_e, \theta_e, \phi_a, \theta_a, \phi_f, \theta_f, \Delta$
\EndProcedure
\end{algorithmic}
\caption{One-qubit and two-qubit flag decompositions. $\smallflag(\theta_Z, \theta_Y, q)$ denotes a single-qubit flag with angles $\theta_{Y,Z}$ acting on qubit $q$. The subroutine \textsc{asymmetric\_decomp} is described in \cref{algo:asymmetric_two_qubit_decomp}, and \textsc{Euler} is the standard $ZYZ$ Euler decomposition. See \cref{eq:flag_decomp_su4_with_variables} for the usage of the computed variables.\hspace*{\fill}}\label{algo:base_case_flag_decomps}
\end{algorithm}

\section{Resource counts for phase gradient decompositions}
\label{app:phase_gradient_resources}
Here we collect decompositions and implementation details for the phase gradient decomposition of our unitary synthesis and MPS preparation circuits.

\subsection{QROM cost}\label{app:qrom_cost}

The Quantum Read-Only Memory (QROM) is an operator that stores bitstrings of size \( b \), each associated with a unique index, such that
\begin{equation}
    \text{QROM}\ket{i}\ket{0} = \ket{i}\ket{a_i},
\end{equation}
where each \( a_i \in \{0,1\}^b \) denotes a classical bitstring. Throughout this work, following~\cite{berry2024rapid}, we adopt a resource-efficient variant of QROM that reduces overall cost by forgoing the uncomputation of auxiliary qubits. Specifically, the QROM operator acts as

\begin{equation}\label{eq:QROM-forward-computation}
   \text{QROM} \ket{i} \ket{0} \ket{0} = \ket{i} \ket{a_i} \ket{v_i}, 
\end{equation}
where \( \ket{v_i} \) is an auxiliary state whose content is irrelevant to the computation. The decomposition introduced in~\cite{LKS2024} follows the \emph{SelectSwap} structure, illustrated below:

\begin{equation}
\label{eq:select_swap_circuit}
\resizebox{\linewidth}{!}{%
\includegraphics{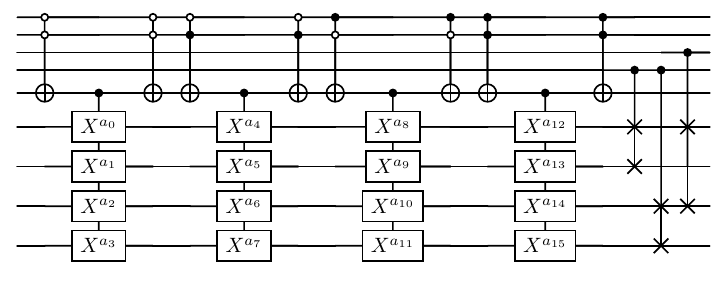}
}.
\end{equation}
In this circuit, the first four qubits represent the input state \(\ket{x}\). The fifth qubit is an auxiliary qubit used in the decomposition process. The following register corresponds to the target qubits, where we aim to load the bitstring \(\ket{a_x}\). The remaining registers consist of additional auxiliary qubits used throughout the computation.

This technique, described in~\cite{GuillermoAlonso2024}, distributes the loading of bitstrings across a two-dimensional memory layout. It partitions the control register into two sets: the first set determines the column in which the bitstring \( \ket{a_i} \) resides, while the second set identifies the corresponding row.

As depicted in the circuit, the QROM operator is divided into two main components. The Select block loads a full column of bitstrings and stores them across the target and auxiliary registers. For example, for the specific case \( \ket{i} = \ket{14} \), the operator acts as

\[
\text{Select} \ket{14} \ket{0} \big(\ket{0}\ket{0}\ket{0}\big) = \ket{14} \ket{a_{12}} (\ket{a_{13}} \ket{a_{14}} \ket{a_{15}}),
\]
where the relevant bitstring \( \ket{a_{14}} \) has been loaded, but not in the target position. The SwapUp block then reorders the bitstrings such that \( \ket{a_{14}} \) is brought to the target register,

\[
\text{SwapUp} \ket{14} \ket{a_{12}} (\ket{a_{13}} \ket{a_{14}} \ket{a_{15}}) = \ket{14} \ket{a_{14}} (\ket{v_{14}}),
\]
where \( \ket{v_{14}} \) denotes a reordered product of the remaining bitstrings, which are no longer used.

To implement the Select block efficiently,~\cite{Babbush_2018} introduced a construction based on TemporaryAnd gates. It was shown that a QROM storing \( 2^n \) entries can be implemented using \( 2^n - 1 \) Toffoli gates.
TemporaryAnd gates are a resource-saving alternative to standard Toffoli gates. They simulate the AND of two control qubits by temporarily storing the result in an auxiliary qubit. These additional qubits are uncomputed later using only Clifford operations, which reduces the overall cost without sacrificing reversibility:

\begin{align*}
\resizebox{\linewidth}{!}{%
\includegraphics{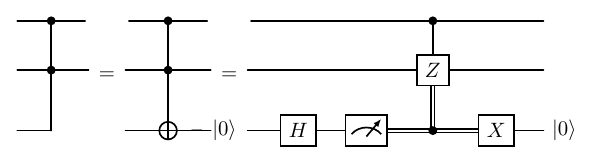}.
}
\end{align*}

The cost of the SelectSwap decomposition depends on the parameter \( \Lambda \), which specifies the number of rows used to arrange the bitstrings. Accordingly, the control register is split into two parts: \( \log_2(\Lambda) \) qubits for the controlled-Swap operations in the $\text{SwapUp}$ block, and \( \log_2(2^n / \Lambda) \) qubits for the Select block.

Given this structure, the Select block requires \( 2^n / \Lambda - 1 \) Toffoli gates, as it acts over a reduced address space, and the $\text{SwapUp}$ block requires \( b(\Lambda - 1) \) Toffoli gates, since each controlled-Swap can be implemented using one Toffoli and two CNOT gates. Hence, the total Toffoli cost of the QROM implementation is given by:
\[
C_{\text{QROM}} = \frac{2^n}{\Lambda} + (\Lambda - 1)b - 1.
\]

\subsection{QROM-based decomposition of \texorpdfstring{$R_Z$}{RZ} multiplexer}
\label{appendix:rz}

The $R_Z$ rotation multiplexer can be decomposed using a QROM, an adder, and a sequence of $\cnot$ gates. This decomposition is illustrated in the following circuit, where we introduce an auxiliary $b$-qubit register onto which the angles are loaded, and another $b$-qubit register that carries a phase gradient state $|\nabla_Z\rangle$:

\begin{align}\label{eq:appendix:RZmultiplexerQROM}
\resizebox{\linewidth}{!}{%
\includegraphics{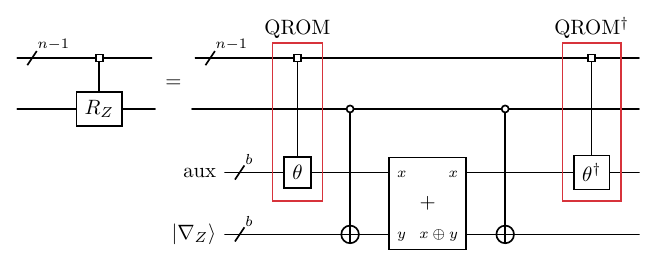}
}.
\end{align}
Here we are using a phase gradient resource state $|\nabla_Z\rangle = \frac{1}{\sqrt{2^b}} \sum_{k=0}^{2^b - 1} e^{-2\pi i k / 2^b} \ket{k},$
and \( b \) denotes the number of bits used to represent the rotation angles of the multiplexer in binary. The circuit leverages the following key properties\footnote{Note that the two sub-systems considered here correspond to the auxiliary qubits and those carrying the phase gradient state in \cref{eq:appendix:RZmultiplexerQROM}. In particular, we write $(I \otimes X) := (I^{\otimes b} \otimes X^{\otimes b})$ for brevity.}:

\begin{enumerate}
    \item[\textbf{(0)}]  \(\text{Add}\ket{x}\ket{y} = \ket{x}\ket{x + y}\)
    \item[\textbf{(1)}] \(\text{Add}\ket{x}\ket{\nabla_Z} = e^{2\pi i x / 2^b} \ket{x}\ket{\nabla_Z}\)
    \item[\textbf{(2)}]  \((I \otimes X)\,\text{Add}(I \otimes X)\ket{x}\ket{y} = \ket{x}\ket{y - x}\)
    \item[\textbf{(3)}]  \((I \otimes X)\,\text{Add}(I \otimes X)\ket{x}\ket{\nabla_Z} = e^{-2\pi i x / 2^b} \ket{x}\ket{\nabla_Z}\)
\end{enumerate}
Property \textbf{(0)} is simply the definition of the \text{Add} operator, which can be used to derive property \textbf{(1)}. This shows how specific phases can be applied using basic
arithmetic.

Another fundamental identity is property \textbf{(2)}, which demonstrates that conjugating the \text{Add} with Pauli-X gates on the target register effectively performs a subtraction. This can be proven by observing that applying an \( X \) gate to \( \ket{y} \) produces \( \ket{1-y} \). After applying the addition, the target register becomes \( \ket{1-y + x} \), and applying \( X \) again yields \( \ket{y - x} \).
Using this identity, we can derive property \textbf{(3)}, which shows that the negative phase can be introduced by conjugating the \text{Add} with \( X \) gates when the target is in the state \( \ket{\nabla_Z} \).

Using the properties described above, we can now demonstrate the equivalence between the arithmetic-based construction and the direct application of multiplexed \( R_Z \) rotations. Assume we begin with an initial state of the form $\sum_i \alpha_i \ket{i} \ket{\psi},$ and our goal is to obtain the state
\[
\sum_i \alpha_i \ket{i} R_Z(\theta_i)\ket{\psi}.
\]
To achieve this, we employ a QROM to load the binary-encoded values of \( \frac{2^b \theta_i}{4\pi} \) into an auxiliary register. That is, we prepare the state
\[
\sum_i \alpha_i \ket{i} \ket{\psi} \ket{\tfrac{2^b \theta_i}{4\pi}}.
\]

Let us write the data qubit as \( \ket{\psi} = \beta\ket{0} + \gamma\ket{1} \). By applying the X gates in the target register controlled on the data qubit, we perform an addition when the data qubit is in \( \ket{1} \), and a subtraction when it is in \( \ket{0} \). This results in the state
\[
\sum_i \alpha_i \ket{i} \left( e^{-2\pi i \left(\tfrac{2^b \theta_i}{4\pi}\right)/2^b} \beta\ket{0} + e^{2\pi i \left(\tfrac{2^b \theta_i}{4\pi}\right)/2^b} \gamma\ket{1} \right) \ket{\tfrac{2^b \theta_i}{4\pi}}.
\]
Simplifying the phase terms, we obtain
\[
\sum_i \alpha_i \ket{i} \left( e^{-i \theta_i / 2} \beta\ket{0} + e^{i \theta_i / 2} \gamma\ket{1} \right) \ket{\tfrac{2^b \theta_i}{4\pi}}.
\]

Finally, we uncompute on the final register, restoring it to \( \ket{0} \), and leaving the state
\[
\sum_i \alpha_i \ket{i} R_Z(\theta_i) \ket{\psi},
\]
which is exactly the transformation we aimed to implement. More information on this technique can be found in the PennyLane compilation hub~\cite{PennyLane-PhaseGradient}.

\subsection{Adder Operator}
\label{appendix:adder}
The adder operator, defined as $\text{Add}\ket{x}\ket{y} = \ket{x}\ket{y + x}$, is a well-studied operation, with notable implementations such as in~\cite{Gidney2018halving}, where the decomposition relies on the use of \text{TemporaryAnd} operators. In our analysis, we count each \text{TemporaryAnd} as a Toffoli gate, as it performs an equivalent logical function. However, we do not count the uncomputation of \text{TemporaryAnd}, since it can be implemented using only Clifford gates and measurements. In the decomposition shown below, for the case of adding numbers with $b = 5$ digits, it can be observed that $(b - 1)$ Toffoli gates are required:

\begin{align*}
\resizebox{\linewidth}{!}{%
\includegraphics{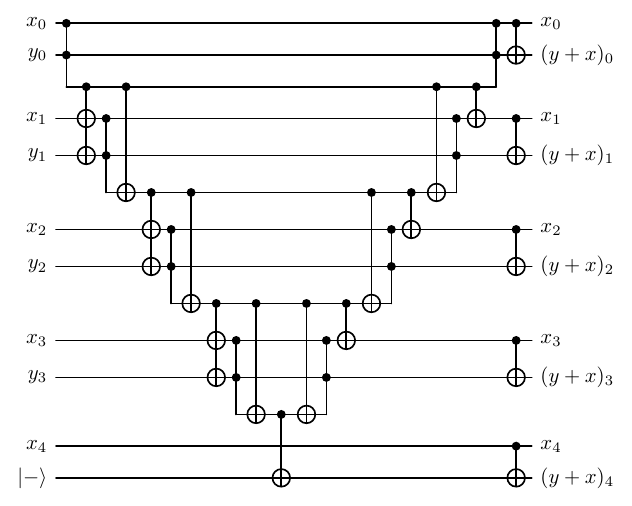}
}.
\end{align*}
We see how this realizes the operation $\text{Add}\ket{x}\ket{y} = \ket{x}\ket{y + x}$ when representing $x_0$ as the least significant bit of $x$, and $x_4$ as the most significant bit. For example, $|x = 3\rangle = |x_4 x_3 x_2 x_1 x_0\rangle = |00011\rangle$.
Note that in the qubit $y_4$, we take the input state to be $\ket{-}$. This simplification comes because of two helpful observations. First, for our usage, the second input register will always contain a phase gradient state, i.e., $|y_4 y_3 y_2 y_1 y_0\rangle = |\nabla_Z\rangle$. Second, the most significant wire of a phase gradient state will always be in a $\ket{-}$ state, regardless of size.
This allows for a simplification of the adder circuit, as there are two CNOT gates targeting this wire. When these are activated, they effectively flip the sign of the state, which is equivalent to replacing the CNOT gates with a single $Z$ gate acting on the control qubits:

\begin{align*}
\resizebox{\linewidth}{!}{%
\includegraphics{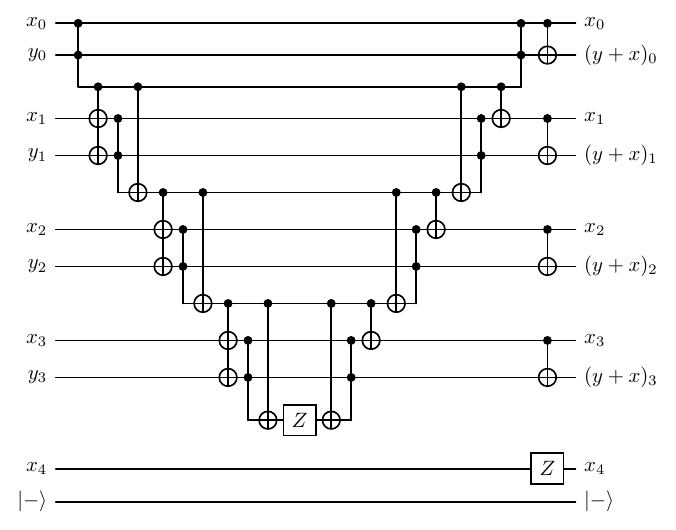}
}.
\end{align*}

Finally, as presented in~\cite{Gidney2018halving}, the circuit can be simplified by removing one of the \text{TemporaryAnd}s:

\begin{align*}
\resizebox{\linewidth}{!}{%
\includegraphics{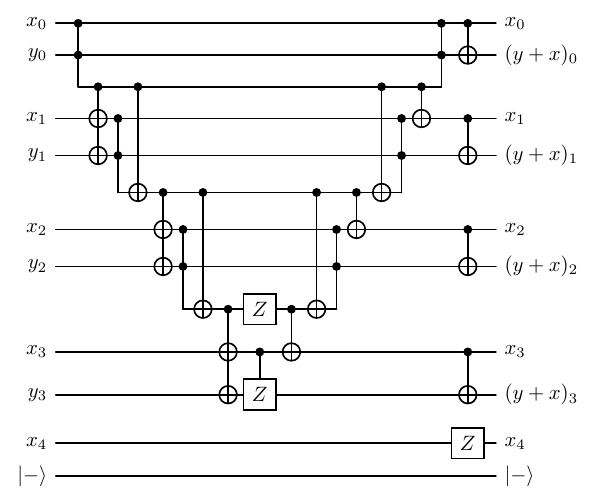}
}.
\end{align*}
In aggregate, the Toffoli gate cost is reduced to $(b - 2)$, a value that will be used throughout this work, and also used in \cite{Sanders_2020}.

\subsection{Multiplexer of 1-qubit flag operators}
\label{appendix:flag_mul}

In this section, we introduce the decomposition of single-qubit flag-operator multiplexers, that is, multiplexers over gates of the form \( R_Y(\phi) \cdot R_Z(\theta) \). To this end, we exploit the identity \( R_Y = H_Y R_Z H_Y^\dagger \), where \( H_Y = S H \), together with the decomposition of the \( R_Z \) multiplexer derived in the previous section. The main idea is to load, in parallel, the binary representations of \( \theta \) and \( \phi \) using a QROM. Then, two adders are applied between these representations together with a phase gradient resource, followed by an uncomputation of the QROM:

\begin{align*}
\resizebox{\linewidth}{!}{%
\includegraphics{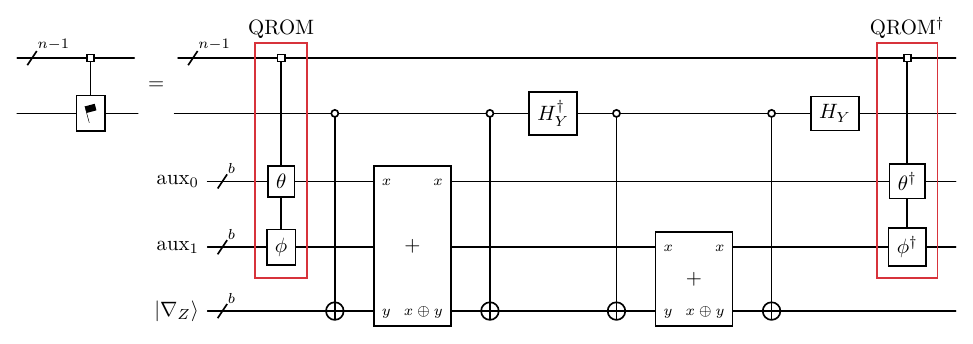}
}.
\end{align*}
The last operation, $\text{QROM}^\dagger$, corresponds to reversing the action described in \cref{eq:QROM-forward-computation}. We can achieve this by measuring all target qubits on the auxiliary register in the $X$ basis and, based on the measurement outcomes, applying phase corrections to the control qubits:

\begin{align*}
\resizebox{\linewidth}{!}{%
\includegraphics{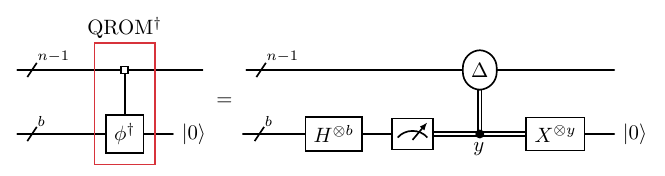}%
}.
\end{align*}
Here, $X^{\otimes y} = \bigotimes_j X^{y_j}$ is a shorthand notation for applying $X$ whenever the measurement result $y_j=1$ on the respective qubit to reset it to $|0\rangle$.

Our goal here is to map any input state $\ket{i} \ket{\phi_i} \ket{\psi_i}$ back to $\ket{i} \ket{0} \ket{\psi_i}$, where $\ket{\phi_i}$ is the $i$th QROM bitstring and $\ket{\psi_i}$ is an arbitrary state that may be entangled with these qubits.

Measuring the auxiliary qubits in the $X$ basis is equivalent to applying a Hadamard gate followed by a measurement in the $Z$ basis. After applying the Hadamard gate, the system evolves into the state\footnote{Here, $\odot$ refers to bitwise multiplication.}
\[
\sum_i \alpha_i \ket{i} \left( \tfrac{1}{\sqrt{2^b}}\sum_x (-1)^{\phi_i \odot x} \ket{x} \right) \ket{\psi_i}.
\]
Assuming that the measurement yields the bitstring $\ket{y}$, the post-measurement state becomes
\[
\sum_i \alpha_i (-1)^{\phi_i \odot y} \ket{i} \ket{y} \ket{\psi_i}.
\]
Therefore, by uncomputing the register containing $\ket{y}$ (e.g., using $X$ gates), and applying a diagonal operator $\Delta$ such that $\Delta_{ii} = (-1)^{\phi_i \odot y}$, we obtain the desired state\footnote{Auxiliary wires used in the QROM decomposition can also be uncomputed by applying the same measurement trick.}.

The diagonal operator $\Delta$ consists of $\pm 1$ entries, which allows it to be merged with the subsequent multiplexers in the circuit. For this reason, we do not account for its cost within this block. Based on this, the cost of the flag multiplexers consists of the initial QROM (using $2b$ target qubits and $n-1$ control qubits), together with two adders, each with cost $b - 2$, as detailed in ~\cref{appendix:adder}. This results in a total cost of:

$$\left(\frac{2^{n-1}}{\Lambda} + (\Lambda - 1)2b - 1\right) + 2(b-2) = \frac{2^n}{2\Lambda} + 2\Lambda b - 5.$$

\subsection{Phase gradient decomposition of a diagonal operator}
\label{appendix:diagonal}

We now want to show how to implement a diagonal operator using a special phase gradient decomposition.
We first note that one can identify a diagonal operator as a multiplexed $RZ$ rotation with an additional qubit initialized in $|0\rangle$ as described in \cref{eq:remove_RZ}. The same logic applies for an initial state in $|1\rangle$, such that we get

\begin{align}\label{eq:artifically_introduce_qubit_for_diagonal}
\resizebox{.5\linewidth}{!}{%
\includegraphics{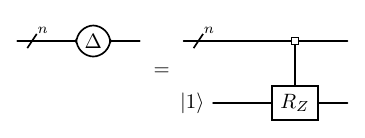}.
}
\end{align}
For diagonal elements $\Delta_{ii} = \exp(i \theta_i)$, the angles of the multiplexed $R_Z$ rotation are given by $2\theta_i$.
This equivalence holds because applying the multiplexed $R_Z$ with angles $2\theta_j$ to a basis state $\ket{j}\ket{1}$ yields $\ket{j}R_Z(2\theta_j)\ket{1} = e^{i \theta_j} \ket{j}\ket{1}$\footnote{Recall that $R_Z(\theta) = e^{-i\tfrac\theta2 Z} = \text{diag}\left(e^{-i\tfrac\theta2}, e^{i\tfrac\theta2}\right)$.}. Thus, a diagonal matrix $\Delta$ acting on the control qubits such that $\Delta\ket{j} = e^{i \theta_j} \ket{j}$, is equivalent to the multiplexed $R_Z$ rotation with angles $2\theta_j$ in \cref{eq:artifically_introduce_qubit_for_diagonal}.

We can now use \cref{eq:appendix:RZmultiplexerQROM} to realize the multiplexed $R_Z$ gate and overall obtain

\begin{align*}
\resizebox{\linewidth}{!}{%
\includegraphics{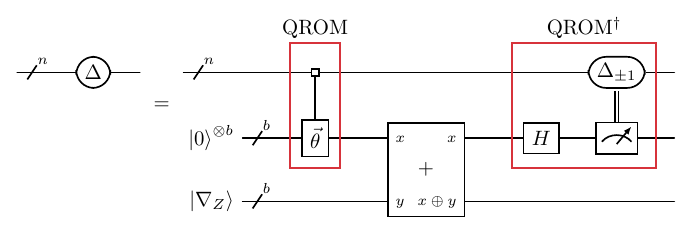}
}.
\end{align*}
Due to the fixed initial state $|1\rangle$ on the target qubit of the multiplexed $R_Z$ rotation, the two CNOT gates in ~\cref{eq:appendix:RZmultiplexerQROM} controlled on the target wire become obsolete. Hence, we can remove the qubit we artificially introduced previously in \cref{eq:artifically_introduce_qubit_for_diagonal} again.

The diagonal correction in the adjoint QROM consists of $\pm 1$ elements and cannot be merged into any subsequent operation in our unitary synthesis scheme, as the full diagonal operator itself appears at the end of the circuit. Therefore, we must describe how to apply the sign corrections. In particular, we will use the circuit introduced in~\cite{Berry_2019}, which employs a QROM to store whether a sign correction should be applied to the $i$th element. This is done by loading a $\ket{1}$ into the target qubit if a correction is needed, or leaving it in the $\ket{0}$ state otherwise. 

By leveraging the Select block and the Swap block (also called $\text{SwapUp}$~\cite{PennyLane-SwapNetwork}) described in the QROM construction~\cite{LKS2024}, we can implement the sign correction through the circuit identity
\begin{align*}
\resizebox{\linewidth}{!}{%
\includegraphics{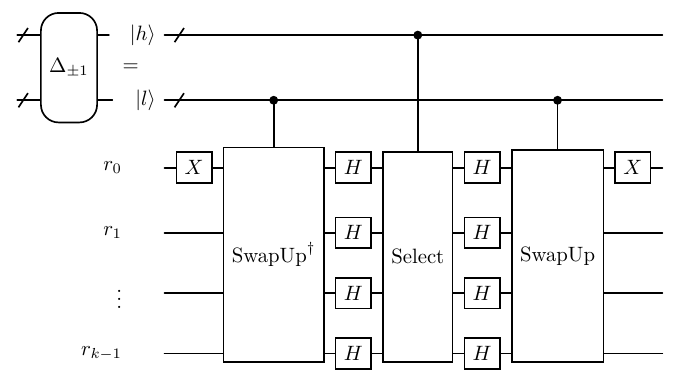}
},
\end{align*}
where $\ket{h}$ and $\ket{l}$ are the two subsets of wires that define the QROM control wires as described in~\cref{app:qrom_cost}, $r_0$ is the target qubit used to store the signs, and the remaining qubits $r_1, \ldots, r_{k-1}$ are auxiliary qubits used for the encoding and manipulation of the QROM data. In this circuit, the size of the register will always be $b=1$, since this is enough to store the target sign. Therefore $a_i$ is a single bit instead of a bitstring.
To understand the behavior of the circuit, let us assume that all the $a_i$ bits are arranged in a $4 \times 4$ matrix in the circuit, distributed in depth and width, as shown in the SelectSwap circuit in \cref{eq:select_swap_circuit}, and that all auxiliary qubits are initially in the $\ket{0}$ state. Let us walk through an example with $k=4$ and suppose we input the state $\ket{1110} = \ket{14}$, i.e., $\ket{h}=\ket{11}=\ket{3}$ and $\ket{l}=\ket{10}=\ket{2}$. After applying the first $X$ gate, the state becomes
\[
(\ket{h}\ket{l})(\ket{r_0}\ket{r_1}\ket{r_2}\ket{r_3})=(\ket{11}\ket{10})(\ket{1}\ket{0}\ket{0}\ket{0}).
\]
Applying the first $\text{SwapUp}^\dagger$ block places the $\ket{r_0} =\ket{1}$ into the row where $a_{14}$ will be encoded,
\[
(\ket{11}\ket{10})(\ket{0}\ket{0}\ket{1}\ket{0}),
\]
which is the second position since $l=2$.
Next, Hadamard gates are applied, yielding the state
\[
(\ket{11}\ket{10})(\ket{+}\ket{+}\ket{-}\ket{+}).
\]
When the third column data is loaded ($h=3$) using the Select block, we apply the operator $X^{a_{12}}\otimes X^{a_{13}}\otimes X^{a_{14}}\otimes X^{a_{15}}$ onto the target register. This yields a negative phase from only the $\ket{-}$ component, as $|+\rangle$ is unaffected by $X$. Therefore, we obtain
\[
(\ket{11}\ket{10})(\ket{+}\ket{+} (-1)^{a_{14}}\ket{-} \ket{+}).
\]
After reapplying Hadamard gates, undoing the $\text{SwapUp}$ with $\text{SwapUp}^\dagger$, and executing the final $X$ on the first auxiliary qubit, we arrive at
\[
(-1)^{a_{14}} (\ket{11}\ket{10}) (\ket{0}\ket{0}\ket{0}\ket{0}).
\]
Therefore, the $i$th bit stored in the QROM determines the sign of the $i$th element of the diagonal operator.

The cost of the \text{SwapUp} block in this construction is $(\Lambda' - 1)$, where $\Lambda'$ denotes the number of rows used in this QROM (with $b = 1$). The cost of the Select operation is given by $2^n / \Lambda' - 1$, resulting in a total cost for this block of
\[
\frac{2^n}{\Lambda'} + \Lambda' - 2.
\]
Note that we do not assign any additional cost to the $\text{SwapUp}^\dagger$ block since, as shown in~\cite{Berry_2019}, it can be implemented without Toffoli gates: all Toffolis involved in this uncomputation can be replaced by their corresponding \text{TemporaryAnd} adjoints. In \cref{appendix:undo_swap} we detail this decomposition.

Combining all these components, we deduce that the cost of implementing the diagonal operator consists of three main contributions. The first comes from the initial QROM, which requires
\[
\frac{2^n}{2\Lambda} + (2\Lambda - 1)b - 1
\]
Toffoli gates. This corresponds to a QROM with $2\Lambda$ rows, which is feasible here since we have more qubits available (we do not need to load two angles in parallel).

The second contribution comes from the adder, with cost $b - 2$, and finally, the third contribution comes from the sign-correction block described previously, whose cost is
\[
\frac{2^n}{\Lambda'} + \Lambda' - 2.
\]
Adding these terms together, the total Toffoli cost becomes
\[
\frac{2^n}{2\Lambda} + 2\Lambda b + \frac{2^n}{\Lambda'} + \Lambda' - 5.
\]

\subsection{Uncomputing Swap block}

\label{appendix:undo_swap}
In the context of the QROM, a \text{SwapUp} block is employed to re-arrange a bitstring across auxiliary qubits. This operation is defined as follows:

\begin{equation}
\resizebox{\linewidth}{!}{%
\includegraphics{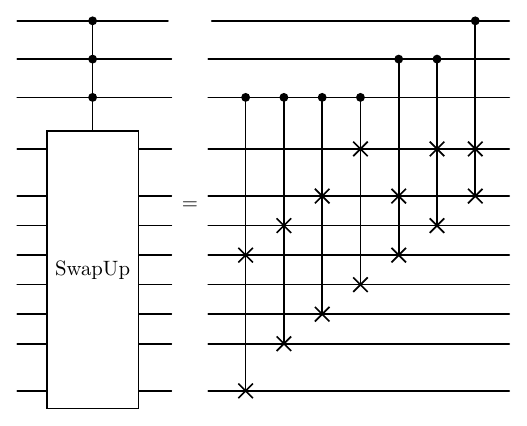}
}.
\end{equation}
Note that the control nodes on the left-hand side are there to indicate the wires that the right-hand side circuit is controlling on\footnote{In particular, the \text{SwapUp} box is not a well-defined operation that we control, but rather the box together with the control nodes define the operation.}.
When this operation is applied at the end of a circuit to uncompute qubits, i.e., we know the output will be $\ket{0}$, it can be implemented without the use of any Toffoli gates. Consider the following decomposition, in which each controlled-Swap gate is expressed solely in terms of CNOT gates and Toffoli gates:

\begin{align}
\resizebox{\linewidth}{!}{
\includegraphics{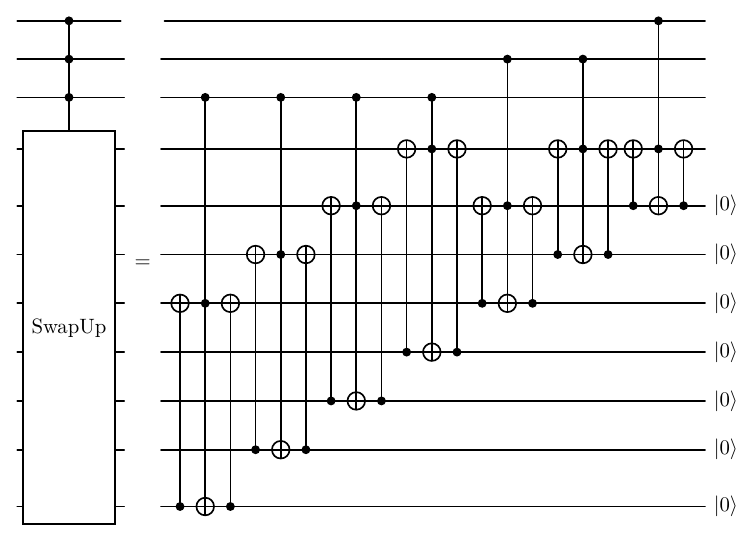}}.
\end{align}
Next, as shown in \cite{Berry_2019}, the CNOT gate after each Toffoli gate can be eliminated, and the Toffoli gates can be replaced by the adjoints of \text{TemporaryAnd} operations:

\begin{align}
\resizebox{\linewidth}{!}{
\includegraphics{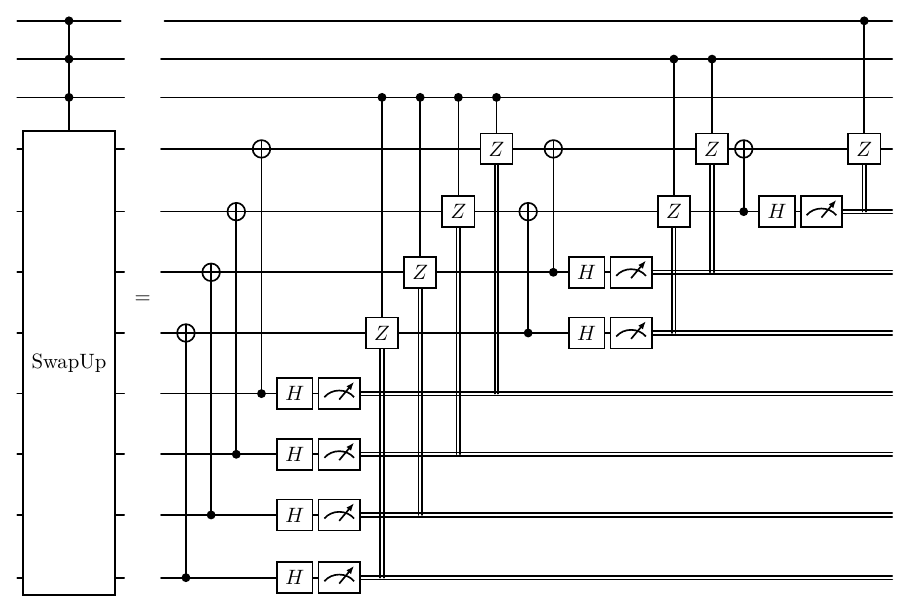}
}.
\end{align}

\section{Recursive Cosine-Sine decomposition}\label{sec:recursive_csd}

Here, we review the optimized recursive cosine-sine decomposition (CSD) as introduced by Bergholm et al. in 2004~\cite{bergholm2005quantumuniformlycontrolled}.
We aim to provide a modernized perspective on this decomposition, putting it into context with the flag decomposition in \cref{eq:flag_decomposition}, and providing detailed computational instructions alongside an optimized implementation in just-in-time compilable Python code~\cite{our_repo}.

In \cref{sec:recursive_csd:overview}, we begin with an overview of the circuit structure and gate counts without considering the details of the synthesis. Then we walk through the decomposition and optimization process in detail in \cref{sec:recursive_csd:details}. Finally, we provide some commentary on the mathematical structure of the recursive CSD and its relationship to the flag decomposition from the main text.

\subsection{Overview of recursive CSD} \label{sec:recursive_csd:overview}
The CSD is a commonly used tool for matrix factorizations. It decomposes a unitary matrix $U$ into two block-diagonal matrices and a block matrix of diagonals. The former can be decomposed recursively by applying CSDs to each subblock, while the block of diagonals is left intact.
Showing the example of $n=3$ qubits, this yields the circuit 

\begin{align}\label{eq:recursive_csd_circ_0}
\resizebox{\linewidth}{!}{%
\includegraphics{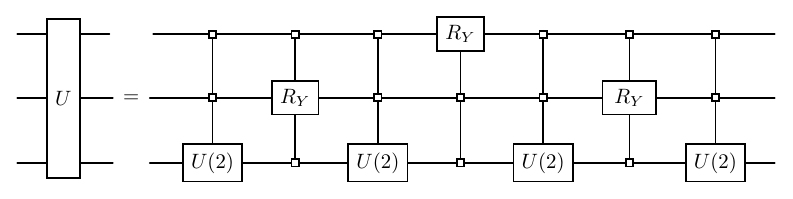}.
}
\end{align}
Here, we performed two recursion steps of the decomposition, and the multiplexed $R_Y$ rotations implement the block matrix of diagonals.
Afterwards, we perform a ``sweep" through the circuit, from left to right, in which we simultaneously decompose the multiplexers above into our target gate set and remove the overparametrization caused by the recursive CSD. If we are targeting a QROM architecture, we a) merge a diagonal from the left into a multiplexed $U(2)$ gate and decompose it, or b) split a diagonal into a part that is combined with a multiplexed $R_Y$ rotation into a new multiplexed $U(2)$ gate, and a part that can be commuted through to the right.
This leads to the following structure:

\begin{align}\label{eq:recursive_csd_circ_1}
\resizebox{\linewidth}{!}{%
\includegraphics{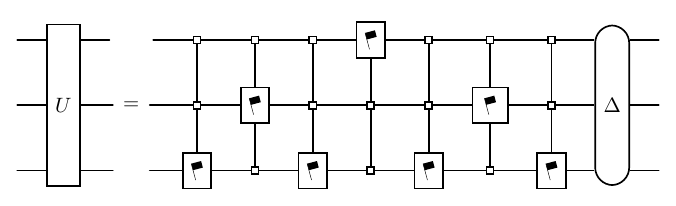}.
}
\end{align}
In this format, the synthesized circuit is particularly suited for the phase gradient decomposition; see \cref{sec:main_flag_decomposition}. It costs $2^n-1$ multiplexed single-qubit flags with $n-1$ multiplexing nodes each, and one $n$-qubit diagonal gate.

Alternatively, when targeting the gate set $\{R_Y, R_Z, \cnot\}$, the sweep can make use of a single step, treating multiplexed $R_Y$ gates just like a multiplexed $U(2)$ gate. In this step, we decompose multiplexed $U(2)$ gates into a diagonal and (non-multiplexed) single-qubit flag circuits interleaved with $\cnot$ gates.
See \cref{sec:recursive_csd:details} for details on the sweeping steps.

We may choose the order of the multiplexing qubits between neighbouring multiplexers such that the synthesized circuit becomes a bit less sparse. In a last step, the diagonal obtained at the right end of the circuit can be decomposed into multiplexed $R_Z$ rotations~\cite{bullock2003smaller,PennyLane-diagonal-unitary-decomp}, which in turn decompose into $R_Z$ rotations and $\cnot$ gates. For $n=3$, we arrive at the following synthesized circuit for the gate set $\{R_Y, R_Z, \cnot\}$:

\begin{widetext}

\begin{align}\label{eq:recursive_csd_circ_2}
\resizebox{\linewidth}{!}{%
\includegraphics{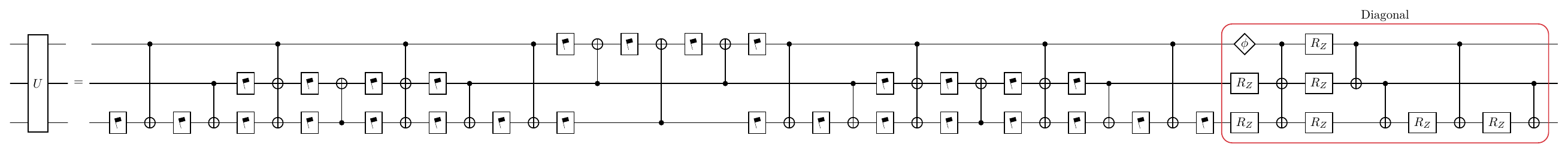}.
}
\end{align}
    
\end{widetext}

Doing this, we decomposed each of the $2^n-1$ multiplexed flags into $2^{n-1}$ single-qubit flags and $2^{n-1}-1$ $\cnot$ gates. Further, we decomposed the diagonal into $2^n-1$ $R_Z$ rotations, $2^n-2$ $\cnot$ gates and a global phase $\phi$ (diamond shape); see \cref{sec:gate_count_calculations:diagonal}. 
This leads to total gate counts of
\begin{align}
    N_{\cnot}(n) &= (2^n-1)(2^{n-1}-1)+2^n-2\nonumber\\
    &=\frac{1}{2}4^n-\frac{1}{2} 2^n - 1,\\
    N_{\text{rot}}(n) &= (2^n-1)(2^{n-1}\cdot 2)+2^n-1\nonumber\\
    &= 4^n-1.
\end{align}

The rotation count matches the lower bound for $U(2^n)$\footnote{Note that we refer to the rotation gate count here, which is one less than the parameter count due to the global phase.}. The $\cnot$ count is slightly above the best known count from the Block-ZXZ or QSD decompositions, and asymptotically a factor $2$ larger than the lower bound.
We were not able to find the opportunity to cancel one more $\cnot$ that is being hinted at in~\cite{bergholm2005quantumuniformlycontrolled}, so we report a count that is larger by $1$.

\subsection{Instructions for CSD compilation} \label{sec:recursive_csd:details}
We now describe the recursive CSD compilation process in detail, aiming for a complete, comprehensive description that also serves as additional documentation for our implementation~\cite{our_repo}.

\subsubsection{Recursive CSD decomposition}

A CSD decomposes a $d\times d$ unitary matrix $U$ into two block-diagonal unitary matrices $K_{0,1}$ and a $2\times 2$ block matrix of real-valued diagonals whose squares sum to $2\id_d$:

\begin{align*}
    U&=K_0 A K_1 = 
    \begin{pmatrix}
        K_{00} & 0\\
        0 & K_{01}
    \end{pmatrix}
    \begin{pmatrix}
        C & -S\\
        S & C
    \end{pmatrix}
    \begin{pmatrix}
        K_{10} & 0\\
        0 & K_{11}
    \end{pmatrix}\\
    K_{ij}&\in \mathcal{U}(d/2)\\
    C&=\operatorname{diag}\left(\cos(\varphi_0), \dots \cos(\varphi_{d/2})\right)\nonumber\\
    S&=\operatorname{diag}\left(\sin(\varphi_0), \dots \sin(\varphi_{d/2})\right).
\end{align*}

While the block sizes in $K_{0}$ and $K_1$ could differ in general, leading to a more fractured structure of $A$ (see e.g. \cref{sec:boundary-effects}), we here limited ourselves to an even dimension $d$ and a specific CSD that produces $d/2$ rotation angles $\vec\varphi$ and block matrices of size $d/2\times d/2$. 
As we work with unitaries represented on qubits, we specifically set $d=2^n$, and represent the CSD as a quantum circuit, c.f. \cref{eq:AIII_flag}:

\begin{align}\label{eq:csd_0}
    \resizebox{\linewidth}{!}{%
\includegraphics{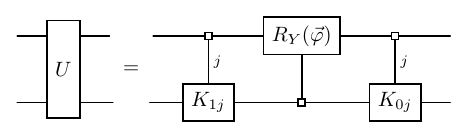}.%
    }
\end{align}

The CSD is readily available in code via standard linear algebra libraries, such as \texttt{scipy} in Python~\cite{virtanen2020scipy}.
Conveniently, we separately obtain the unitary matrix blocks $K_{ij}$ and the angles $\vec\varphi$\footnote{In \href{https://docs.scipy.org/doc/scipy/reference/generated/scipy.linalg.cossin.html}{\texttt{scipy.linalg.cossin}}, this is available via the setting \texttt{separate=True}.}.
In code, we represent each multiplexed unitary as a list of complex matrix blocks, or a complex $3$-tensor, and each multiplexed $R_Y$ gate as a vector storing the angles $\vec\varphi$ together with an integer denoting the target qubit index.

Next, we may apply the CSD to each of the two multiplexed unitaries, by applying it to each block separately and combining the results into two doubly-multiplexed unitaries on $n-2$ qubits and one $n-1$-multiplexed $R_Y$ rotation:

\begin{align}
    \resizebox{\linewidth}{!}{%
\includegraphics{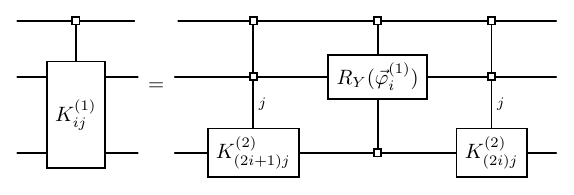}.%
    }
\end{align}

Here we denoted the recursion level with a superscript, so that $U=K_{00}^{(0)}$ for the input $U$ and $K_{ij}=K_{ij}^{(1)}$ for the blocks from \cref{eq:csd_0}.
These blocks in general have 5 indices;
one for the recursion depth (superscript), one for the index of the multiplexed unitary at a given recursion depth (first subscript), the block index along the diagonal within a multiplexed unitary (second subscript), and finally the two indices into the matrix block itself (not printed explicitly). Note that the superscript has an effect on the range of all other indices and we only require the data from one recursion level to describe the full circuit. As such, the superscript can be seen as an index over time during the compilation process.
At recursion level $r$, a full circuit description consists of $2^{r}$ $r$-multiplexed $(n-r)$-qubit unitaries (represented by one complex $3$-tensor of shape $(2^r, 2^{n-r}, 2^{n-r})$ each) and $2^r-1$ $(n-1)$-multiplexed $R_Y$ rotations (a real vector of length $2^{n-1}$ and an integer, each). The total number of real parameters stored at this level is thus 
\begin{align}\label{eq:recursive_csd_overparametrization_in_code}
    2\cdot 2^{r+r+2(n-r)}+(2^r-1) 2^{n-1}=2\cdot 4^n + (2^r-1) 2^{n-1},
\end{align}
where the overhead factor $2$ in the leading term stems from the generic representation of unitary matrices as complex matrices, and the second term directly points at the overparametrization introduced during the recursive CSD.

At recursion level $r=n-1$, we arrive at the circuit shown in \cref{eq:recursive_csd_circ_0}, with an overparametrization of 
\begin{align}\label{eq:recursive_csd_overparametrization_in_code_final}
    4^{n-1}-2^{n-1},
\end{align}
an overhead of about $25\%$ over the $4^n$ parameters that uniquely determine $U$.
The next step, which is concerned with removing these excess parameters and decomposing into a preferred gate set, depends on that target gate set. We describe it for a QROM-based architecture that directly uses \cref{eq:recursive_csd_circ_1} together with the implementation of multiplexed flags described in \cref{sec:main_flag_decomposition}, and for the more traditional gate set $\{R_Y, R_Z, \cnot\}$ without using auxiliary qubits, arriving at \cref{eq:recursive_csd_circ_2}.

\subsubsection{Phase gradient decomposition}
If the target gate set consists of multiplexed flags and diagonals, we are almost at the desired circuit after performing the recursive CSD, but we still need to get rid of the massive overparametrization.
For this, we will need two subroutines: pulling a diagonal unitary from left to right (in circuit ordering) through a multiplexed $U(2)$ gate while reducing it to a multiplexed flag, and pulling a diagonal unitary from left to right through a multiplexed $R_Y$ rotation while \textit{enhancing} it to a multiplexed flag\footnote{It is perfectly valid to invert the direction, which will result in multiplexed ``right-handed" single-qubit flags.}.

    

The first subroutine is summarized as follows:

\begin{align*}
\vcenter{\hbox{\scalebox{0.7}{%
\includegraphics{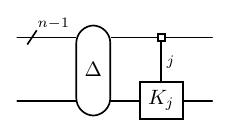}
}}}&= 
\vcenter{\hbox{\scalebox{0.7}{%
\includegraphics{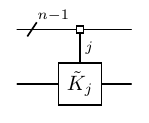}
}}}\\
&=
\vcenter{\hbox{\scalebox{0.7}{%
\includegraphics{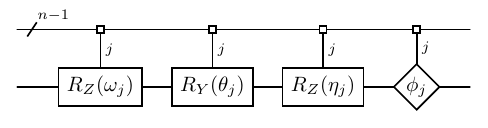}
}}}\\
&=
\vcenter{\hbox{\scalebox{0.7}{%
\includegraphics{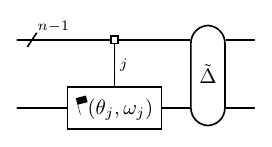}
}}}.
\end{align*}
We are given a diagonal unitary $\Delta$ and a $(n-1)$-multiplexed $U(2)$ gate, represented as $2^{n-1}$ $2\times 2$ matrices $\{K_j\}_j$, whose target we assume to be the least significant qubit. Then, we may first merge $\Delta$ into the multiplexer by computing $\tilde{K}_j= K_j\Delta_{[2j:2j+2]}$, interpreted as matrix multiplication.
Next, we pull off a diagonal from the modified multiplexer to the right.
Computationally, we do this by decomposing each of the $\{\tilde{K}_j\}_j$ into a global phase (diamond shape) and standard Euler angles:

\begin{align}
    \vcenter{\hbox{\scalebox{0.8}{%
\includegraphics{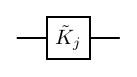}
    }}}
    &=
    \vcenter{\hbox{\scalebox{0.8}{%
\includegraphics{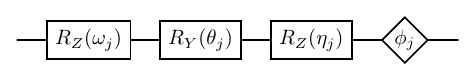}
    }}},
    \nonumber
\end{align}
grouping the resulting angles into vectors $\vec{\phi}$, $\vec{\eta}$, $\vec{\theta}$ and $\vec{\omega}$ of length $2^{n-1}$, and processing the vectors $\vec{\eta}$ and $\vec{\phi}$ into the diagonal unitary
\begin{align}
    \tilde{\Delta} = e^{-i(\vec\phi+\vec\eta/2)} \underset{n-1}{\oplus} e^{-i(\vec\phi-\vec\eta/2)}.
\end{align}
Here we introduced the notation $\underset{k}{\oplus}$ to denote the direct sum between two equally-sized vectors rearranged such that the split of the total vector space corresponds to that arising from the split $|0\rangle \oplus|1\rangle$ on the $k$th out of all qubits (using zero-based indexing).
We illustrate the notation with some examples:
\begin{align}
    (a, b, c, d)\underset{0}{\oplus} (e, f, g, h) &= (a, b, c, d, e, f, g, h)\\
    (a, b, c, d)\underset{1}{\oplus} (e, f, g, h) &= (a, b, e, f, c, d, g, h)\\
    (a, b, c, d)\underset{2}{\oplus} (e, f, g, h) &= (a, e, b, f, c, g, d, h).
\end{align}
The Euler decomposition can be computed in code with simple (inverse) trigonometric functions.
We thus moved the diagonal unitary $\Delta$ through the multiplexer $\{K_j\}_j$, resulting in the modified diagonal unitary $\tilde{\Delta}$ and the multiplexed flag with angles $\vec{\omega}$ and $\vec{\theta}$ for its $R_Z$ and $R_Y$ rotations, respectively.

The second subroutine is summarized as follows:

\begin{align}
\vcenter{\hbox{\scalebox{0.8}{%
\includegraphics{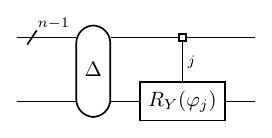}
}}}
&=
\vcenter{\hbox{\scalebox{0.8}{%
\includegraphics{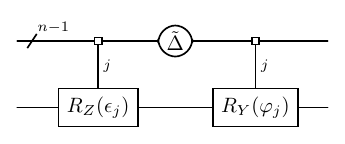}
}}}\nonumber \\
&=
\vcenter{\hbox{\scalebox{0.8}{%
\includegraphics{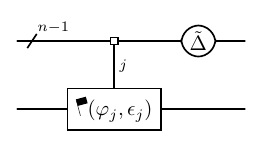}
}}}.
\end{align}
We are given a diagonal unitary $\Delta$ and a multiplexed $R_Y$ gate, defined by $2^{n-1}$ angles $\vec\varphi$ and a target qubit index $k$. 
We split $\Delta$ into a diagonal on all qubits except the one with index $k$, and a $(n-1)$-multiplexed $R_Z$ rotation with target qubit index $k$.
For this, we compute the complex arguments of $\Delta$, and split them into two vectors $\vec\delta_{0,1}$ of length $2^{n-1}$ such that $\Delta\eqqcolon e^{i\vec\delta_0}\underset{k}{\oplus}e^{i\vec\delta_1}$. We may rewrite this as
\begin{align}
    \Delta &= e^{i\vec\delta_0}\underset{k}{\oplus}e^{i\vec\delta_1}\\
    &=e^{i(\vec\delta_0+\vec\delta_1)/2}e^{i(\vec\delta_0-\vec\delta_1)/2}\underset{k}{\oplus} e^{i(\vec\delta_0+\vec\delta_1)/2} e^{-i(\vec\delta_0-\vec\delta_1)/2}\\
    &=\left(e^{i(\vec\delta_0+\vec\delta_1)/2\ \underset{k}{\otimes}\ (1, 1)} \right)\left(e^{(i(\vec\delta_0-\vec\delta_1)/2\ \underset{k}{\otimes}\ (1, -1)}\right)\\
    &\eqqcolon(\tilde{\Delta}\underset{k}{\otimes}\id_2)\ \text{MUX-}R_Z(\vec{\epsilon},k),
\end{align}
where we extended the subscript notation from $\oplus$ to $\otimes$ to denote a $((n-1)\otimes 1)$-qubit tensor product with the single qubit at qubit index $k$.
Now we may commute the diagonal $\tilde{\Delta}$ through the given $R_Y$ multiplexer controls. In turn, this multiplexer combines with the $R_Z$ multiplexer to form a multiplexed flag, completing the subroutine.

With the two subroutines in our hands, we simply sweep from left to right through the circuit obtained by the recursive CSD, applying the subroutines alternatingly. For the left-most multiplexer, there is no diagonal to merge from the left, so we start immediately with the Euler decomposition step of the first subroutine.
After the sweep, we obtain the circuit in \cref{eq:recursive_csd_circ_1}, which can then be compiled further into QROMs and adders, using auxiliary qubits and phase gradient states.

\subsubsection{Auxiliary-free decomposition}

For the classical gate set $\{R_Y, R_Z, \cnot\}$ and without auxiliary wires, we follow a similar strategy as for the QROM architecture, but we will directly integrate the decomposition of multiplexers with the reduction of the overparametrization.
For this, we use a single subroutine that we then apply to both the multiplexed $U(2)$ gates and the multiplexed $R_Y$ rotations alike. It proceeds as follows:

\begin{multline*}
\resizebox{0.65\linewidth}{!}{%
\includegraphics{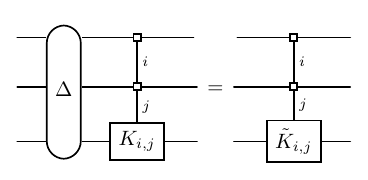}%
} \\
\resizebox{0.85\linewidth}{!}{%
$=
\vcenter{\hbox{\includegraphics{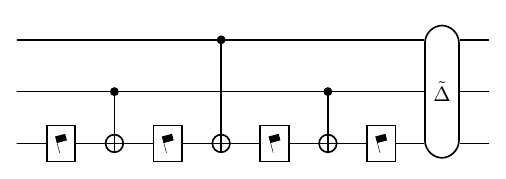}}}$
}.
\end{multline*}
The subroutine compiles a diagonal unitary $\Delta$ followed by a given multiplexed $U(2)$ gate $\{K_j\}_j$ into  single-qubit flags, $\cnot$ gates, and a new diagonal $\tilde{\Delta}$. It was described first in~\cite{bergholm2005quantumuniformlycontrolled}. If we are given a multiplexed $R_Y$ gate, we simply replace it by a multiplexed $U(2)$ gate by converting the given vector of rotation angles into the respective sequence of $2\times 2$ rotation matrices. In either case, we multiply the diagonal into the multiplexer via $\tilde{K}_j=K_j \Delta_{[2j:2j+2]}$, like for the first subroutine in the previous section.

The subroutine then exclusively consists of \textit{$U(2)$ de-multiplexing}, a recursive technique to remove multiplexing nodes from a $U(2)$ gate until we arrive at a parameter-optimal circuit of single-qubit flags and an $n$-qubit diagonal. It can be considered the key innovation unlocking the reduced $\cnot$ count in the optimized CSD decomposition~\cite{bergholm2005quantumuniformlycontrolled}. Further details can also be found in the PennyLane compilation hub~\cite{PennyLane-SelectU2Decomp}.

Let's first consider the simplest case of a single multiplexing node, which also serves as base case for the recursion. We de-multiplex a singly-multiplexed $U(2)$ operator $\{K_0, K_1\}$ with a variant of a type-A Cartan decomposition that is used for de-multiplexing larger operators.
First, compute $X=K_0K_1^\dagger$\footnote{This is almost the relative complex structure with respect to the relevant type-A Cartan involution, which is formed as $X \oplus X^\dagger$.} and denote its matrix as
\begin{align}
    X=e^{i\phi/2} \begin{pmatrix} a & b \\ -\bar{b} & \bar{a} \end{pmatrix}.
\end{align}
Then, define the angles
\begin{align}
    \rho_0 &= \frac{\pi-\phi}{4}-\frac{\arg(a)}{2},\\
    \rho_1 &= \frac{3\pi-\phi}{4}+\frac{\arg(a)}{2},
\end{align}
as well as the matrix $r=\operatorname{diag}(e^{i\rho_0},e^{i\rho_1})$ and compute $Y=rXr$.
The eigenvalues of $Y$ then are guaranteed to be $\pm i$, i.e.,
\begin{align}
    Y = L_0 \operatorname{diag}(i, -i) L_0^\dagger
\end{align}
for a unitary $L_0$.
Finally, define $d=\operatorname{diag}(e^{i\pi/4},e^{-i\pi/4})$ and $L_1=dL_0^\dagger r^\dagger K_1$.
Then we have
\begin{align}
    K_1 &= r L_0 d^\dagger L_1 \nonumber \\
    K_0 &= r^\dagger Yr^\dagger K_1  = r^\dagger L_0 d^2 L_0^\dagger r^\dagger(r L_0 d^\dagger L_1)\nonumber \\
    &=r^\dagger L_0 d L_1\\
    \begin{pmatrix}
        K_0 & 0 \\ 0 & K_1
    \end{pmatrix}&=
    \begin{pmatrix} r^\dagger & 0 \\ 0 & r \end{pmatrix}
    \begin{pmatrix} L_0 & 0 \\ 0 & L_0 \end{pmatrix}
    \begin{pmatrix} d & 0 \\ 0 & d^\dagger \end{pmatrix}
    \begin{pmatrix} L_1 & 0 \\ 0 & L_1 \end{pmatrix}.\nonumber
\end{align}
That is, we decomposed the multiplexer into two single-qubit unitaries $L_0$ and $L_1$, a static diagonal $D=d\oplus d^\dagger=\operatorname{diag}(e^{i\pi/4}, e^{-i\pi/4}, e^{-i\pi/4}, e^{i\pi/4})$, and a diagonal matrix $R=r^\dagger\oplus r$ in the form of a multiplexed $R_Z$ gate.
We can polish this decomposition to suit our purposes by extracting from $D$ an $S^\dagger$ gate on each qubit as well as a global phase, and merging them into $L_1$ and $R$. We then obtain the circuit

\begin{align}\label{eq:de-multiplexing_u2-base-case}
    \resizebox{\linewidth}{!}{%
\includegraphics{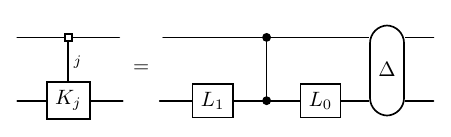}.%
    }
\end{align}

For our case of multiple multiplexing qubits, we choose one of them and pair up the matrices of the multiplexer corresponding to the states $|0\rangle$, $|1\rangle$ of that qubit. For simplicity, we here choose the most significant multiplexing qubit and write the matrix pairs as $\{\tilde{K}_{0j'}, \tilde{K}_{1,j'}\}$. Then we apply the de-multiplexing above to each pair, arriving at the multiplexer-extended version of \cref{eq:de-multiplexing_u2-base-case}:

\begin{align}
    \resizebox{\linewidth}{!}{%
\includegraphics{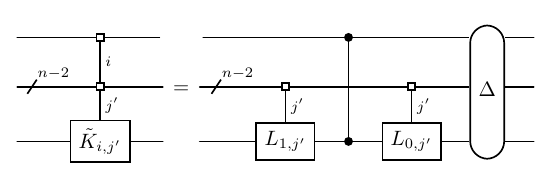}.%
    }
\end{align}

The $(n-2)$-multiplexed $U(2)$ gate $\{L_{1,j'}\}$ is then de-multiplexed again, the resulting diagonal is merged into $\{L_{0,j'}\}$, and the latter is de-multiplexed as well.
This way, we obtain a decomposition of $\{\tilde{K}_j\}$ into $2^{n-1}$ single-qubit $U(2)$ gates, $2^{n-1}-1$ $\cz$ gates, and a diagonal.
At this point, we just need to reduce the parametrization in this decomposition, which is highly redundant. For this, we start at the left of the circuit and decompose the $U(2)$ gate into a global phase and Euler rotations of the form $R_Z R_Y R_Z$. Then we merge the right $R_Z$ gate through the $\cz$ gate to its right into the next $U(2)$ gate, and the global phase into the trailing diagonal $\Delta$.
If $\cnot$ gates are preferred over $\cz$ gates, we can extract a Hadamard gate from the $R_Y$ rotation to the left and from the $U(2)$ gate to the right of the $\cz$ gate to transform it into a $\cnot$.
This procedure is repeated, until we are left with $2^{n-1}$ single-qubit flags ($R_Z$ followed by $R_Y$), $2^{n-1}-1$ $\cz$ (or $\cnot$) gates, and a diagonal.
With this, we complete the subroutine of pulling a diagonal unitary through a multiplexed $U(2)$ gate while decomposing it into elementary gates at the same time.

The coarse-level compilation then proceeds exactly like for the QROM architecture: we repeatedly apply the subroutine to all multiplexers in \cref{eq:recursive_csd_circ_0}, sweeping from left to right. This leaves behind the circuit shown in \cref{eq:recursive_csd_circ_2}, which does not contain any excess parameters anymore.

\subsection{Math commentary}

The CSD of unitary (orthogonal) matrices is a KAK decomposition of type AIII (BDI)~\cite{nakajima2005new,sutton2009computing}, induced by a Cartan decomposition\footnote{See~\cite{wierichs2025recursive} for some remarks about nomenclature and Abelian factors on top of semisimple algebras.} of the corresponding Lie algebra $\mathfrak{u}(d)$ ($\mathfrak{so}(d)$). 
Cartan decompositions can be applied recursively, by further decomposing the subgroup $\mathcal{K}$ from which the elements $K$ in the KAK decomposition originate.
This concept has been studied thoroughly in the past~\cite{Dagli2008,dalessandro2006decompositions,dalessandro2007quantum,wierichs2025recursive}.
For the unitary CSD, this subgroup of the total group $\mathcal{G}=U(d)$ is $\mathcal{K}=U(d/2)\times U(d/2)$, so that its elements $K$ can be decomposed by applying CSDs to each factor individually. This structure manifests in the parallelized, or batched, decomposition of the blocks in the block diagonal matrices.
The element $A$ from the Abelian Cartan subgroup (CSG) $\mathcal{A}$, which corresponds to a maximal torus in the symmetric space $\mathcal{P}\coloneqq\mathcal{G}/\mathcal{K}$, naturally takes the form of a multiplexed $R_Y$ rotation for the conventional basis and involution choices, which in addition interacts favourably with the block diagonal structure of the subgroup(s). The CSG elements from higher recursion levels are just multiplexed $R_Y$ rotations again, with the less significant multiplexer qubits originating in the higher-level CSD itself, and the more significant ones being inherited from the initial and intermediate $K$s.
As reported in~\cite{wierichs2025recursive}, a type-AIII decomposition (with dimensional parameters $p=q=d/2$) leads to an overparametrization of $d/2$. This matches the $2^{n-1}$ parameters in the $R_Y$ multiplexer produced in the first CSD. Each recursion step has to decompose four times as many matrices as the previous one, because there are twice as many subgroup elements, and each of them has an additional multiplexing node attached. Thus, the $r$th recursion step adds $2^{n-1} (\tfrac{4}{2})^{r-1}$ parameters. Summing up all added parameters, we find the same overparametrization count as in \cref{eq:recursive_csd_overparametrization_in_code_final}:
\begin{align}
    \sum_{r=1}^{n-1} 2^{n+r-2} = 2^{n-2}(2^n-2)=4^{n-1}-2^{n-1}.
\end{align}

Pulling a diagonal off a unitary matrix corresponds to the flag decomposition introduced in \cref{eq:flag_decomposition} in the main text, and it is the key step in the recursive CSD that allows us to remove the excess parametrization after the fact.
As such, the flag decomposition allows us to exploit a particularly simple degree of freedom in KAK decompositions: elements from the commutant $\mathcal{K}^{\mathcal{A}}$ of the CSG $\mathcal{A}$ within the subgroup $\mathcal{K}$ can move freely between the two $K$s. For the first-level CSD with the standard choice of bases, we have 
\begin{align}
    \mathcal{K}^\mathcal{A}=\exp(\operatorname{span}_{i\mathbb{R}}\{\id_2\otimes \{\id_2, Z\}^{n-1}\}),
\end{align}
i.e., the elements that can be merged through the $R_Y$ multiplexer into the other group element are the $(n-1)$-qubit diagonals obtained as part of the flag decomposition. Note that the diagonal group whose generators have a Pauli-$Z$ on the first qubit instead are not part of the commutant.
This emphasizes the compatibility between these two concepts, lifting the circuit decomposition technique by Bergholm et al.~\cite{bergholm2005quantumuniformlycontrolled} to a mathematically founded technique that can readily be generalized to other Cartan decompositions.
For example, in selective de-multiplexing, which we present in \cref{sec:selective_de_multiplexing}, the second step is a type-A Cartan decomposition of $U(d/2)\times U(d/2)$, which leads to an overparametrization of $d/2$. The CSG is $\mathcal{A}=\exp(\operatorname{span}_{i\mathbb{R}}\{Z\otimes \{\id_2, Z\}^{n-1}\})$, the group of $R_Z$ multiplexers targeting the most significant qubit. The commutant of this CSG within the subgroup $\id_2 \otimes U(d/2)$ is
\begin{align}
    \mathcal{K}^\mathcal{A} = \exp(\operatorname{span}_{i\mathbb{R}}\{\id_2 \otimes \{\id_2, Z\}^{n-1}\}),
\end{align}
i.e., the (Abelian) group of diagonals acting on all but the most significant qubit.
This is what allows us to leverage a flag decomposition into removing overparametrization, just like for the type-AIII Cartan decompositions above.

Flag decompositions thus form a useful additional tool when using Cartan decompositions in the context of parameter optimality, extending the very limited repertoire of \textit{unique} (recursive) Cartan decompositions that do not introduce overparametrization at all~\cite{wierichs2025recursive}.

\section{Gate count calculations for \{Clifford + Rot\}}\label{sec:gate_count_calculations}

Here we present gate count calculations for important subroutines and gate families. We summarize them in \cref{tab:gate_counts_subroutines} and collect comments for how to obtain those counts in the following.

\bgroup
\def\arraystretch{1.5}%
\begin{table*}[]
    \centering
    \begin{tabular}{ccccl}
        Subroutine & Rot & Clifford & $\ \ \Delta\ \ $ & comment \\\hline
        $1Q$-flag & $2$ & $0$ & $0$ & \cref{eq:flag_decomp_n-1} \\
        $2Q$-flag & $12$ & $2$ & $0$ & \cref{eq:flag_decomp_n-2} \cite{shende2006synthesis} \\
        mux$_k$-Rot & $2^k$ & $2^k$ & $0$ & $k\geq 1$ \cite{Mottonen2004quantumgeneralmultiqubit} \\
        sym-mux$_k$-Rot & $2^k$ & $2^k-1$ & $0$ & $k\geq 0$ \cite{shende2006synthesis} \\
        mux$_k$-$1Q$-flag & $2^{k+1}$ & $2^k-1$ & $1$ & $k\geq 0$ \cite{bergholm2005quantumuniformlycontrolled} \\
        mux$_k$-$2Q$-flag & $3\cdot 2^{k+2}$ & $6\cdot2^k-4$ & $1$ & $k\geq 0$ \cref{eq:flag_decomp_n-3_b-2} \\
        $nQ$-flag & $4^n-2^n$ & $\tfrac{1}{2}4^n-\tfrac{3}{2}2^n+1$ & $1$ & recursive flag decomp/CSD with  $b=1$; \cref{sec:recursive_csd}, \cite{bergholm2005quantumuniformlycontrolled}\\
        & $4^n-2^n$ & $\frac{1}{2}4^n-\frac{n+12}{8}2^n+1$ & $1$ & $n\geq 3$ \\
        mux$_k$-$nQ$-flag & $(4^n-2^n)2^k$ & $\frac{1}{2}(4^n-2^n)2^k-\frac{5}{4} 2^n+1$ & $1$ & $n\geq 2, k>0$ \\
        $n$-qubit diagonal & $2^n-1$ & $2^n-2$ & $0$ & \\
        $n$-qubit unitary & $4^n-1$ & $\frac{1}{2}4^n-\frac{3}{8}(n+2)2^n+n-1$ & $0$ & selective de-multiplexing \\
        MPS prep & $(2L+1)4^n$ & $\tfrac12(2L+1)4^n-\tfrac{(2L+3)n+(8L+6)}{8}2^n+n-1$ & $0$ & can be reduced further; \cref{sec:boundary-effects} \\
    \end{tabular}
    \caption{Gate counts for elementary subroutines used throughout this work, in the context of \{Clifford + Rot\} decompositions. The column $\Delta$ indicates whether an additional diagonal unitary is needed in the decomposition. All reported decompositions are parameter-optimal, except for parameters in the potentially required diagonal. We only report the cheapest decompositions, with an exception for the $n$-qubit flag for which we also show the cost by Bergholm et al., for comparison. See \cref{sec:mps_state_prep} for details about MPS state preparation.\hspace*{\fill}}
    \label{tab:gate_counts_subroutines}
\end{table*}
\egroup

\subsection{One- and two-qubit flags} 
The one-qubit flag decomposition can be obtained from an Euler decomposition, as discussed in \cref{eq:flag_decomp_n-1}. The flag circuit consists of a single $R_Z$ and a single $R_Y$ gate.

The two-qubit flag decomposition, shown in \cref{eq:flag_decomp_n-2}, is reported in~\cite[Thm.~14]{shende2006synthesis} and~\cite[Prop.~10]{shende2004incompletely}, without the mathematical context of flag manifolds. The flag circuit itself is given in terms of $2$ $\cnot$ gates and $12$ Rot gates. Note that no diagonal is required for the flag circuit itself.

\subsection{Multiplexed one- and two-qubit gates}\label{sec:gate_count_calculations:mux-one-two-qubit-ops}
\textbf{mux$\mathbf{_k}$-$\mathbf{Rot}$}
A multiplexed Rot gate about the Pauli operator $P\neq X$ can be decomposed into an alternating sequence of $\cnot$ and Rot gates~\cite{mottonen2004transformation,shende2006synthesis,PennyLane-SelectU2Decomp}. For a single multiplexing node, there are $2$ $\cnot$ and two Rot gates, and their number doubles with each additional multiplexing node. This leads to $2^k$ $\cnot$ and Rot gates each, for $k$ multiplexing nodes. Note that this formula does not reproduce the cost of non-multiplexed rotations, as it predicts them to require a $\cnot$ gate.

A slight generalization of the above, which allows us to treat $R_X$ gates as well, is the decomposition of a multiplexed $R_P$ gate, with $P$ a Pauli word, into an alternating sequence of $\text{C}Q$ and $R_P$ gates, where $Q$ is another Pauli word with $\{P, Q\}=0$. This generalization holds because the only property of $\cnot$ and $R_Y$ (or $R_Z$) gates required for its derivation is the anti-commutativity between $X$ and $Y$ (or $X$ and $Z$).
For a given single-qubit operator $P$, there are only two options (up to an inconsequential phase) for anti-commuting single-qubit operators, say $Q_0$ and $Q_1$. We then know the following decomposition, depicted for $n=3$ for simplicity:
\begin{align}\label{eq:generalized_mottonen_decomp}
\resizebox{\linewidth}{!}{%
\includegraphics{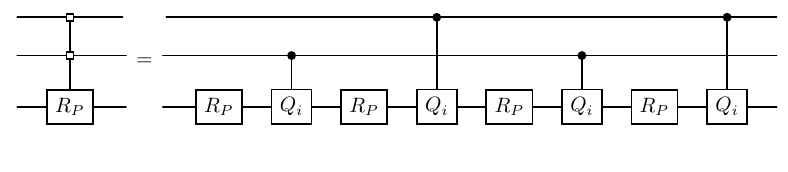}.%
}
\end{align}
As a matter of fact, we can be even more flexible in the decomposition and choose the operator $Q$ to differ at every location; for this, we pick the right-most $Q_i$ to our liking, and create the decomposition in \cref{eq:generalized_mottonen_decomp}. Afterwards, we can use $R_P(\pm\tfrac{\pi}{2})Q_0R_P(\mp\tfrac{\pi}{2})=Q_1$, where the sign choice depends on the assignment of the labels $Q_{0,1}$, to transform any of the other $\text{C}Q_i$ into the alternative option. The inserted $R_P$ gates are then simply absorbed into the present gates by modifying their angles. 
One example that we use in selective de-multiplexing in \cref{sec:selective_de_multiplexing} is the following decomposition:
\begin{align}\label{eq:Möttönen-generic}
\resizebox{\linewidth}{!}{%
\includegraphics{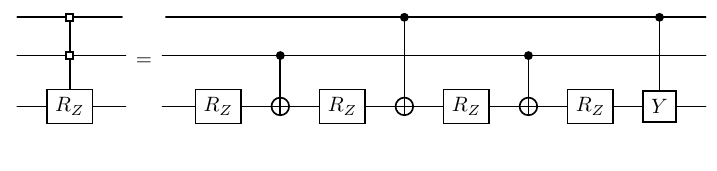}.%
}
\end{align}

The advantage of this flexibility is that we may adjust the decomposition of a multiplexed rotation such that the trailing (or leading, if preferred) entangling gate can be absorbed into neighbouring gate families, which leads to the symmetrized M\"ott\"onen decomposition in \cref{eq:symmetrized_Möttönen}:
\begin{align}\label{eq:symmetrized_Möttönen_app}
\resizebox{\linewidth}{!}{%
\includegraphics{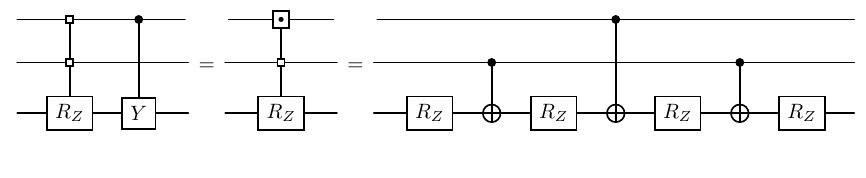}.%
}
\end{align}

Formally, we need the multiplexed Rot gate to neighbour a parametrized gate that contains the space
\begin{align}
    \mathcal{E}=\operatorname{span}_{i\mathbb{R}}\{Z^{(c)}\otimes Q_i^{(t)}, Z^{(c)}\otimes \id^{(t)}, \id^{(c)}\otimes Q_i^{(t)}\},
\end{align}
in the span of its generators, where $Q_i^{(t)}$ is either of $Q_{0,1}$, acting on the target qubit $t$, and $c$ is one of the multiplexing qubits of the gate we decompose\footnote{The space $\mathcal{E}$ here is a possible dynamical Lie algebra (DLA) of the entangling gate $\mathrm{CP}$ with respect to the Pauli basis. One can also generalize the merging result by choosing other valid DLAs for the entangler.}.

\textbf{mux$\mathbf{_k}$-$\mathbf{1Q}$-flag.}
A multiplexed single-qubit flag can be decomposed into a diagonal and an alternating sequence of single-qubit flags (with two Rot gates each) and $\cnot$ gates (see~\cite{bergholm2005quantumuniformlycontrolled,PennyLane-SelectU2Decomp} and \cref{sec:recursive_csd}). For example, for $n=3$ we obtain
\begin{align}
    \resizebox{\linewidth}{!}{%
\includegraphics{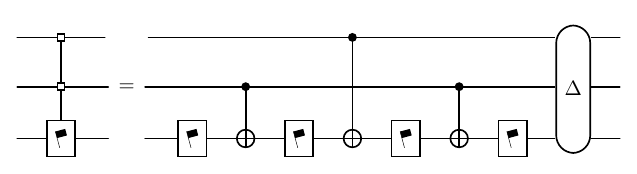}.
    }
\end{align}
This decomposition results in $2^{k+1}$ Rot gates and $2^k-1$ $\cnot$ gates for $k$ multiplexing nodes.
This rule correctly reproduces the count of non-multiplexed one-qubit flags.

\textbf{mux$\mathbf{_k}$-$\mathbf{2Q}$-flag.}
Applying the MEP to the two-qubit flag decomposition, we find that a $k$-multiplexed two-qubit flag decomposes into $6$ $k$-multiplexed single-qubit flags and two $\cnot$ gates; see \cref{eq:flag_decomp_n-2}.
We may then decompose each multiplexed flag as described above, starting from the left. After each decomposition, we can pull the resulting diagonal through the neighbouring $\cnot$ gate\footnote{Diagonals commute with the control node of the $\cnot$. To commute with the target node, the $\cnot$ can first be converted to a $\cz$ gate, with the required $H$ gates being absorbed by the flag to its left and the multiplexed flag to the right, which can be treated as a more general multiplexed $SU(2)$ gate.}, resulting in a single diagonal at the right end of the decomposition.
Overall, we find $6\cdot 2^{k+1}$ Rot gates, and $2+6(2^k-1)$ $\cnot$ gates.
This rule correctly reproduces the count of non-multiplexed two-qubit flags.

\subsection{(Multiplexed) \texorpdfstring{$n$}{n}-qubit flags}\label{sec:gate_count_calculations:flags}
To decompose an $n$-qubit flag circuit parameter-optimally, the literature only knows of the recursion relation (see \cref{sec:flag_decomposition}) 
\begin{align}\label{eq:flag_decomp_restate}
\resizebox{\linewidth}{!}{%
\includegraphics{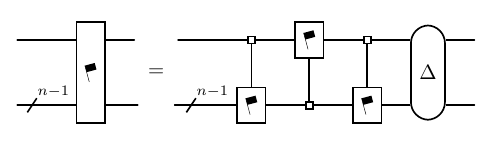}%
},
\end{align}
which leads to the recursive CSD decomposition as shown in \cref{sec:recursive_csd} if multiplexed single-qubit flags are used as base case.
Here, as above, the diagonal is a result of the decomposition process, but as we assume it to be merged into other circuit components, we do not count its degrees of freedom towards the total parameter count of the circuit on the right hand side.
The decomposition in \cref{eq:flag_decomp_restate} adds $2^{k-1}$ $(n-1)$-multiplexed single-qubit flags at the recursion step $k$ (starting with $k=1$), and doubles the number of multiplexed multi-qubit flags with each step. As we perform $n-1$ steps to arrive at single-qubit flags only, we find $(2^{n-1}-1)+2^{n-1}=2^n-1$ $(n-1)$-multiplexed single-qubit flags.

If we instead make use of the improved decomposition of multiplexed two-qubit flags, i.e., use the base case $n_b=2$ above, we simply perform one step less, arriving at $2^{n-2}-1$ $(n-1)$-multiplexed single-qubit flags and $2^{n-2}$ $(n-2)$-multiplexed two-qubit flags. Inserting the counts for those components from above, the cost is
\begin{align}
    C_{\text{rot}}^{\text{flag}, n_b=2} &= (2^{n-2}-1)2^n+2^{n-2}\cdot3\cdot2^{n}\\
    &=4^n-2^n,  \nonumber\\
    C_{\cnot}^{\text{flag}, n_b=2} &= (2^{n-2}-1)(2^{n-1}-1)+2^{n-2}(6\cdot2^{n-2}-4)\nonumber\\
    &=\frac{1}{2}4^n-\frac{7}{4}2^n+1. \nonumber
\end{align}

Inspired by the QSD, which introduces de-multiplexing to reduce $\cnot$ counts, we presented selective de-multiplexing in \cref{sec:selective_de-mux}, leading to the modified recursion relation
\begin{align}\label{eq:selective_de-mux_restate}
\resizebox{\linewidth}{!}{%
\includegraphics{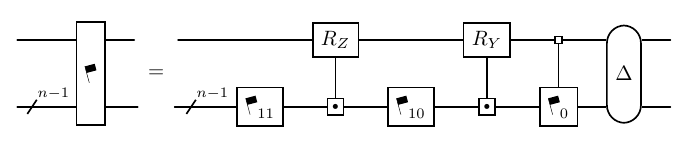}.%
}
\end{align}

Which of the two options for decomposing flags is cheaper will depend on the relative cost of $1$-multiplexed $(n-1)$-qubit flags, non-multiplexed $(n-1)$-qubit flags, and $(n-1)$-multiplexed single-qubit flags and rotations.
More precisely, the cost for \cref{eq:flag_decomp_restate}, and in general for its $k$-multiplexed variant, is
\begin{align}
    C_{\cnot}^{\text{flag,\ref{eq:flag_decomp_restate}}}(n,k)&=
    2C_{\cnot}^{\text{flag}}(n-1,1+k)\\
    &+C_{\cnot}^{\text{flag}}(1,n-1+k)
\end{align}
where we denote the target and multiplexing qubit counts as arguments to $C_{\cnot}$. 
In contrast, \cref{eq:selective_de-mux_restate} implies the cost relationship
\begin{align}
    C_{\cnot}^{\text{flag,\ref{eq:selective_de-mux_restate}}}(n,k)&=C_{\cnot}^{\text{flag}}(n-1,1+k)\\
    &+2C^{\text{Rot}}_{\cnot}(n-1+k)+2C_{\cnot}^{\text{flag}}(n-1,k),
\end{align}
where the multiplexed rotations are neighbouring suitable flags so that we may subtract one CNOT from their cost using the technique in \cref{eq:symmetrized_Möttönen_app}, leading to
\begin{align}
    C^{\text{Rot}}_{\cnot}(k) = 2^{k}-1.
\end{align}

To settle which option is cheaper, we first inspect the cost at small qubit counts.
For ($k$-multiplexed) two-qubit flags, we above found the optimal decomposition to contain $6\cdot 2^k-4$ $\cnot$s.
With this, the cost to decompose a $k$-multiplexed $3$-qubit flag is given by
\begin{align}
    C_{\cnot}^{\text{flag,\ref{eq:flag_decomp_restate}}}(3,k)&=2\cdot (6\cdot 2^{k+1}-4)+(2^{k+2}-1)\\
    &=28\cdot 2^k-9,\nonumber\\
    C_{\cnot}^{\text{flag,\ref{eq:selective_de-mux_restate}}}(3,k)&=6\cdot 2^{k+1}-4+2(2^{k+2}-1)+2(6\cdot 2^{k}-4)\nonumber\\
    &=32\cdot 2^k-14,\nonumber
\end{align}
where we used that $k$-multiplexed flags and (symmetrized) $k$-multiplexed Rot both cost $2^k-1$ $\cnot$s.
For non-multiplexed $3$-qubit flags, we find the second option to be preferable, with a $\cnot$ count of $18$.
In contrast, $k$-multiplexed $3$-qubit flags are optimally decomposed with the first option, leading to $28\cdot2^k-9$ $\cnot$s.
This implies that when decomposing larger objects, we will choose a different strategy for $3$-qubit flags and their multiplexed variant.
We can combine their cost into 
\begin{align}
    C_{\cnot}^\text{flag,min}(3,k)=28\cdot2^k-9-\delta_{k0}.
\end{align}

We can repeat this calculation for $4$-qubit flags, and find
\begin{align}
    C_{\cnot}^{\text{flag,\ref{eq:flag_decomp_restate}}}(4,k)=&2\cdot (28\cdot 2^{k+1}-9-\delta_{k+1,0})+(2^{k+3}-1)\nonumber\\
    =&120 \cdot 2^k-19,\\
    C_{\cnot}^{\text{flag,\ref{eq:selective_de-mux_restate}}}(4,k)=&(28\cdot 2^{k+1}-9-\delta_{k+1,0}) +2(2^{k+3}-1)\nonumber\\
    &+\,2\cdot (28\cdot 2^k -9-\delta_{k0})\nonumber\\
    =&128\cdot 2^k-29-2\delta_{k0}.\nonumber
\end{align}
We find the same behaviour as for $n=3$, with selective de-multiplexing being the better option for $k=0$ only, where it saves $4$ $\cnot$s. The coerced cost function is thus 
\begin{align}
    C_{\cnot}^\text{flag,min}(4,k)=120 \cdot 2^k-19-4\delta_{k0}.
\end{align}

At this point it seems wise to distill some recursion relations from the observed behaviour.
Denote the minimized cost for a $k$-multiplexed $n$-qubit flag by 
\begin{align}
C_{\cnot}^\text{flag,min}(n,k)=A_n2^k+B_n-C_n\delta_{k0}.
\end{align}
We will prove by induction that $A_n=2^{n-1}(2^n-1)$, $B_n=1-5\cdot 2^{n-2}$ and $C_n=(n-2)2^{n-3}$.
The induction hypothesis is that it holds for some $n$, which we have seen to be true for $n=2$:
\begin{align}
    A_2&=6=2\cdot3=2^{2-1}(2^{2}-1),\\
    B_2&=-4=1-5=1-5\cdot 2^{2-2},\\
    C_2&=0=(2-2)2^{2-3}.
\end{align}

Then we make the induction step; for an $(n+1)$-qubit flag, the two options yield the cost 
\begin{align}
    C_{\cnot}^{\text{flag,\ref{eq:flag_decomp_restate}}}(n+1,k)=&2\cdot (A_n 2^{k+1}+B_n-C_n\delta_{k+1,0})\nonumber\\
    &+\,(2^{k+n}-1)\\
    =&(4(2^{n-1}(2^n-1))+2^{n})2^k\nonumber\\
    &+\,(2(1-5\cdot 2^{n-2})-1)\nonumber\\
    =&2^{n}(2^{n+1}-1)2^k + (1-5\cdot 2^{n-1}),\nonumber\\
    C_{\cnot}^{\text{flag,\ref{eq:selective_de-mux_restate}}}(n+1,k)=&(A_n2^{k+1}+B_n-C_n\delta_{k+1,0}) \nonumber\\
    &+\,2(2^{k+n}-1)\nonumber\\
    &+\,2\cdot (A_n2^k+B_n-C_n\delta_{k0})\nonumber\\
    =&(4(2^{n-1}(2^n-1))+2^{n+1})2^k\nonumber\\
    &+\,(3(1-5\cdot 2^{n-2})-2) \nonumber\\
    &-\,2(n-2)2^{n-3}\delta_{k0}\nonumber\\
    =&2^{2n+1+k}+(1-15\cdot 2^{n-2})\nonumber\\
    &-\,(n-2)2^{n-2}\delta_{k0}.\nonumber
\end{align}
Here we used the induction hypothesis to fill in $A_n, B_n$ and $C_n$.
The second option has additional cost over the first of
\begin{align}
    2^{n+k}&-\left(\frac{15}{4}-\frac{5}{2}\right)2^n-(n-2)2^{n-2}\delta_{k0}\\
    =&\begin{cases}
        -\frac{n-1}{4}2^n & k=0\\
        2^n(2^k-\tfrac{5}{4}) & k>0
    \end{cases}\ ,
\end{align}
such that it is cheaper for $k=0$ and more expensive for $k>0$. This means that the minimized cost for the $(n+1)$-qubit flag is given by
\begin{align}
    2^{n}(2^{n+1}-1)2^k + (1-5\cdot 2^{n-1})-(n-1)2^{n-2}\delta_{k0},
\end{align}
which is exactly the statement of the induction for $n+1$.

We thus may conclude for all $n$ that the selective de-multiplexing step is favourable for non-multiplexed $n$-qubit flags, and the recursive flag decomposition/CSD step is preferable for multiplexed flags, with the final $\cnot$ cost
\begin{align}\label{eq:minimal_flag_cost}
    C_{\cnot}^{\text{flag}}(n)&=\frac{1}{2}4^n-\frac{n+12}{8}2^n+1,\\
    C_{\cnot}^{\text{mux}\text{-flag}}(n,k)&=\frac{1}{2}(4^n-2^n)2^k-\frac{5}{4} 2^n+1.
\end{align}

\subsection{Diagonal unitaries}\label{sec:gate_count_calculations:diagonal}
A diagonal unitary can be decomposed into a sequence of (multiplexed) $R_Z$ rotations and a global phase, denoted by a diamond-shaped gate, by iterating \cref{eq:balance_diagonal_circ}:
\begin{align}
\includegraphics{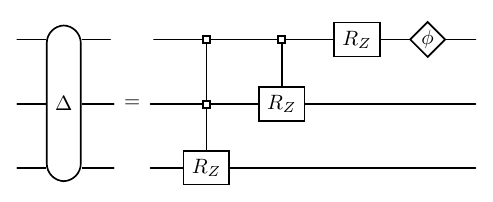}.
\end{align}
As we noted above, a $k$-multiplexed $R_Z$ rotation requires $2^k$ $\cnot$ gates, and for each of $1\leq k\leq n-1$ we get one multiplexer, so that the diagonal in total costs
\begin{align}
    C_{\cnot}^{\text{diag}}(n) = \sum_{k=1}^{n-1}2^k=2^n-2.
\end{align}

\subsection{Arbitrary unitaries}
We may now compute the $\cnot$ cost of a unitary.
At the first level, we can choose between no de-multiplexing (pure CSD, \cref{eq:flag_decomp_parametrized}), selective de-multiplexing (\cref{eq:selective_de-mux}), and full de-multiplexing (QSD, \cref{eq:parameter_optimal_qsd}).
This results in the cost
\begin{align}
    C_{\cnot}^{U,\text{CSD}}(n)=&2C_{\cnot}^{\text{mux}\text{-flag}}(n-1,1)\nonumber\\
    &+C_{\cnot}^{\text{mux}\text{-flag}}(1,n-1)+C^\text{diag}_{\cnot}(n)\nonumber\\
    =&\frac{1}{2}4^n-\frac{3}{4}2^n-1,\\
    C_{\cnot}^{U,\text{selective}}(n)=&2C_{\cnot}^{\text{flag}}(n-1)+2C_{\cnot}^{\text{sym-mux-Rot}}(n-1)\nonumber\\
    &+\,C_{\cnot}^{\text{mux}\text{-flag}}(n-1,1)+C^\text{diag}_{\cnot}(n)\nonumber\\
    =&\frac{1}{2}4^n-\frac{n+4}{8}2^n-1,\\
    C_{\cnot}^{U,\text{QSD}}(n)=&3C_{\cnot}^{\text{flag}}(n-1)+2C_{\cnot}^{\text{sym-mux-Rot}}(n-1)\nonumber\\
    &+\,C_{\cnot}^{\text{mux-Rot}}(n-1)+C_{\cnot}^{U}(n-1).
\end{align}

Here, we used the results of the previous sections and employed the reduction of the cost of multiplexed rotations by one where appropriate, according to \cref{sec:selective_de_multiplexing} and \cref{eq:po_qsd_cnot_trick} in particular. Note that the relations use the decomposition of (multiplexed) flags that come with constraints of $n\geq3$ ($n\geq2$).
For the QSD step, we do not have the explicit cost yet. However, we can compute the difference between the selective de-multiplexing and QSD, and by bounding $C_{\cnot}^U(n-1)$ using the cost with selective de-multiplexing itself, we find that QSD will perform better in general.
This tells us that unitaries should be de-multiplexed fully, which was not possible in a parameter-optimal way for flag circuits in the previous section.
For three qubits, we reproduce the best known $\cnot$ count of $19$ through our techniques:
\begin{align}
    C_{\cnot}^{U,\text{SDM}}(3)=3\cdot 2+2(2^2-1)+2^2+3=19,
\end{align}
where we used that a two-qubit flag (unitary) costs $2$ ($3$) $\cnot$s and that two of the multiplexed rotations can be implemented at reduced cost due to the trick in \cref{eq:po_qsd_cnot_trick}.

In order to generalize, we need to resolve the recursion relation
\begin{align}
    A(n)=&3C_{\cnot}^{\text{flag}}(n-1)+2C_{\cnot}^{\text{sym-mux-Rot}}(n-1)\nonumber\\
    &+\,C_{\cnot}^{\text{mux-Rot}}(n-1)+A(n-1)\\
    =&\frac{3}{8}4^n-\frac{3}{16}(n+3)2^n+1+A(n-1),
\end{align}
with the base case $A(3)=19$ for three qubits. The solution is
\begin{align}
    C_{\cnot}^{U,\text{SDM}}(n)=\frac{1}{2}4^n-\frac{3}{8}(n+2)2^n+n-1.
\end{align}
Although we derived it for $n\geq 3$, it turns out that this formula also conveniently reproduces the value for $n=2$, which is $3$, correctly.

\section{Details for MPS preparation}

\subsection{Causal ordering for practical synthesis}
\label{sec:mps_causal_ordering}

In order to compensate for the gauge degrees of freedom in \cref{subsec:mps_phase_gradient}, we set out to merge the $n$-qubit flag $\smallflag$ and diagonal $\Delta$ to the left and right, respectively. In particular, in \cref{eq:isometry_before_merging_left_right}, we want to perform 
\begin{equation*}
\resizebox{\linewidth}{!}{%
\includegraphics{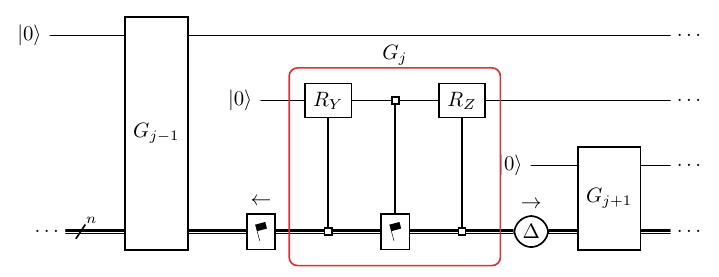} %
},
\end{equation*}
where the arrows above $\smallflag$ and $\Delta$ indicate the merge direction. However, this means that in order to synthesize $G_j$, we already need to have merged the flag circuit from $G_{j+1}$ into $G_j$.
In order to achieve this in practice, we need to be careful about the causal order in which we synthesize the matrices. We will walk through the synthesis step by step in the following.

First, we synthesize each matrix $G_j$ in an overparametrized fashion by performing a standard CSD:

\begin{equation}
\resizebox{\linewidth}{!}{%
\includegraphics{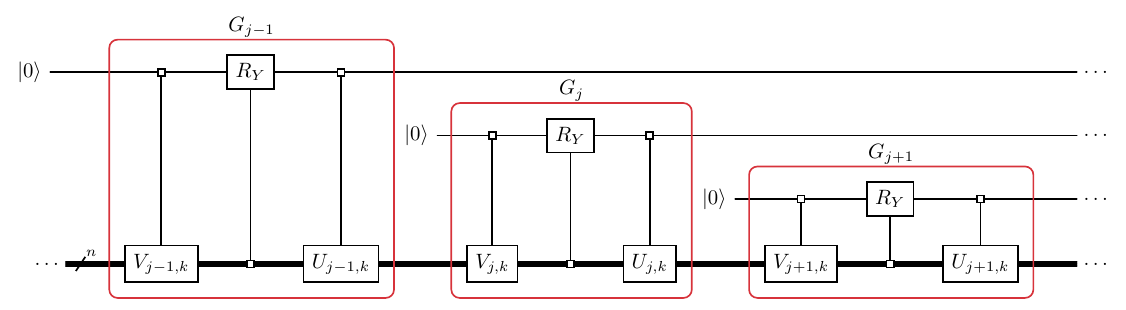}
}.
\end{equation}
The index $k\in\{0,1\}$ indicates the two possible values from the single-qubit multiplexing.
We then simply remove the multiplexer node attached to the $V$ matrices, flag-decompose the resulting $V = \Delta\smallflag$\footnote{Keep in mind that the ordering of matrix multiplication is reversed compared to the circuit diagrams.}, and merge the diagonal $\Delta$ into the multiplexed $U$ on the right, yielding $U':=U \cdot \Delta$:

\begin{equation}
\resizebox{\linewidth}{!}{%
\includegraphics{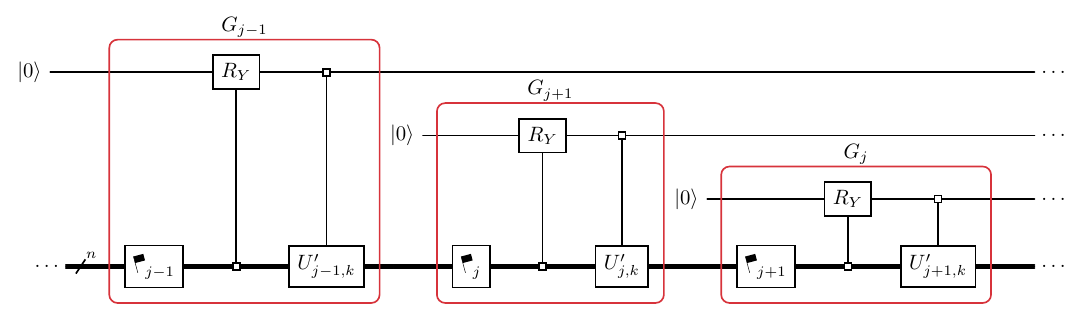}
}.
\end{equation}
In the next step, we merge all leading $\smallflag_j$ unitaries in the previous multiplexed $U_{j-1,k}$, yielding $U''_{j,k} := U_{j,k}\smallflag_{j+1}$:

\begin{equation}
\resizebox{\linewidth}{!}{%
\includegraphics{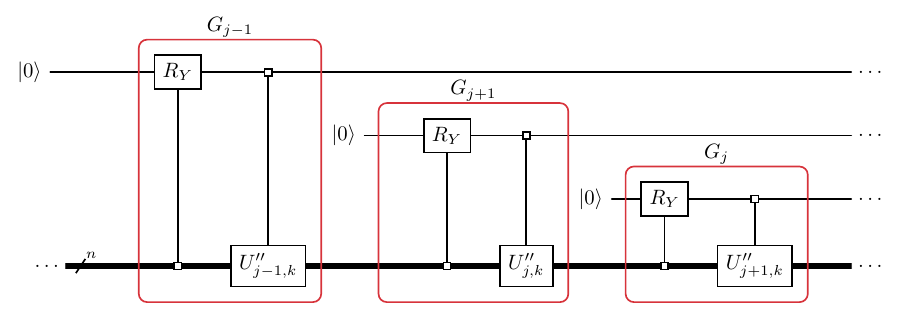}
}.
\end{equation}

Up until this point, everything could be performed in parallel. Now we need to be careful with the causal ordering, moving from left to right in the chain of isometries $G_j$. We flag-decompose the first multiplexed unitary $U_{0,k}'' = \Delta'' \smallflag''$. The diagonal $\Delta''$ is then decomposed into a multiplexed $R_Z$ and a smaller diagonal $\delta''$. This smaller diagonal is then merged with the multiplexed unitary $U_{1, k}''$ on the right. This is sequentially repeated going from left to right in the MPS: 

\begin{equation}
\resizebox{\linewidth}{!}{%
\includegraphics{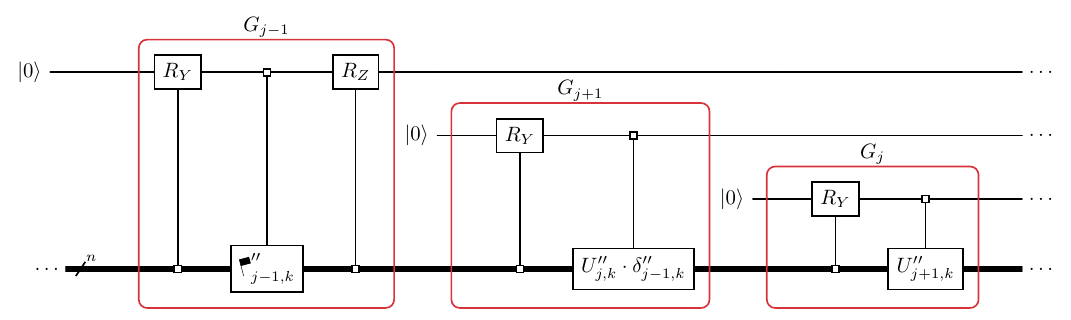}
}.
\end{equation}
Finally, we reach the desired circuit structure of \cref{eq:mps_phase_gradient_skeleton}, where each isometry is parameter-optimal,
\begin{equation}
\resizebox{\linewidth}{!}{%
\includegraphics{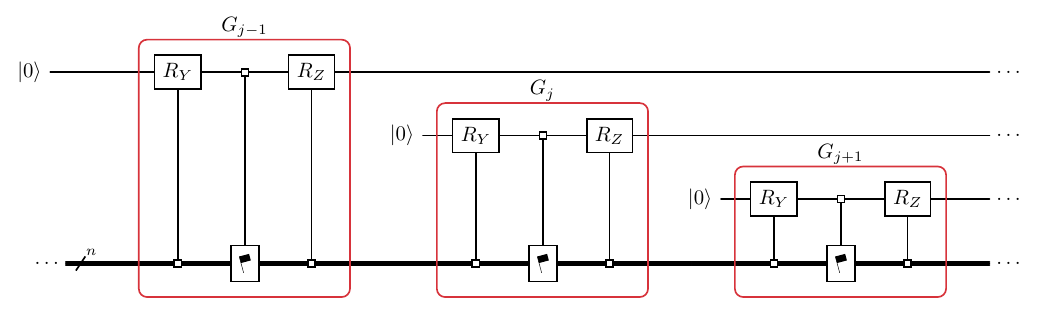}
}.
\end{equation}

\subsection{Boundary effect savings in MPS preparation}
\label{sec:boundary-effects}

The dimensions of MPS tensors at the boundary are growing in steps of $2$ at each site:

\begin{equation*}
\resizebox{\linewidth}{!}{%
\includegraphics{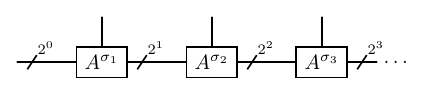}
}.
\end{equation*}
In general, the isometry $A^{\sigma_k}$ has dimensions $2^{k-1} \times 2 \times 2^k$, 
\begin{equation*}
\resizebox{.65\linewidth}{!}{%
\includegraphics{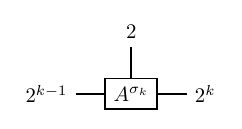}
}.
\end{equation*}
The corresponding unitary completion $G_k$ maps from $k-1$ qubits to $k+1$ qubits, where the reduced input is due to the reduced dimension $2^{k-1}$ for the left horizontal bond:
\begin{equation*}
\resizebox{.45\linewidth}{!}{%
\includegraphics{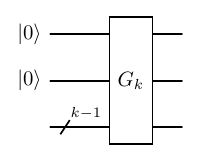}
}.
\end{equation*}
Note that this argument can be reversed for the right boundary, with the isometries being such that they produce $|0\rangle$ states.

Starting from the left boundary, we obtain a circuit structure where a new zeroed qubit is added at the top and then left untouched, and a new zeroed qubit is added for each isometry to the lower qubit register:
\begin{equation}
\resizebox{.9\linewidth}{!}{%
\includegraphics{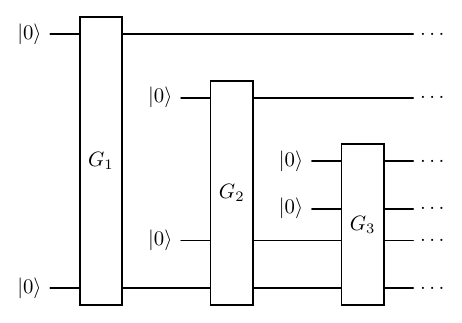}
}.
\end{equation}
This is continued until the maximal bond dimension $\chi = 2^n$ is reached. In particular, we go from $k=1$ until $n$.
The isometries $G_k$ at the boundary live in a Stiefel manifold of dimension $2 (2^{k-1}) \times (2^{k+1}) - (2^{k-1})^2 = \tfrac74 \ \chi_k^2$, where we define $\chi_k = 2^k$ as the local bond dimension. We now need to construct a circuit for $G_k$ that attains this optimal parameter bound. Additionally, we remove the extra $4^{k-1} = \tfrac1  4 \chi_k^2$ gauge degrees of freedom between site $k-1$ and $k$.
Thus, we overall need to attain the \textit{optimal} bound on the degrees of freedom (DOF),
\begin{equation}\label{eq:optimal_parameters_boundary}
    C_\text{DOF} = \tfrac64 \ \chi_k^2.
\end{equation}

Simply repeating the trick from \cref{eq:mps_phase_grad_reduction} does not suffice as it leads to $\tfrac{10}{4} \chi_k^2$ degrees of freedom. We need a better first decomposition to account for the \textbf{two} zeroed inputs. This is achieved by a CSD that allows for differently shaped blocks on either side of the central multiplexer. Concretely, we compute the decomposition 
\begin{equation}
G_k = 
\begin{pNiceArray}{cc|cc}[
    margin, 
    cell-space-limits = 3pt, 
    columns-width = .3cm      
]
  \Block{2-2}{K_{00}} & & \Block{2-2}{\mathbf{0}} & \\
   & & & \\
  \hline
  \Block{2-2}{\mathbf{0}} & & \Block{2-2}{K_{01}} & \\
   & & & 
\end{pNiceArray}
A
\begin{pNiceArray}{c|ccc}[
    margin, 
    columns-width = .15cm      
]
  \Block{1-1}{K_{10}} & \Block{1-3}{\mathbf{0}} & & \\
  \hline
  \Block{3-1}{\mathbf{0}} & \Block{3-3}{K_{11}} & & \\
   & & & \\
   & & & 
\end{pNiceArray}.
\end{equation}
The dimensions are chosen such that $K_{10} \in U(2^k)$, $K_{11} \in U(3\cdot 2^k)$ and $K_{0j} \in U(2\cdot 2^{k})$. The $A$ can be chosen to be an $R_Y$ multiplexer as in the regular CSD\footnote{Half of the angles in this multiplexer take a special value, but this will not be too relevant for us.}.
Now, choosing the two zeroed qubits to be the most significant bits, we write the input quantum state as $|00\rangle\otimes|\psi\rangle$ and find the asymmetrically multiplexed gate $K_{10}\oplus K_{11}$ to act on it as
\begin{align}
    (K_{10}\oplus K_{11})(|00\rangle\otimes|\psi\rangle)
    &=(K_{10}\oplus K_{11})(|\psi\rangle\oplus |0_{3\cdot 2^k}\rangle)\\
    &=|00\rangle\otimes (K_{10}|\psi\rangle).
\end{align}
That is, $K_{11}$ does not have any impact on the state, and we may as well change it, for example to $K_{10}^{\oplus 3}$. This replaces the asymmetric multiplexer by the gate $\id_4\otimes K_{10}$. 
\begin{align}
\resizebox{\linewidth}{!}{%
\includegraphics{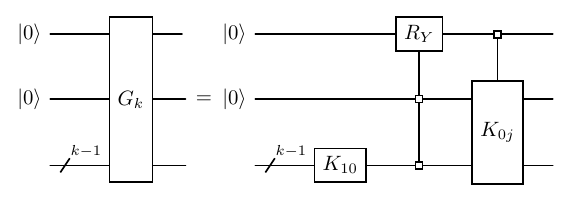}.%
}
\end{align}

At this stage, it is easy to see how we can remove the $4^{k-1} = \tfrac14 \chi_k^2$ gauge degrees of freedom: We simply merge the unitary of exactly this size with the unitary of the same size that is multiplexed on the top qubit (indicated by arrows):

\begin{equation*}
\resizebox{\linewidth}{!}{%
\includegraphics{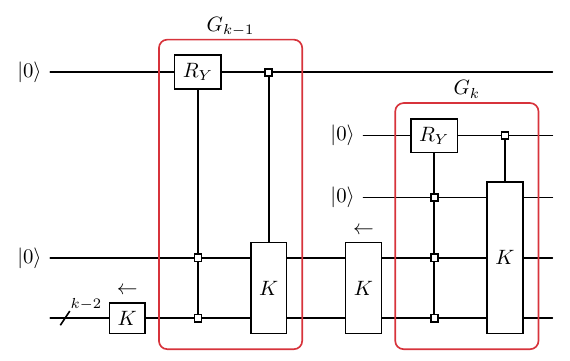}
}.
\end{equation*}
Note that in this diagram we left out the subscripts for the sake of brevity.

Next, we may use the second zeroed qubit to remove the corresponding multiplexing node of the $R_Y$ gate, to obtain

\begin{align}
\resizebox{.65\linewidth}{!}{%
\includegraphics{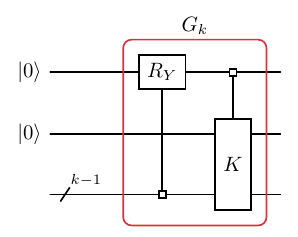}.%
}
\end{align}
The multiplexed unitary $K$ can be either de-multiplexed and then flag-decomposed when targeting the \{Clifford + Rot\} gate set, or directly flag-decomposed when targeting phase gradient decompositions. The steps are analogous to \cref{subsec:mps_clifford_rot} and \cref{subsec:mps_phase_gradient}, respectively. Let us walk through the former explicitly to make sure we have the right number of parameters.

In particular, after flag-decomposing the multiplexed $K_{0j}$ we obtain 
\begin{align}
\resizebox{\linewidth}{!}{%
\includegraphics{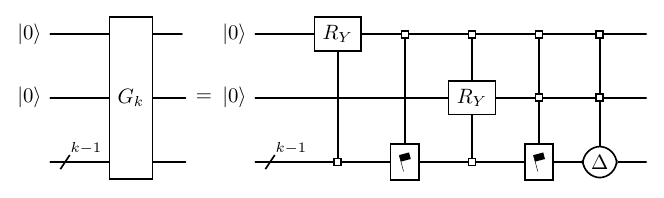}.%
}
\end{align}
Note that we have removed another multiplexer on the second qubit, as well as the leading $R_Z$ rotation due to the fixed input state (using the same argument as in \cref{eq:remove_RZ}).
We can then combine \cref{eq:balance_diagonal_circ,eq:MEP_flag2} to yield the identity
\begin{align}
\resizebox{\linewidth}{!}{%
\includegraphics{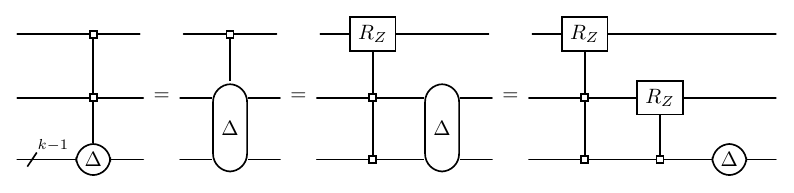}%
}.
\end{align}

With this we arrive at the final circuit for the boundary isometries

\begin{align}
\resizebox{\linewidth}{!}{%
\includegraphics{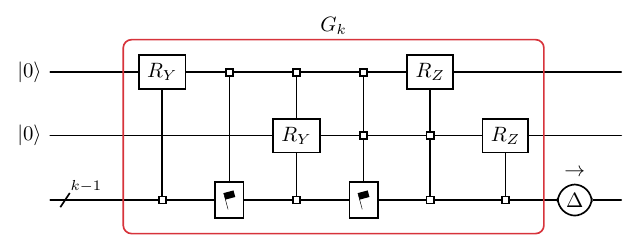},%
}
\end{align}
where the diagonal on the right can be again merged with the following isometry (as discussed in detail in the previous section).

The number of parameters in this circuit is composed of (from left to right)
\begin{align*}
C_\text{DOF} = & 2^{k-1} \text{ (multiplexed $R_Y$)} \\
&+ 2 \ (4^{k-1} - 2^{k-1}) \text{ (multiplexed flag)} \\
&+ 2^k \text{ (multiplexed $R_Y$)} \\
&+ 4 \ (4^{k-1} - 2^{k-1}) \text{ (multiplexed flag)} \\
&+ 2^k \text{ (multiplexed $R_Z$)} \\
&+ 2^{k-1} \text{ (multiplexed $R_Z$)} \\
=& \tfrac64 4^{k}= \tfrac64 \chi_k^2,
\end{align*}
matching the desired optimum from \cref{eq:optimal_parameters_boundary}.

\end{document}